\documentclass[%
 aip,
 amsmath,amssymb,
preprint,%
]{revtex4-2}

\usepackage{graphicx}
\usepackage{dcolumn}
\usepackage{bm}

\usepackage[utf8]{inputenc}
\usepackage[T1]{fontenc}
\usepackage{mathptmx}

\DeclareMathAlphabet\mathbfcal{OMS}{cmsy}{b}{n}
\usepackage{mathrsfs} 

\usepackage{feynmp-auto}

\begin{document}


\title{Plasma physics in strong-field regimes: theories and simulations}

\author{Yuan Shi}
\email{shi9@llnl.gov}
\affiliation{Lawrence Livermore National Laboratory, Livermore, California 94551, USA}

\author{Hong Qin}%
\affiliation{Princeton Plasma Physics Laboratory, Princeton University, Princeton, New Jersey 08543, USA}%
\affiliation{Department of Astrophysical Sciences, Princeton University, Princeton, New Jersey 08544, USA}%

\author{Nathaniel J. Fisch}
\affiliation{Princeton Plasma Physics Laboratory, Princeton University, Princeton, New Jersey 08543, USA}%
\affiliation{Department of Astrophysical Sciences, Princeton University, Princeton, New Jersey 08544, USA}%

\date{\today}

\begin{abstract}
In strong electromagnetic fields, unique plasma phenomena and applications emerge, whose description requires recently developed theories and simulations [Y. Shi, Ph.D. thesis, Princeton University (2018)]. 
In the classical regime, to quantify effects of strong magnetic fields on three-wave interactions, a convenient formula is derived
by solving the fluid model to the second order in general geometry.
As an application, magnetic resonances are exploited to mediate laser pulse compression, using which higher intensity pulses can be produced in wider frequency ranges, as confirmed by particle-in-cell simulations. 
In even stronger fields, relativistic-quantum effects become important, and a plasma model based on scalar quantum electrodynamics (QED) is developed, which unveils observable corrections to Faraday rotation and cyclotron absorption in strongly magnetized plasmas.
Beyond the perturbative regime, lattice QED is extended as a numerical tool for plasma physics, using which the transition from wakefield acceleration to electron-positron pair production is captured when laser intensity exceeds the Schwinger threshold.
\end{abstract}

\maketitle


\section{\label{sec:intro}Introduction}
Strong electromagnetic fields, which are known to exist near compact astrophysical bodies \cite{Becker2009x}, become increasingly accessible in laboratory due to the advance of high energy density (HED) laboratory drivers \cite{NRC03}.
In the astrophysical context, magnetic fields greater than \mbox{$10^{12}$ G} are found near neutron stars. In such strong fields, electron cyclotron energy \mbox{$\hbar|\Omega_e|\sim 10$ keV} is larger than the star's temperature and is nonnegligible compared to electron's rest energy \mbox{$m_ec^2\approx511$ keV}, so quantum and relativistic effects are both important. Describing dynamical  processes near neutron stars and relating them to observable signals remains a challenging task \cite{Burnell2017past}, and will likely require further developments of theoretical models and numerical tools. 
Although the physics is less extreme in \mbox{$10^6$-$10^9$ G} magnetic fields that are available in laboratory, 
modeling magnetization effects in HED plasma remains a largely open research field. 
In addition to modifying hydrodynamics, radiation, and transport properties \cite{Jennings2010simulations,Thornhill15}, strong magnetic fields also changes laser-plasma interactions \cite{Shi2018laser}. Understanding these phenomena is crucial for magneto-inertial fusion concepts \cite{Wurden2016magneto}, which strive to combine both confinement approaches to achieve ignition. 
Apart from quasi-static fields, transient strong fields are available in laboratory during beam-laser \cite{Bula96,Baumann2019probing,Meuren2020seminal} and beam-beam \cite{Yakimenko2019prospect} collisions. Boosted by the high energy of an electron beam, the field of a tightly-focused laser or another highly-charged bunch may approach the Schwinger field \mbox{$E_s=m_e^2c^3/(e\hbar)\sim10^{18}$ V/m}, where quantum electrodynamics (QED) becomes nonperturbative. Studying processes in this regime may shed light on other nonperturbative interactions like the nuclear force, and may also unveil physics beyond the Standard Model.

In this invited paper, we review recent developments of theories and simulations related to the Marshall N. Rosenbluth Outstanding Doctoral Thesis Award\cite{Shi_thesis18}. 
We focus on elucidating the key physics and bring additional insights beyond what is contained in previous publications where detailed results can be found. This paper is organized as follows.  
In Sec.~\ref{sec:3wave}, effects of a strong magnetic field on coherent three-wave interactions are discussed in the classical regime.
Starting from MG fields, scattering of lasers becomes manifestly anisotropic (Sec.~\ref{sec:3wave_intro}). A convenient formula for the three-wave coupling coefficient is derived from the second-order solution of the fluid model in general geometry (Sec.~\ref{sec:3wave_coupling}). Using the formula, special angles are unveiled where laser scattering is either enhanced or suppressed due to magnetization (Sec.~\ref{sec:3wave_example}). For a given coupling coefficient, the three-wave interaction problem is intrinsically quantum, which is shown to be solvable using quantum computers (Sec.~\ref{sec:3wave_quantum}).
As an application, magnetized three-wave interactions are exploited to improve plasma-mediated laser pulse compression (Sec.~\ref{sec:3wave_comp}). Using the additional degrees of control, more intense pulses can be generated in wider frequency ranges, as corroborated by particle-in-cell (PIC) simulations. 
In Sec.~\ref{sec:QED}, to describe plasmas beyond the classical regime, scalar QED is extended to include nontrivial dynamical background fields, which capture effects of collective plasma dynamics (Sec.~\ref{sec:QED_BK}). 
As an example, the background field theory is used to study plasma waves by computing their 1-loop effective action (Sec.~\ref{sec:QED_action}), from which the QED-modified dispersion relation is extracted (Sec.~\ref{sec:QED_disp}). Applying the formulas to neutron-star conditions, anharmonic cyclotron absorption and anomalous Faraday rotation are unveiled, which may also be observable in laboratory.
Beyond the perturbative regime, lattice QED is extended into a plasma simulation tool (Sec.~\ref{sec:lattice}), which is uniquely suited when collective scales overlap with QED scales. A variational algorithm is developed and applied to simulate laser-plasma interactions (Sec.~\ref{sec:lattice_example}), and the transition from wakefield acceleration to electron-positron pair production is demonstrated when the laser intensity exceed the Schwinger threshold.

\section{\label{sec:3wave}Magnetized three-wave interactions}
Coherent three-wave coupling is the lowest-order nonlinear interaction in plasmas \cite{Davidson2012methods}. The presence of a background magnetic field substantially proliferates the ways in which the resonance condition $k_1^\mu=k_2^\mu+k_3^\mu$ can be satisfied, where $k_i^\mu=(\omega_i,\mathbf{k}_i)$ is the four-momentum of the $i$-th participating mode.
In an unmagnetized two-species plasma, although there are only three eigenmodes --- the gapped degenerate electromagnetic (EM) waves, the gapped Langmuir wave (P), and the gapless acoustic wave (S) --- the possible combinations are already numerous \cite{Montgomery2016two}. 
Once the plasma becomes magnetized, there are many more eigenmodes available for resonant nonlinear interactions. For example, in magnetized two-species warm-fluid plasma, there are six linear eigenmodes, including two gapped EM waves, which are now nondegenerate, one gapped hybrid wave, and three gapless waves, which become the three magnetohydrodynamics (MHD) waves in the long wavelength limit. Moreover, in magnetized kinetic plasmas, there are infinite ladders of Bernstein waves, and the possible number of resonant interactions explodes. 
In order to simplify the analysis, previous attempts in the literature were usually restricted to some specific wave triads, and focused on the special geometry where waves propagate either parallel or perpendicular to the background magnetic field \cite{Stenflo1972kinetic,Grebogi1980brillouin,Barr1984raman,Wong1986parametric,Vinas1991parametric,Brodin2012three}. 
However, as we will discuss below, the physics that governs magnetized three-wave interactions is not as complicated as it seems and is in fact generally applicable to arbitrary geometry. Once we understand the underlying processes, we can exploit them to benefit applications such as plasma-mediated laser pulse compression.

\subsection{\label{sec:3wave_intro}Effects of magnetization}
Before introducing any equation, it is helpful to point out qualitative ways in which magnetization affects the strength of nonlinear three-wave coupling. 
First, magnetization changes the wave dispersion relation, and therefore the kinematics of resonant interactions. Since the coupling, which can be related to the scattering cross section, depends on the phase space volume, kinematic changes alone could affect the strength of the coupling. 
Second, magnetization changes the polarization of eigenmodes, and therefore the overlap of incoming and outgoing wave states. Since the coupling depends on the scattering matrix, its strength is affected by changes in polarization, which in general has both longitudinal and transverse components when magnetized.
Finally, and perhaps most importantly, magnetization changes the energy partition between the field and particle degrees of freedom. Since three-wave coupling is nonvanishing only inside changed medium, the response of the medium is what determines the strength of the interactions. Compared to the unmagnetized response, where wave energy is only distributed between the oscillating EM fields and the quivering charged particles, in the magnetized case, gyro motion also carries energy. Therefore, when waves couple via gyrating particles, energy can now flow through additional channels. Since three-wave interaction is a phase-sensitive process, the interference between different channels determines the overall strength of the interaction, which can either be enhanced or suppressed compared to unmagnetized cases.

A manifestation of magnetization effects is the anisotropy of the interaction strength in the absence of any gradient, because the background magnetic field already defines a special direction. In the unmagnetized case, backscattering, where the two smaller wave vectors are antiparallel, usually dominate side and forward scattering. However, this is no longer the case when the plasma becomes magnetized, in which the coupling has intricate angular dependence. As a rule of thumb, magnetized coupling is comparable to its unmagnetized value, with the possibility of being a few times larger, except at special angles where the following three mechanisms suppress the coupling \cite{Shi2018laser}. 
First, the coupling may be polarization suppressed, which occurs when the incoming and outgoing wave states have poor overlap so that every interaction channel is weak. 
Second, the coupling may be interference suppressed, which occurs when different interaction channels, albeit strong individually, destructively interfere to cancel the overall scattering. 
Third, the coupling may be energy suppressed, which occurs when the energy participation radio is small in the sense that the wave energy is largely carried by degrees of freedom that provide little coupling.
%

\subsection{\label{sec:3wave_coupling}Computing coupling coefficient}
To compute the three-wave coupling coefficient in the weak-coupling regime, a systematic approach is to solve the plasma model to the second order in the perturbation series \cite{Shi2017three}. This perturbative picture respects the permutation symmetry between the participating waves, and allows profound simplifications that are obscured in the parametric picture, in which a pump wave is singled out on whose background the nonlinear dispersion relation of decay products is calculated. In either picture, the lowest-order solutions are linear waves. Suppose the first-order electric field is expanded by $\mathbf{E}_1=\frac{1}{2}\sum_{\mathbf{k}\in\mathbb{K}_1}\mathbfcal{E}_{1,\mathbf{k}} e^{i\theta_{\mathbf{k}}}$, 
where the summation is over a discrete spectrum $\mathbb{K}_1$ and $\theta_{\mathbf{k}}=\mathbf{k}\cdot\mathbf{x}-\omega_\mathbf{k} t$ is the wave phase, then the Fourier amplitude satisfies the matrix equation
\begin{equation}
    \label{eq:DE}
    \mathbb{D}_\mathbf{k} \mathbfcal{E}_{1,\mathbf{k}}=\mathbf{0}.
\end{equation}
In order for this equation to have nontrivial solutions, $\det\mathbb{D}=0$ must be degenerate, from which the linear dispersion relation $\omega=\omega(\mathbf{k})$ and the unit polarization vector $\mathbf{e}=\mathbf{e}(\mathbf{k})$, with $\mathbfcal{E}=\mathcal{E}\mathbf{e}$ and $\mathbf{e}^\dagger\mathbf{e}=1$, can be found. 
The key quantity for linear waves is therefore the dispersion tensor
\begin{equation}
	\label{eq:Dtensor}
	\mathbb{D}_\mathbf{k}^{ij}=(\omega_\mathbf{k}^2-c^2\mathbf{k}^2)\delta^{ij}+c^2k^i k^j-\sum_s\omega_{ps}^2\mathbb{F}_{s,\mathbf{k}}^{ij}.
\end{equation}
Here, $\delta^{ij}$ is the Kronecker delta, $\omega_{ps}^2=e_s^2n_{s0}/\epsilon_0m_s$ is the plasma frequency of species $s$, and the forcing operator $\mathbb{F}$ is related to the linear susceptibility $\chi$ by $\omega_\mathbf{k}^2 \chi_{s,\mathbf{k}}=-\omega_{ps}^2\mathbb{F}_{s,\mathbf{k}}$. While $\chi$ is more frequently used for linear waves, $\mathbb{F}$ is more useful for nonlinear interactions, because it captures the response of individual particles to the wave electric field. Notice that in a cold unmagnetized plasma, $\mathbb{F}=\mathbb{I}$ is simply the identity operator. Therefore, magnetization and thermal effects are manifested by a forcing operator that deviates from unity.

In the perturbative picture, the second-order electric-field equation also has an elegant form. Due to nonlinearities in plasma models, the usual perturbation series, in which only fields are expanded, is plagued by secular terms that grow indefinitely instead of oscillating with bounded amplitudes. To remove this pathological behavior, multi-scale perturbative expansion can be used, which additionally expands spacetime $x^\mu=x^\mu_{(0)}+x^\mu_{(1)}/\epsilon +x^\mu_{(2)}/\epsilon^2+\dots$. Here, the small parameter $\epsilon$ is the same auxiliary parameter that is used to expand, for example, the electric field $\mathbf{E}=\mathbf{E}_0+\epsilon \mathbf{E}_1+\epsilon^2\mathbf{E}_2+\dots$, and $x^\mu_{(n+1)}$ is the scale slower than $x^\mu_{(n)}$ by a factor of $\epsilon$. The scale separation is assumed to be sufficiently large such that one can treat $\partial_\nu^{(a)}x^\mu_{(b)}=\delta_\nu^\mu\delta^{(a)}_{(b)}$. The multi-scale expansion technique has an intuitive interpretation: Assuming nonlinearities are weak, then linear waves maintain their dispersion relation and polarization, but their amplitudes slowly evolve. After tedious but otherwise rigorous manipulation of the equations, the second-order electric-field equation can be put into a simple form \cite{Shi2019three}
\begin{eqnarray}
\label{eq:E2eq}
\sum_{\mathbf{k}\in\mathbb{K}_2}\mathbb{D}_\mathbf{k} \mathbfcal{E}_{2,\mathbf{k}} e^{i\theta_\mathbf{k}}
+i\sum_{\mathbf{k}\in\mathbb{K}_1}
\Big(\frac{\partial\mathbb{D}_\mathbf{k}}{\partial\omega_\mathbf{k}}\partial_{t}^{(1)} 
-\frac{\partial \mathbb{D}_\mathbf{k}}{\partial\mathbf{k}}\cdot\nabla^{(1)}\Big)\mathbfcal{E}_{1,\mathbf{k}}e^{i\theta_\mathbf{k}}
=\frac{i}{2}\sum_{\mathbf{p},\mathbf{q}\in\mathbb{K}_1}\mathbf{S}_{\mathbf{p},\mathbf{q}} e^{i\theta_\mathbf{p}+i\theta_\mathbf{q}}.
\end{eqnarray}
The first term has the same structure as linear waves, except that the dispersion tensor $\mathbb{D}_\mathbf{k}$ is nondegenerate for second-order Fourier amplitude $\mathbf{E}_2=\frac{1}{2}\sum_{\mathbf{k}\in\mathbb{K}_2}\mathbfcal{E}_{2,\mathbf{k}} e^{i\theta_{\mathbf{k}}}$, which has a different spectrum $\mathbb{K}_2$. 
The second term captures the slow advection of wave envelope $\mathbfcal{E}_{1,\mathbf{k}}$ on the $x^\mu_{(1)}$ scale.
Notice that the advection transports wave energy, which is manifested by the appearance of the linear-wave Hamiltonian $\mathbb{H}=\frac{1}{\omega}\frac{\partial\mathbb{D}}{\partial\omega}$. 
Also notice that the advection projects out changes of the polarization vector and preserves $\mathbb{D}\mathbfcal{E}_1=\mathbf{0}$, because $d\mathbb{D}/d\mathbf{k}=\mathbf{v}_g \partial\mathbb{D}/\partial\omega+\partial\mathbb{D}/\partial\mathbf{k}$, where $\mathbf{v}_g=\partial\omega/\partial\mathbf{k}$ is the wave group velocity.
Finally, the term on the right-hand side (RHS) is the scattering strength, which encapsulates key physics of nonlinear three-wave interactions.

While the vector form of the scattering strength may look nonintuitive, 
its scalar form is much more insightful. To obtain the scalar equations, which are known as the three-wave equations \cite{Davidson2012methods}, it is important to recognize the exact action conservation law
$\mathbfcal{E}_{1}\cdot\mathbf{S}_{\bar{2},\bar{3}}/\omega_{1}^2 
=\mathbfcal{E}_{\bar{2}}\cdot\mathbf{S}_{\bar{3},1}/\omega_{2}^2
=\mathbfcal{E}_{\bar{3}}\cdot\mathbf{S}_{1,\bar{2}}/\omega_{3}^2$
when the three waves satisfy the resonance condition $k_1^\mu=k_2^\mu+k_3^\mu$. Here, we have used the notation $\bar{k}^\mu=-k^\mu$ for four-momentum and $\bar{\mathbf{Z}}=\mathbf{Z}^\dagger$ for complex vectors.
The action conservation law can be proven rigorously using the nontrivial identity \cite{Shi2017three}  
$(\omega_2-\omega_1)\mathbb{F}_1\mathbb{F}_2=\omega_2\mathbb{F}_1-\omega_1\mathbb{F}_2$ for the cold-fluid forcing operator.
Then, multiplying $\bar{\mathbfcal{E}}$ on both sides of Eq.~(\ref{eq:E2eq}), resonant wave triads satisfy the canonical form of three-wave equations: $d_t a_1=-\Gamma a_2 a_3/\omega_1$, $d_t a_2=\Gamma^* a_1 a_3^\dagger/\omega_2$, and $d_t a_3=\Gamma^* a_1 a_2^\dagger/\omega_3$, where $d=\partial_t+\mathbf{v}_g \cdot \nabla$ is the convective derivative at respective wave group velocities, $a=e\mathcal{E}u^{1/2}/(m_e c\omega)$ is the normalized scalar wave amplitude, 
and the coupling coefficient is given by the concise formula \cite{Shi2017three}
\begin{equation}
\label{eq:coupling}
\Gamma=\sum_s\frac{Z_s\omega_{ps}^2\Theta^s}{4M_s(u_1u_2u_3)^{1/2}}.
\end{equation}
Here, $Z_s=e_s/e$ and $M_s=m_s/m_e$ are the normalized charge and mass of species $s$, and $u=\frac{1}{2}\mathbf{e}^\dagger \mathbb{H}\mathbf{e}$ is the dimensionless wave energy coefficient. The scalar electromagnetic scattering strength $\Theta^s$ contains six permutations $\Theta^s=\Theta^s_{1,\bar{2}\bar{3}}+\Theta^s_{\bar{2},\bar{3}1}+\Theta^s_{\bar{3},1\bar{2}}+\Theta^s_{1,\bar{3}\bar{2}}+\Theta^s_{\bar{3},\bar{2}1}+\Theta^s_{\bar{2},1\bar{3}}$, which corresponds to the six ways of performing Wick contractions when computing the Feynman diagram 
\begin{fmffile}{w3}
	\begin{eqnarray}
	\begin{gathered}
	\begin{fmfgraph*}(40,40)
	\fmfkeep{w3}
	\fmfleft{i1}
	\fmfright{o2,o3}
	\fmf{photon}{i1,v1}
	\fmf{photon}{v2,o2}
	\fmf{photon}{v3,o3}
	\fmf{fermion}{v1,v3}
	\fmf{plain}{v1,v2}
	\fmfdot{v2,v3}
	\fmfv{label=$1$,label.angle=-120,label.dist=6}{v1}
	\fmfv{label=$2$,label.angle=-120,label.dist=6}{v2}
	\fmfv{label=$3$,label.angle=120,label.dist=6}{v3}
	\end{fmfgraph*}
	\end{gathered}
	&=& i\frac{e_s\omega_{ps}^2}{2m_sc}\Theta_{1,\bar{2}\bar{3}}^s, \\
	\Theta_{i,jl}^s&=&\frac{1}{\omega_j}(c\mathbf{k}_i\cdot\mathbf{f}_{s,j})(\mathbf{e}_i\cdot\mathbf{f}_{s,l}).	
	\end{eqnarray}
\end{fmffile}In the above formula, $\mathbf{f}_{s,j}=\mathbb{F}_{s,j}\mathbf{e}_j$, and the Feynman diagrams arise from the interaction Hamiltonian $H_{I, s}=P_s^i(\partial_iA_l)J^l_s$ due to the second-order response of species $s$, where $\mathbf{P}$ is the displacement operator, $\mathbf{A}$ is the vector potential, and $\mathbf{J}$ is the current operator. Notice that the strength of each scattering channel is complex, and the total strength $\Theta^s$ is determined by the interference between all six channels. 
In warm-fluid plasma, additional scattering channels arise due to thermal nonlinearities, and the pure electromagnetic scattering $\Theta^s$ is replaced by $\Theta^s\rightarrow\Theta^s+\Phi^s$, where $\Phi^s$ arises entirely from fluid nonlinearities \cite{Shi2019three}. Since $\Phi^s\propto u_s^2/c^2$, where $u_s$ is the thermal speed of species $s$, the correction is usually negligible at nonrelativistic temperature.
The above formula is of course also applicable to unmagnetized cases: For cold Raman scattering, the species-agnostics $\Theta^s=-\frac{ck_3}{\omega_3}(\mathbf{e}_2^\dagger\cdot\mathbf{e}_1)$, where waves ``1'' and ``2'' are transverse EM waves and wave ``3'' is the Langmuir wave. Similarly, for warm Brillouin scattering, $\Theta^s+\Phi^s=\frac{ck_3}{\omega_3}\hat{\gamma}_{s,3}^2 (\mathbf{e}_2^\dagger\cdot\mathbf{e}_1)$, where ``3'' is now the acoustic wave and $\hat{\gamma}_{s,3}^2=1/(1-\hat{\beta}_{s,3}^2)$ with $\hat{\beta}_{s,3}^2=u_s^2k_3^2/\omega_3^2$.
Notice that unmagnetized couplings are zero when the EM modes $\mathbf{e}_1$ and $\mathbf{e}_2$ are orthogonal. In contrast, magnetized couplings are usually nonzero for orthogonal modes because what enters the formula is $\mathbf{f}_{s,j}$ instead of $\mathbf{e}_{j}$, which is due to the fact that gyration introduces additional velocity components to charged particles. 

\begin{figure}[t]
	\centering
	\includegraphics[width=0.48\textwidth]{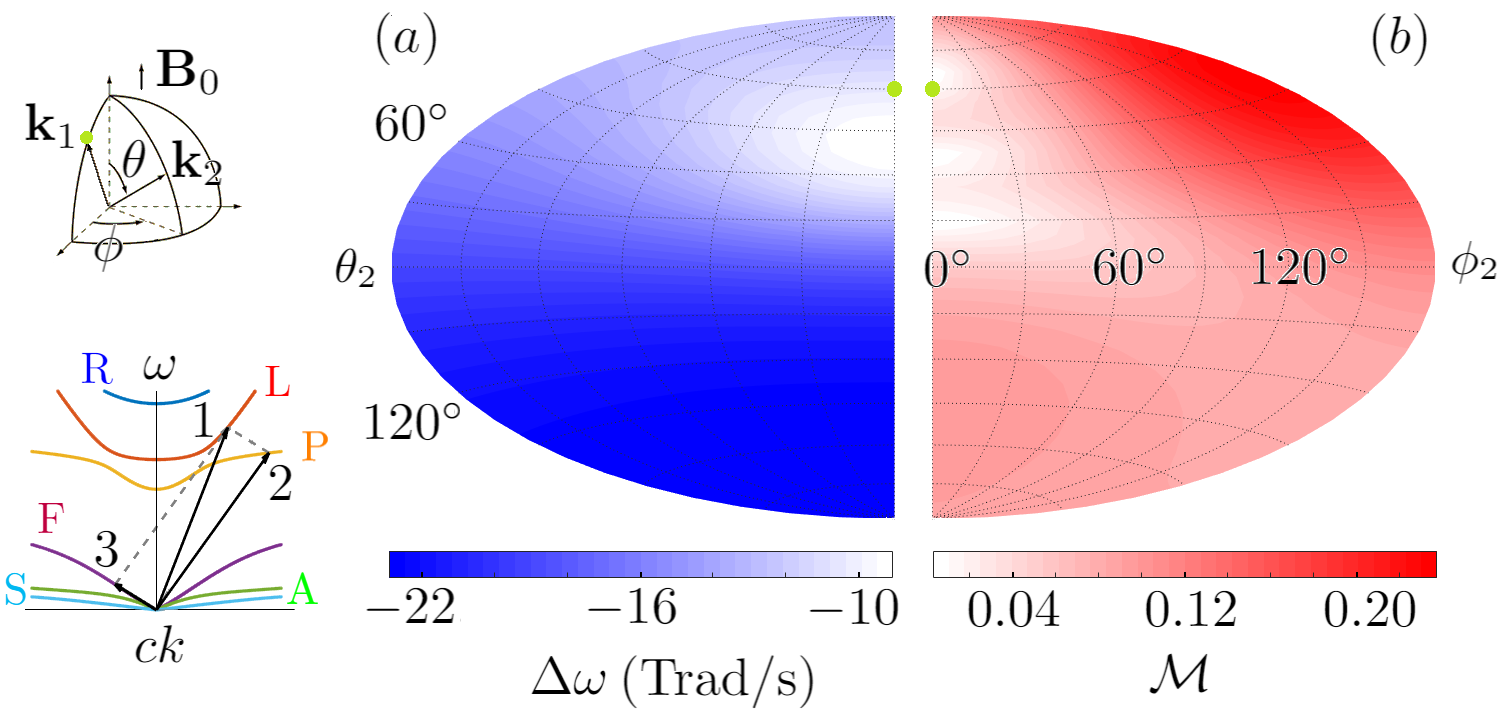}
	\caption{Frequency downshift (a) and normalized growth rate (b) when an L wave ($\omega_{pe}/\omega_1\approx0.75$) decays to P and F daughter waves (insets) in a plasma with $|\Omega_e|/\omega_{pe}\approx 0.8$ and $v_A/c_s\approx 4$, where $v_A$ is the Alfv\'en speed and $c_s$ is the sound speed.  
	The L wave propagates at $\langle\mathbf{k}_1,\mathbf{B}_0\rangle=30^\circ$, while the P wave propagates at polar angle $\theta_2$ and azimuthal angle $\phi_2$. Notice that backscattering is not the strongest, and special angles exist where the coupling is zero.
	[Y. Shi, Phys. Rev. E \textbf{99}, 063212 (2019).]
	}
	\label{fig:map}
\end{figure}

\subsection{\label{sec:3wave_example}Examples at oblique angles}
The coupling coefficient can be evaluated at arbitrary angles of wave propagation once the three resonant waves are specified. The inputs for the general formula [Eq.~(\ref{eq:coupling})] are the wave frequencies, the wave vectors, and the wave polarization vectors. In a given coordinate system, one can then evaluate the forcing operator, which is proportional to the linear susceptibility tensor, from which the scattering strength $\Theta^s$ and the wave energy coefficients $u$ can be determined. 
The only demanding step during the evaluation is matching the resonance condition, which requires numerical root finding. Consider scattering experiments, where the pump frequency $\omega_1$, the pump direction $\hat{\mathbf{k}}_1$, and the probe direction $\hat{\mathbf{k}}_2$ are given by the experimental setup. To determine the kinematics, we first solve for the pump wave vector $k_1$, such that the wave dispersion relation  $\omega_1=\omega_1(k_1\hat{\mathbf{k}}_1)$ is satisfied. Second, we match the resonance condition by numerically solving for $k_2$ from the equation $\omega_1=\omega_2(k_2\hat{\mathbf{k}}_2)+\omega_3(k_1\hat{\mathbf{k}}_1-k_2\hat{\mathbf{k}}_2)$. 
Finally, we compute the unit polarization vectors by solving the degenerate matrix equation $\mathbb{D}\mathbf{e}=\mathbf{0}$. Notice that the complex phase of $\mathbf{e}$ is indeterminate, so is the phase of the coupling coefficient $\Gamma$. Nevertheless, only the relative phases between the three waves are of physical importance, and the relative phase is controlled by the experimental setup.

From the coupling coefficient, one can compute observable quantities such as the linear growth rate. Assuming pump depletion is negligible, then a resonant seed wave grows exponentially at a rate $\gamma_0=|\Gamma a_1|/\sqrt{\omega_2\omega_3}$. This expression can be symmetrized by normalizing the growth rate with that of unmagnetized Raman backscattering $\gamma_R=\sqrt{\omega_1\omega_p}|a_1|/2$. Then, $\gamma_0=\gamma_R\mathcal{M}$, where the dimensionless $\mathcal{M}$ is given by
\begin{equation}
\label{eq:growthM}
\mathcal{M}=2\frac{|\Gamma|}{\omega_p^2}\Big(\frac{\omega_p^3}{\omega_1\omega_2\omega_3}\Big)^{1/2}.
\end{equation} 
In the absence of damping, $\mathcal{M}$ is invariant when scaling \mbox{$\omega\rightarrow\xi\omega$} and \mbox{$\mathbf{k}\rightarrow\xi\mathbf{k}$}, if we also scale the plasma density by \mbox{$n_s \rightarrow \xi^2 n_s$} and scale the magnetic field by \mbox{$B_0\rightarrow\xi B_0$} while keeping the plasma temperature constant \cite{Shi2019amplification}. 
As an example, consider a pump laser with $\omega_1=75$ Trad/s and propagating at $\langle\mathbf{k}_1,\mathbf{B}_0\rangle=30^\circ$ in the left-handed elliptically polarized eigenmode (L). One scattering mode [Fig.~\ref{fig:map}, inset] is that the laser excites an upper-hybrid-like plasma wave (P) and decay into the whistler-like fast wave (F). 
The frequency downshift $\Delta\omega=\omega_2-\omega_1$ and the normalized growth rate $\mathcal{M}$ are shown in Fig.~\ref{fig:map}, where the magnetic field is \mbox{$B_0=2.5$ MG}, the plasma density is $n_e=n_i=10^{18} \;\text{cm}^{-3}$, the plasma temperature is $T_e=T_i=3.2$ keV, and the warm-fluid polytropic index is adiabatic. Here, the mass ratio $m_i/m_e=5$ is artificial such that all frequencies are on the same scale. For realistic ion masses, numerical root finding may take additional iterations to converge, but there is no additional difficulty when evaluating the formulas.

\subsection{\label{sec:3wave_quantum}Simulation on quantum computer}
That the coupling coefficient can be computed using Feynman diagrams hints at the quantum origin of three-wave interactions.
Although plasmas are usually considered classical, they possess the same cubic nonlinearity that is known to give rise to nonlinear quantum optical phenomena \cite{Loudon2000quantum}. In quantum optics, the degree of quantumness, which can be measured by the extent to which wave correlations violate the classical Cauchy–Schwarz inequality, increases as the flux decreases towards the single photon limit. Although the low-flux limit is likely overwhelmed by background noise in hot plasmas, the cubic interaction remains an intrinsically quantum process.
In the quantum version, the complex wave amplitudes are promoted to operators $\sqrt{\omega_i}a_i\rightarrow \hat{A}_i$ that satisfy the canonical commutation relations $[\hat{A}_i(t), \hat{A}_j^\dagger(t')]=\delta_{ij}\delta(t-t')$, where we have used the units $\hbar=1$. The three-wave equations are then the Heisenberg equations $d_t\hat{A} = i[\hat{H}, \hat{A}]$ with a cubic Hamiltonian $\hat{H} = \int dt(ig^*\hat{A}_1\hat{A}_2^\dagger \hat{A}_3^\dagger-ig \hat{A}_1^\dagger \hat{A}_2 \hat{A}_3)$, where $g=\Gamma/\sqrt{\omega_1\omega_2\omega_3}$.
The quantum problem approaches the classical problem in the limit of large photon numbers.

\begin{figure}[t]
	\begin{center}
		\includegraphics[width=0.48\textwidth]{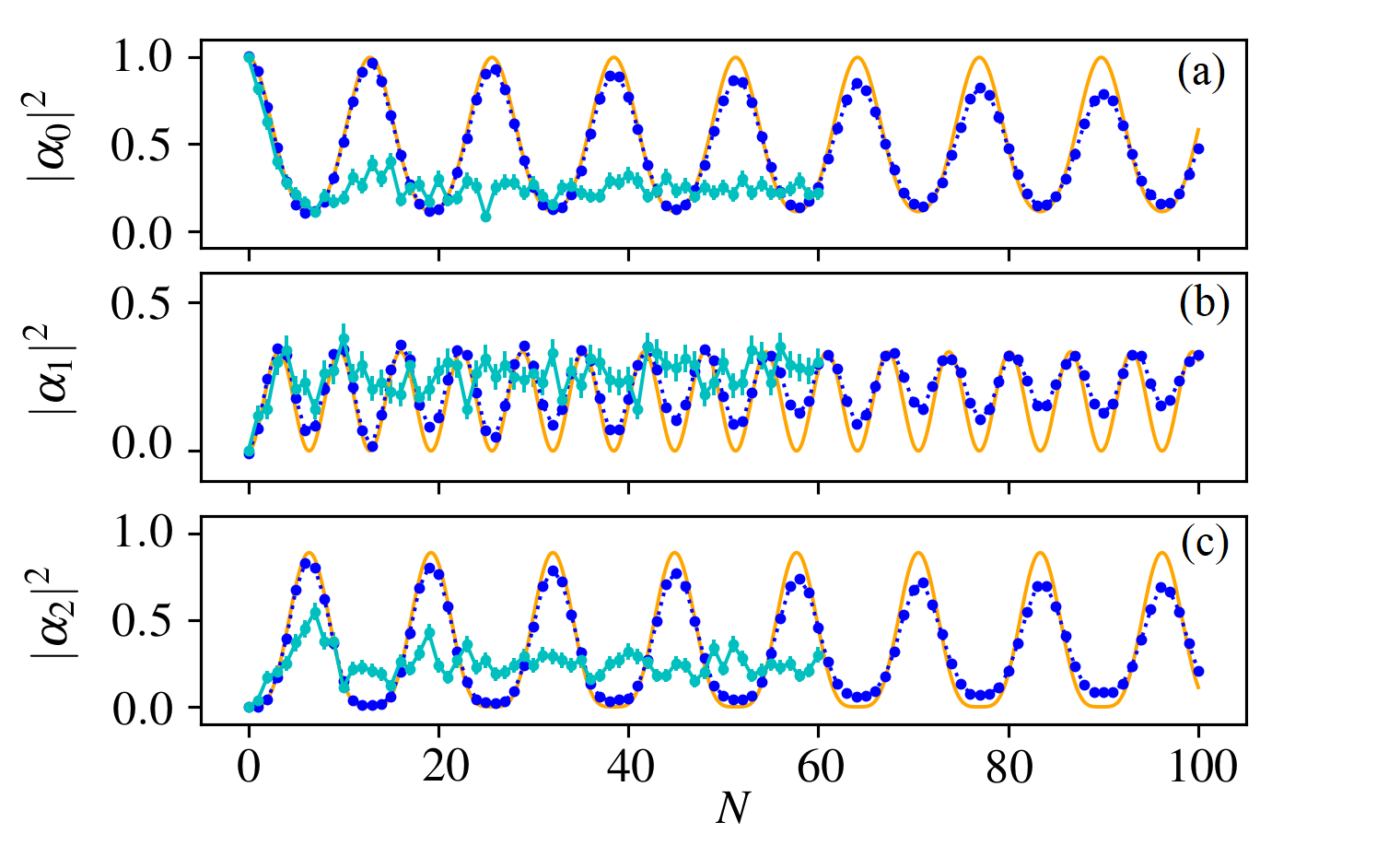} 
		\caption{Occupations of basis states after $N$ simulation steps when solving the quantum three-wave problem in a three-dimensional block. When $U$ is compiled as a sequence of standard gates (cyan), results follow the exact solutions (orange) up to $N\lesssim10$. The quantum coherence is more efficiently utilized when $U$ is compiled as a single customized gate (blue), using which $N\gtrsim100$ steps can be carried out before results are corrupted by noise. [Adapted from Y. Shi \textit{et al.}, arXiv:2004.06885 (2020).]}
		\label{fig:Results}
	\end{center}
\end{figure}

It turns out that quantum computers can be programmed to simulate the three-wave interaction problem \cite{Shi2020quantum}, a first-of-the-kind example that plasma physics may benefit from advances in quantum information science. 
To map the problem to quantum computers, a convenient basis is spanned by the simultaneous eigenspaces of the action operators $\hat{S}_2=\hat{n}_1+ \hat{n}_3$ and $\hat{S}_3=\hat{n}_1+ \hat{n}_2$, where $\hat{n}=\hat{A}^\dagger\hat{A}$ are the number operators. In this basis, since $\hat{S}_2$ and $\hat{S}_3$ commute with $\hat{H}$, the nonlinear three-wave interaction problem is mapped to a Hamiltonian simulation problem where the Hamiltonian matrix is block tridiagonal. Within each block, the subspace is finite dimensional, and can be readily mapped to the memory of quantum computers. 
As a proof-of-principle demonstration, a three dimensional block is recently implemented on quantum hardware to solve the Schr\"odinger equation $i\partial_t|\psi\rangle=H|\psi\rangle$, where $|\psi\rangle=\alpha_0|2, s-2, 0\rangle +\alpha_1|1, s-1, 1\rangle +\alpha_2|0, s, 2\rangle$ and $|n_1,n_2,n_3\rangle$ is the Fock state of the three waves. The quantum hardware is programmed to evolve the quantum states according to the unitary operator $U=\exp(-iH\Delta t)$ where $\Delta t$ is the time step size. 
Using the standard approach, the unitary operator is realized by a sequence of standard gates, and the results (Fig.~\ref{fig:Results}, cyan) track the exact solutions (Fig.~\ref{fig:Results}, orange) up to $N\lesssim10$ time steps. 
As a more efficient approach, the unitary operator is also compiled as a single customized gate, and the results (Fig.~\ref{fig:Results}, blue) are significantly improved, making it promising to use near term quantum hardware to simulate problems of physical interest.

\subsection{\label{sec:3wave_comp}Magnetized laser pulse compression}
As an application, magnetized three-wave interactions can be exploited to improve laser pulse compression \cite{Shi2017laser}. Similar to unmagnetized cases \cite{Malkin99,Andreev2006short}, magnetized plasma waves can mediate the energy transfer from a pump laser to a seed pulse via stimulated scattering. The seed pulse is amplified and shortened in the pump depletion regime, thereby achieving effective compression of the pump laser. 
Plasma-based pulse compression can in principle produce pulses that are much more intense than what is achievable using solid-state media \cite{Mourou1998ultrahigh}, where the intensity is limited by ionization and thermal damage. In comparison, the much higher intensity attainable in plasmas is limited by relativistic effects \cite{Malkin2016extended}, in the absence of which three-wave coupling remains the dominant nonlinearity. 
However, realistic plasma sources suffer from nonuniformities and laser heating, which spoil the ideal phase matching conditions and hence limits the performance attainable in experiments \cite{Ping00,Ping2004amplification,Cheng2005reaching,Ren2007new,Pai08, Ping2009development,Kirkwood2011observation,Lancia16,Vieux2017ultra}. Moreover, even for an ideal plasma target, three-wave interactions compete with other processes, including modulational instability, wave breaking, and wave damping. These competing processes limit the parameter regime where pulse compression can be carried out efficiently \cite{Clark2003operating,Trines2011simulations}.
Now, with the addition of a magnetic field, plasma conditions may be better regulated, and competing effects may be better controlled. With the extra degrees of freedom, magnetization can potentially improve the performance of laser pulse compressors, and produce higher intensity pulses in wider frequency ranges.

\begin{figure}[t]
	\begin{center}
		\includegraphics[width=0.48\textwidth]{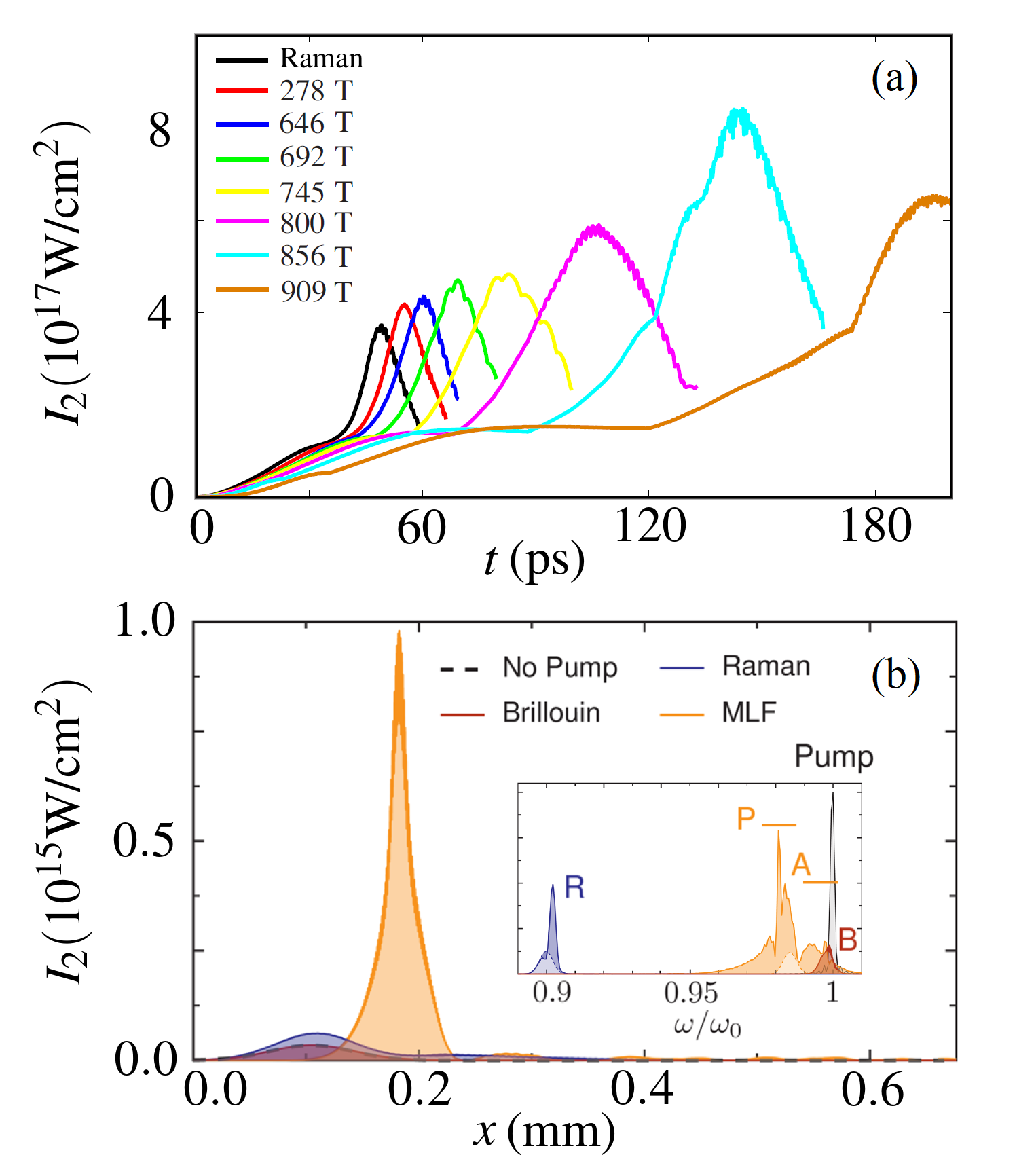} 
		\caption{Compression of 1-$\mu$m laser using (a) upper-hybrid wave mediation with $\omega_3/\omega_1\sim0.1$. The peak intensity of the seed pulse grows slower, but the saturation due to modulational instability is delayed. The final intensity increases with $B_0$ before it reaches the optimal value (cyan), after which the performance degrades due to wavebreaking and wakefield excitation. [Adapted from J. Qin \textit{et al.}, Phys. Plasmas \textbf{24}, 093103 (2017).] 
		(b) Using magnetized low frequency (MLF) waves to mediate pulse compression (orange), the amplification rate is significantly larger than unmagnetized Raman (blue) and Brillouin (red) when the laser frequency is close to the electron gyro frequency. Due to synergistic actions of MLF waves, the bandwidth is ultra wide (inset), so the attainable pulse duration is significantly shorter. [Adapted from M. R. Edwards \textit{et al.}, Phys. Rev. Lett. \textbf{123}, 025001 (2019).]}
		\label{fig:PIC}
	\end{center}
\end{figure}

Let us take pulse compression mediated by the upper-hybrid (UH) wave as an example to elucidate the benefits of magnetization \cite{Shi2017laser}. The UH mediation is the magnetized version of Raman compression, which is mediated by the Langmuir wave. 
Consider experiments where the pump laser (wave ``1'') and seed laser (wave ``2'') are fixed, and one adjusts the plasma conditions to optimize the output intensity. In order to resonantly couple the two lasers, plasma parameters need to be selected such that the detuning $\omega_3=\omega_1-\omega_2$ matches the frequency of a plasma eigenmode (wave ``3''). For UH mediation, the two lasers propagate perpendicular to the magnetic field, and $\omega_3=\sqrt{\omega_p^2+\Omega_e^2}$ is the upper-hybrid frequency. Then, as one ramps up the magnetic field $B_0$, the plasma density $n_0$ needs to be ramped down to keep $\omega_3$ fixed. 
When the plasma density is reduced, the three-wave coupling coefficient decreases, so the amplification rate $\gamma_0=\frac{\omega_p}{2}\sqrt{\frac{\omega_1}{\omega_3}}|a_1|$ is smaller. This may seem disadvantageous, because a longer plasma is then required to achieve the same amplification. However, at the expense of a longer plasma, the output pulse intensity can be increased beyond the saturation value limited by the modulational instability, which introduces unwanted phase shift that spoils the phase matching conditions. 
The key physics that enables a larger saturation value is that the growth rate of the modulational instability $\gamma_M=\frac{\omega_p^2}{8\omega_2}|a_2|^2$ is proportional to $n_0$, unlike $\gamma_0\propto n_0^{1/2}$. In other words, as the density decreases, $\gamma_M$ decreases faster than $\gamma_0$, so the demanded pulse compression process gains a relative advantage and the unwanted modulational instability is relatively suppressed. As corroborated by 1D PIC simulations \cite{Jia2017kinetic}, the net consequence of magnetization is that the seed pulse grows slower, but the saturation occurs later at higher intensity (Fig.~\ref{fig:PIC}a). This mechanism is also at play when lasers propagate at other angles with respect to the background magnetic field \cite{Li2020boosting}.

While substituting $n_0$ by a moderate $B_0$ increases the output intensity, pulse compression is inefficient in a plasma that is too tenuous \cite{Jia2017kinetic}. A major limitation is the excitation of wakefield by the amplified seed pulse: When the plasma density is too low, it becomes too easy for the ponderomotive force to expel plasma electrons. This process destroys the coherent eigenmode structure, which is required to mediate efficient pulse compression. 
Even before ponderomotive expulsion kicks in, the UH wave may already loss coherence due to wave breaking \cite{Shi2017laser}. Wave breaking occurs when the quiver velocity of electrons $v_q= eE_3\omega_3/m_e\omega_p^2=ca_3\omega_3/\omega_p$ exceeds the phase velocity $v_p=\omega_3/k_3$ of the plasma wave. Since $v_p$ is approximately fixed, decreasing the plasma density makes the plasma wave easier to break. Using the Manley-Rowe relation for three-wave interactions, the maximum plasma wave amplitude $a_3=\sqrt{\omega_1/\omega_3}a_1$. Therefore, in order to avoid breaking the plasma wave, the pump intensity is limited to $a_1\leq\frac{\omega_p}{\sqrt{\omega_1\omega_3}}\frac{v_p}{c}$.
Therefore, as the plasma density decreases, the pump intensity is capped at an ever smaller value. When the pump becomes too weak, the growth no longer dominates damping, and pulse compression becomes inefficient and eventually cease to work.

When the magnetic field strength is chosen appropriately, UH mediation not only increasing the output intensity, but also allows for the compression of shorter wavelength lasers by alleviating wave damping \cite{Shi2017laser}. 
Notice that damping is more severe for shorter wavelength lasers, because in order to avoid wave breaking, one must increase $\omega_p$ for larger $\omega_1$. However, in a denser plasma, damping is stronger, which now becomes the major limiting effect.
For collisional damping, the coherent quiver motion of charged particles is randomized by interspecies collision. Consequently, the fraction of wave energy that is carried by particles is dissipated to heat. The dissipation rate is proportional to the collision frequency, which scales as $n_0^2$ due to the binary nature of collisions. Therefore, when substituting $n_0$ by $B_0$, UH mediation suffers significantly less collisional damping. 
The other way to reduce collisional damping is increasing the plasma temperature $T_0$, because the collision frequency roughly decreases as $1/v_T^3$ in thermal plasmas, where $v_T$ is the electron thermal speed. However, increasing $T_0$ unwittingly increases collisionless damping, a major mechanism that dissipates the plasma wave. 
For collisionless damping, trapped particles, whose velocity is near $v_p$, irreversibly exchange energy with the wave through phase mixing. The wave is damped when more particles gain energy than loss energy, which occurs as long as the distribution function falls off near $v_p$. Although the presence of $B_0$ complicates the process \cite{Sagdeev73,Karney78,Karney79,Dawson83,Winjum2018mitigation}, the single-species nature of phase mixing causes the collisionless damping to roughly scale with $n_0$, which is reduced at lower density.
Therefore, compared with unmagnetized case, where both $n_0$ and $T_0$ can neither be too large nor too small, UH mediation opens up the $n_0$-$T_0$ parameter window, in which efficient pulse compression may be achieved. The benefit of magnetization is more pronounced for shorter wavelength lasers, and the window for UH mediation remains open until the x-ray range, beyond the reach of unmagnetized schemes \cite{Malkin2007compression,Malkin2009quasitransient,Edwards2017x}.

In addition to the UH wave, a large variety of other magnetized waves can also be used to mediate pulse compression. To select a mediating wave, one chooses laser propagation angles with respect to $\mathbf{B}_0$, and detunes the seed frequency from the pump to match plasma resonances. By choosing a different wave, what is essentially changed is the energy partition between field and particle degrees of freedom. Consequently, the coupling coefficient, the damping rates, and the growth rates of competing instabilities are changed, so is their relative importance.  
For example, as an analogue to unmagnetized Brillouin compression, which is mediated by the ion acoustic wave or quasimode, pulse compression can be mediated by magnetized low frequency (MLF) waves \cite{Edwards2019laser}, which are the short-wavelength extensions of MHD waves. An immediate advantage of using low frequency plasma waves is that the required laser detuning is small, so the pump and seed beams can be derived from a single oscillator, which simplifies the experimental setup. However, in the unmagnetized case, the growth rate of Brillouin amplification is substantially smaller than that of Raman, roughly by a factor of $(m_e/m_i)^{1/2}$ where $m_i$ is the lightest ion mass, because nearly half of the acoustic wave energy is carried by slow ions, which are ineffective for providing three-wave coupling. In other words, Brillouin is usually not competitive with Raman in electron-ion plasmas \cite{Jia2016distinguishing,Edwards2016short}. 
The situation is drastically changed with the addition of an oblique magnetic field, whose strength is such that $\omega_{1,2}\sim|\Omega_e|\gg\omega_p$. In this regime, a large fraction of the pump and seed energy is carried by the resonantly-driven electron gyro motion. In other words, for the same laser intensity, electrons now quiver at much higher velocity, which compensates for the reduced energy share of electrons in low frequency waves. This effect greatly enhances the three-wave coupling via MLF waves, which provide large amplification rates that are even grater than Raman.   
Moreover, mediation via MLF waves has significantly larger bandwidth, which allows the seed pulse to reach much shorter duration. The ultra-wide bandwidth partly comes from the large growth rates. But perhaps more importantly, it comes from the synergistic action of all three branches of MLF waves, which can be excited simultaneously when their frequencies are close (Fig.~\ref{fig:PIC}b). The frequencies of the three branch are roughly $\omega_p\cos\theta_3, c_s k_3$, and $\Omega_i$, where $c_s$ is the sound speed. 
Therefore, at near perpendicular angle, all three branches have comparable frequencies, which allows their bandwidths to overlap, giving rise to an ultra-wide total bandwidth. It is worth noting that $\theta_3$ can not be too close to $90^\circ$, because at perpendicular angle, energy in MLF waves are largely spent on compressing the magnetic field, so the three-wave coupling is energy suppressed and the amplification rates diminish.

\section{\label{sec:QED}Scalar QED plasma model}
The megagauss-level magnetic fields required to affect three-wave coupling is large, but still keep the plasma in the classical regime, which may no longer be the case in even larger fields.
Experimentally, gigagauss-level magnetic fields have been reported \cite{Wagner2004laboratory,Fujioka2013kilotesla}.
There, the quantized electron perpendicular energy \mbox{$\hbar|\Omega_e|\sim10$ eV} is larger than the photon energy of optical lasers, so the quantum nature of electrons starts to manifest. At the same time, the intense lasers accelerate electrons to high energy, so the plasma is relativistic in addition to being quantum. These conditions created in laboratory are thus reminiscent of extreme astrophysical environments, for which the classical plasma model may no longer be sufficient. 

The classical plasma model, 
which assumes point-like charged particles and wave-like EM fields, breaks down when the wave nature of charged particles and the particle nature of EM excitations are resolved.    
In the later regime, a different modeling framework is required, especially when the wave-particle duality of both electrons and EM fields are important. 
For a few electrons and photons, it is well known that the fundamental theory is quantum electrodynamics (QED). The question is then how to extend QED to describe collective effects.  
As a toy model, we consider scalar QED, which is described by the Lagrangian density
\begin{equation}
\label{eq:sQED}
\mathcal{L}=
(D_{\mu}\phi)^{*}(D^{\mu}\phi)-m^2\phi^{*}\phi
-\frac{1}{4}F_{\mu\nu}F^{\mu\nu},
\end{equation}
where $\phi$ is a complex scalar field with mass $m$ and charge $e$,  $D_{\mu}=\partial_{\mu}-ieA_{\mu}$ is the covariant derivative, and $F_{\mu\nu}=\partial_{\mu}A_{\nu}-\partial_{\nu}A_{\mu}$ is the field strength tensor. Here, we have used the natural units $\hbar=c=1$, and omitted the self coupling $\frac{\lambda}{2}(\phi^{*}\phi)^2$.
The complex scalar field describes \mbox{spin-0} charged bosons, instead of \mbox{spin-1/2} fermions. Nevertheless, the usual plasma physics takes no account of the spin-statistics of charged particles, and the scalar QED model is sufficient to demonstrate that quantum field theories (QFT) can be extended to incorporate collective many-body effects. With the implied understanding that charged particles are bosons, we will refer to them as electrons and positrons for simplicity. 
The approach we discuss below, which has since been adapted to spinor QED plasmas \cite{Chen2019gauge,Wu2020background}, are complementary to other approaches including finite temperature field theories \cite{Rojas1979polarization,Inagaki2005proper}, statistic Green's function theories \cite{Bezzerides1972quantum,Melrose2012quantum}, and quantum hydrodynamics \cite{Brodin2008quantum,Haas2011quantum}.

\subsection{\label{sec:QED_BK}Field theory with nontrivial backgrounds}
To incorporate plasma effects in to scalar QED, we decompose the bosonic fields into classical backgrounds and quantum excitations: $\phi=\phi_{0}+\varphi$ and $A_{\mu}= \bar{A}_{\mu}+\mathcal{A}_{\mu}$. In usual QFT, $\phi_{0}=0$ and $\bar{A}_{\mu}=0$, and one studies fluctuations on the vacuum background. 
Now to describe plasmas, $\phi_{0}$ and $\bar{A}_{\mu}$ are nonzero. 
These background fields, which are not necessarily constants nor Bose–Einstein condensates, can be dynamical and the only conditions they need to satisfy is the classical field equations
\begin{eqnarray}
\label{eq:EOM_phi}
(\bar{D}_{\mu}\bar{D}^{\mu}+m^2)\phi_{0}&=&0,\\
\label{eq:EOM_A}
\partial_{\mu}\bar{F}^{\mu\nu}=\bar{J}_{0}^{\nu}.
\end{eqnarray}
Here, $\bar{D}_{\mu}=\partial_{\mu}-ie\bar{A}_{\mu}$ is the background gauge covariant derivative, $\bar{F}_{\mu\nu}=\partial_{\mu}\bar{A}_{\nu} -\partial_{\nu}\bar{A}_{\mu}$ is the background field strength, and $\bar{J}_{0}^{\mu}=\frac{e}{i}[\phi_0^{*}(\bar{D}^{\mu}\phi_0)-\text{c.c.}]$ is the total background current, which is summed over all particles when the model include multiple charged species.
It is obvious that the classical vacuum $\phi_{0}=0$ and $\bar{A}_{\mu}=0$ is a trivial solution. In more general cases where the solutions are nontrivial, the fields $\bar{F}_{\mu\nu}$ are the usual classical EM fields, whereas the bosonic field $\phi_{0}$ is formally related to the symmetrized many-body wave function by $\phi_0(x)=\int\sqrt{V}\Phi_0(x,x_2,\dots,x_N)$, where $V=d^4x_2\wedge\dots\wedge d^4x_N$ is the volume form.

On a given nontrivial background, quantum excitations $\varphi$ and $\mathcal{A}_{\mu}$ interacts via additional vertices that are absent in vacuum field theories \cite{Shi2016effective}. To obtain the Lagrangian that describes the excitations, we use classical field equations [Eqs.~(\ref{eq:EOM_phi}) and (\ref{eq:EOM_A})] in the action integral $S=\int d^4x\mathcal{L}$. After integration by part, the Lagrangian density can be decomposed as $\mathcal{L}=\mathcal{L}_\varphi+\mathcal{L}_\mathcal{A}+\mathcal{L}_I$, where
\begin{eqnarray}
\label{eq:L_phi}
\mathcal{L}_\varphi &=& (\bar{D}_{\mu}\varphi)^{*}(\bar{D}^{\mu}\varphi)-m^2\varphi^{*}\varphi,\\
\label{eq:L_A}
\mathcal{L}_{\mathcal{A}}&=&-\frac{1}{4}\mathcal{F}_{\mu\nu}\mathcal{F}^{\mu\nu}+e^2\phi_{0}^{*}\phi_{0}\mathcal{A}_{\mu}\mathcal{A}^{\mu},\\
\label{eq:L_I}
\mathcal{L}_I&=&-(\bar{j}^{\mu}+\bar{\eta}^{\mu})\mathcal{A}_{\mu}+e^2(\phi_{0}\varphi^{*}+\phi_{0}^{*}\varphi +\varphi^{*}\varphi)\mathcal{A}_{\mu}\mathcal{A}^{\mu}.
\end{eqnarray}
The first term $\mathcal{L}_\varphi$ describes the free $\varphi$ field. Here, free is in the sense that $\varphi$ does not interact with $\mathcal{A}_{\mu}$, but $\varphi$ clearly feels the influence of the classical field $\bar{A}_{\mu}$ through the background gauge covariant derivative. 
Likewise, the second term $\mathcal{L}_\mathcal{A}$ describes the free $\mathcal{A}_{\mu}$ field, with $\mathcal{F}_{\mu\nu}=\partial_{\mu}\mathcal{A}_{\nu}-\partial_{\nu}\mathcal{A}_{\mu}$. The $\mathcal{A}_{\mu}$ field gains a mass term through $\phi_0$, which can have spacetime dependencies \cite{Shi2019force}. 
Finally, $\mathcal{L}_I$ includes all interaction terms. In particular, $\bar{j}^{\mu}=\frac{e}{i}(\varphi^{*}\bar{D}^{\mu}\varphi-\text{c.c.})$ is the current due to vacuum excitation, and 
$\bar{\eta}^{\mu}=\frac{e}{i}(\phi_{0}^{*}\bar{D}^{\mu}\varphi+\varphi^{*}\bar{D}^{\mu}\phi_{0}-\text{c.c.})$ is the currents due to plasma excitation. 
The background field theory defined by Eqs.~(\ref{eq:L_phi})-(\ref{eq:L_I}) is an extension to Furry's picture that is commonly used in strong-field QED \cite{Furry51}, which includes $\bar{A}_{\mu}$ but not $\phi_0$. 

\subsection{\label{sec:QED_action}Wave effective action}
Using the background field theory, we study properties of $\mathcal{A}_{\mu}$ excitations in scalar QED plasmas \cite{Shi2016effective}. 
One way to do this is to use path integral to integrate out charged degrees of freedom. What remains is then the effective action for $\mathcal{A}_{\mu}$, which includes plasma dressing effects. Using perturbation theory, the effective action can be written as 
\begin{equation}
S_\mathcal{A}=\frac{1}{2}\int d^4x \mathcal{A}_{\mu}(x)(\partial^2g^{\mu\nu}-\partial^{\mu}\partial^{\nu})\mathcal{A}_{\nu}(x)+\frac{1}{2}\int d^4xd^4x'\mathcal{A}_{\mu}(x)\Sigma_2^{\mu\nu}(x,x')\mathcal{A}_{\nu}(x')+O(e^3),
\end{equation}
where $g^{\mu\nu}$ is the Minkowski tensor, and the higher order terms describe three-wave interactions and so on. 
To $e^2$ order, plasma and vacuum responses decouple, and the linear response tensor is $\Sigma_2^{\mu\nu}=\Sigma_{2,\text{bk}}^{\mu\nu}+\Sigma_{2,\text{vac}}^{\mu\nu}$. 
The background plasma response is due to the excitation of on-shell particles:
\begin{fmffile}{bk}
	\begin{eqnarray}
	\label{eq:bk}
	\nonumber
	\Sigma_{2,\text{bk}}^{\mu\nu}(x,x')
	&=&\quad
	\begin{gathered}
	\begin{fmfgraph*}(40,15)
	\fmfkeep{mass}
	\fmfleft{i}
	\fmfright{o}
	\fmf{photon}{i,v}
	\fmf{photon}{v,o}
	\fmfdot{v}
	\fmfv{label=$x$,label.angle=90,label.dist=6}{v}
	\fmfv{label=$\mu$,label.dist=0.2}{i}
	\fmfv{label=$\nu$,label.dist=0.2}{o}
	\end{fmfgraph*}
	\end{gathered}
	\quad+\quad
	\begin{gathered}
	\begin{fmfgraph*}(40,15)
	\fmfkeep{line}
	\fmfleft{i}
	\fmfright{o}
	\fmf{plain}{v1,v2}
	\fmfdot{v1,v2}
	\fmfv{label=$x$,label.angle=90,label.dist=8}{v1}
	\fmfv{label=$x'$,label.angle=90,label.dist=8}{v2}
	\fmfv{label=$\mu$,label.dist=0.5}{i}
	\fmfv{label=$\nu$,label.dist=0.5}{o}
	\fmf{photon}{i,v1}
	\fmf{photon}{v2,o}
	\end{fmfgraph*}
	\end{gathered}\\
	&=&\!2e^2\phi_0\phi^*_0\delta(x-x')g^{\mu\nu}\!+i\Pi^{\mu\nu}_{2,\text{bk}}(x,x').
	\end{eqnarray}
\end{fmffile}The first term is the photon mass term, and the second term is the plasma polarization tensor  $\Pi^{\mu\nu}_{2,\text{bk}}(x,x')=\langle \bar{\eta}^{\mu}(x) \bar{\eta}^{\nu}(x') \rangle =e^2\big[\phi_0^*\bar{D}^{\mu}-(\bar{D}^{\mu}\phi_{0})^*\big]\big[\phi_0'\bar{D}^{'*\nu}-(\bar{D}^{'\nu}\phi_{0}')\big]G-\text{c.c.}$.
For simplicity, we have used the notation $\phi_0=\phi_0(x)$, $\phi_0'=\phi_0(x')$, $G=G(x,x')$, $G'=G(x',x)$, and so on.  
Here, $G(x,x')$ is the electron propagator, which satisfies $[\bar{D}_{\mu}(x)\bar{D}^{\mu}(x)+m^2]G(x,x')=-i\delta(x-x')$. 
In comparison, the vacuum response is due to virtual electron-positron pair excitation, and
\begin{fmffile}{vac}
	\begin{eqnarray}
	\label{eq:vac}
	\nonumber
	\Sigma_{2,\text{vac}}^{\mu\nu}(x,x')
	&=&\quad
	\begin{gathered}
	\begin{fmfgraph*}(40,20)
	\fmfkeep{hairpin}
	\fmfleft{i}
	\fmfright{o}
	\fmf{photon}{i,v}
	\fmf{photon}{v,o}
	\fmf{plain}{v,v}
	\fmfdot{v}
	\fmfv{label=$x$,label.angle=-90,label.dist=6}{v}
	\fmfv{label=$\mu$,label.dist=0.2}{i}
	\fmfv{label=$\nu$,label.dist=0.2}{o}
	\end{fmfgraph*}
	\end{gathered}
	\quad+\quad
	\begin{gathered}
	\begin{fmfgraph*}(50,20)
	\fmfkeep{bubble}
	\fmfleft{i}
	\fmfright{o}
	\fmf{plain,left=1,tension=0.3}{v1,v2}
	\fmf{plain,right=1,tension=0.3}{v1,v2}
	\fmfdot{v1,v2}
	\fmfv{label=$x$,label.angle=120,label.dist=8}{v1}
	\fmfv{label=$x'$,label.angle=60,label.dist=8}{v2}
	\fmfv{label=$\mu$,label.dist=0.5}{i}
	\fmfv{label=$\nu$,label.dist=0.5}{o}
	\fmf{photon}{i,v1}
	\fmf{photon}{v2,o}
	\end{fmfgraph*}
	\end{gathered}\\
	&=&2e^2\langle\varphi\varphi*\rangle\delta(x-x')g^{\mu\nu}+i\Pi^{\mu\nu}_{2,\text{vac}}(x,x').
	\end{eqnarray}
\end{fmffile}The first term gives rise to photon mass renormalization, and the second term is the vacuum polarization tensor $\Pi^{\mu\nu}_{2,\text{vac}}(x,x')=\langle \bar{j}^{\mu}(x) \bar{j}^{\nu}(x') \rangle =e^2\big[G'\bar{D}^{\mu}-(\bar{D}^{*\mu}G')\big](\bar{D}^{'*\nu}G)+\text{c.c.}$.
Notice that to $e^2$ order, $\phi_0$ does not directly affect $\Sigma_{2,\text{vac}}^{\mu\nu}$, and only enters indirectly via $\bar{A}_{\mu}$ through the self-consistency conditions [Eqs.~(\ref{eq:EOM_phi}) and (\ref{eq:EOM_A})].
Using the above gauge-independent formulas, the $e^2$-order effective action can in principle be evaluated once the background fields $\phi_0$ and $\bar{A}_{\mu}$ are specified.

As an example, we obtain an explicit formula for the dispersion tensor in a cold uniformly magnetized plasma \cite{Shi2016effective}.
Using the transnational symmetry, $\Sigma(x,x')=\Sigma(r)$ only depends on the coordinate separation $r=x-x'$, so the Fourier-space effective action is simplified: $S_\mathcal{A}=\frac{1}{2}\!\int\!\frac{d^4k}{(2\pi)^4} \hat{\mathcal{A}}_{\mu}(-k)\mathbb{D}^{\mu\nu}(k)\hat{\mathcal{A}}_{\nu}(k) +O(e^3)$. Here, $\mathbb{D}^{\mu\nu}(k)=k^{\mu}k^{\nu}-k^2g^{\mu\nu}+\Sigma_2^{\mu\nu}(k)$ is the dispersion tensor, and $\hat{\Sigma}_2^{\mu\nu}(k)=\int d^4re^{ikr}\Sigma_2^{\mu\nu}(r)$. 
To evaluate Eqs.~(\ref{eq:bk}) and (\ref{eq:vac}), we use the symmetric gauge for $\bar{A}_\mu$ and solve for the Green's function $G$. Additionally, we specify the plasma background by constructing a symmetrized many-body wave function $\Phi_0$ from single-particle wave functions.
The wave function $\Phi_0$ depends on the plasma distribution function, which is now Landau quantized in the direction perpendicular to the background magnetic field $\mathbf{B}_0$. For neutron star magnetosphere \cite{Becker2009x} where the plasma temperature $k_BT\lesssim\frac{1}{4}\hbar\Omega_e$ is cold, all particles are in the lowest Landau level. We further assume for simplicity that all particles stream along $\mathbf{B}_0$ at the same velocity. Then, in the rest frame of the plasma, and in the coordinate system $x^\mu$ where $\mathbf{B}_0$ is along the $x^3$ direction, components of the plasma response tensors are \cite{Shi2016effective}
\begin{eqnarray}
\label{MagPol}
\hat{\Sigma}^{\lambda\sigma}_{2,\text{bk}}(k)&=&\frac{m\omega_p^2}{m_0}\Big[g^{\lambda\sigma}-\frac{1}{2}\sum_{\varsigma=\pm 1}(\kappa+\varsigma\varrho)^{\lambda}(\kappa+\varsigma\varrho)^{\sigma}K_\varsigma^{(0)}\Big],\\
\hat{\Sigma}^{ab}_{2,\text{bk}}(k)&=&\frac{m\omega_p^2}{2m_0}\sum_{\varsigma=\pm 1}\big\{\varepsilon^{ac}\varepsilon^{bd}\kappa^c\kappa^d(2K_\varsigma^{(1)}-K_\varsigma^{(0)})
-\kappa_\varsigma^2[\delta^{ab}K_\varsigma^{(1)}\pm i\varsigma\varepsilon^{ab}(K_\varsigma^{(1)}-K_\varsigma^{(0)})]\big\},\\
\hat{\Sigma}^{\lambda a}_{2,\text{bk}}(k) &=&\hat{\Sigma}^{a\lambda}_{2,\text{bk}}(-k)=\frac{m\omega_p^2}{2m_0}\sum_{\varsigma=\pm 1}(\kappa+\varsigma\varrho)^{\lambda}
\big[-\kappa^aK_\varsigma^{(1)}\pm i\varsigma\varepsilon^{ab}\kappa^b(K_\varsigma^{(1)}-K_\varsigma^{(0)})\big],
\end{eqnarray}
where $\lambda,\sigma=0,3$ and $a,b=1,2$. 
Here, $m_0=\sqrt{m^2+|eB_0|}$ is the ground state mass, and the summation over $\varsigma=\pm 1$ corresponds to the $s$-channel and the $t$-channel Feynman diagrams. 
Since $\mathbf{B}_0$ breaks the symmetry, two-dimensional Levi-Civita symbol $\varepsilon^{ab}$ appears, and the upper (lower) sign of $\pm$ corresponds to $eB_0>0$ ($eB_0<0$). 
Effects of Landau quantization are encapsulated in special functions $K_\varsigma^{(n)}= K(\kappa_\varsigma^2-n,\bm{\kappa}^2)$, 
where the $K$-function is related to the confluent hypergeometric function by $K(x,z)=\frac{1}{x}{}_1F_1(1;1-x;-z)$.
The arguments of the $K$-function are $\kappa_\varsigma^2=\kappa_0^2-\kappa_3^2+\varsigma\varrho_0\kappa_0$ and $\bm{\kappa}^2=\kappa_1^2+\kappa_2^2$, where $\kappa^{\mu}=r_0k^{\mu}/2$ and $\varrho^{\mu}=r_0(m_0,0,0,0)$ are normalized by the magnetic de Broglie wavelength $r_0=\sqrt{2/eB_0}$.
The above formulas make intuitive sense, because heuristically from the Feynman diagram, the plasma polarization is $\hat{\Pi}_{2,\text{bk}}\sim e^2|\phi_0|^2(k/M)^2$, where $\phi_0\sim\sqrt{n_0/m}$, $k$ is the energy scale of $\mathcal{A}_\mu$ excitations, and $M$ is the energy scale of the electron propagator. In magnetized plasmas, the two intrinsic energy scales are the electron mass $m$ and the gyro frequency $\Omega=eB_0/m$. We see unless $B_0$ approaches the Schwinger field $B_c=m^2/e$ or the photon energy approaches $m$, plasma response usually dominates the vacuum response, which heuristically scales as $\hat{\Pi}_{2,\text{vac}}\sim e^2 k^6/M^4$ and is given exactly by the Euler–Heisenberg effective Lagrangian \cite{Heisenberg1936consequences,Dunne2012heisenberg}.

\subsection{\label{sec:QED_disp}Modified wave dispersion relations}
From the wave effective action, the dispersion relation can be extracted. The dispersion tensor has complementary interpretations in QFT and in plasma physics. 
In QFT, the propagator of the $\mathcal{A}$ field is schematically $i/\mathbb{D}$ up to some gauge fixing condition.
The poles of $\mathbb{D}$ then give rise to peaks in cross sections. The energy of each peak is usually identified as the mass a particle in the field theory, and the width of the peak is associated with the life time of the particle.
In comparison, in plasma physics, one looks for nontrivial $\mathcal{A}$ that satisfies $\mathbb{D}^{\mu\nu}\hat{\mathcal{A}}_{\nu}=0$. Due to gauge invariance, only three components of this matrix equation are linearly independent. In temporal gauge, Eq.~(\ref{eq:DE}) is recovered, and the dispersion relation $\omega=\omega(\mathbf{k})$ is again solved from $\det \mathbb{D}_{ij}=0$, where $\mathbb{D}_{ij}$ is the spatial block of the Lorentz-covariant $\mathbb{D}^{\mu\nu}$. For a real-valued $\mathbf{k}$, $\omega$ may have an imaginary part, which is usually identified as the wave damping rate.
The above field theoretical picture and the plasma physics picture are consistent if we regard plasma waves as quasi particles.

\begin{figure}[t]
	\renewcommand{\figurename}{FIG.}
	\includegraphics[angle=0,width=0.48\textwidth]{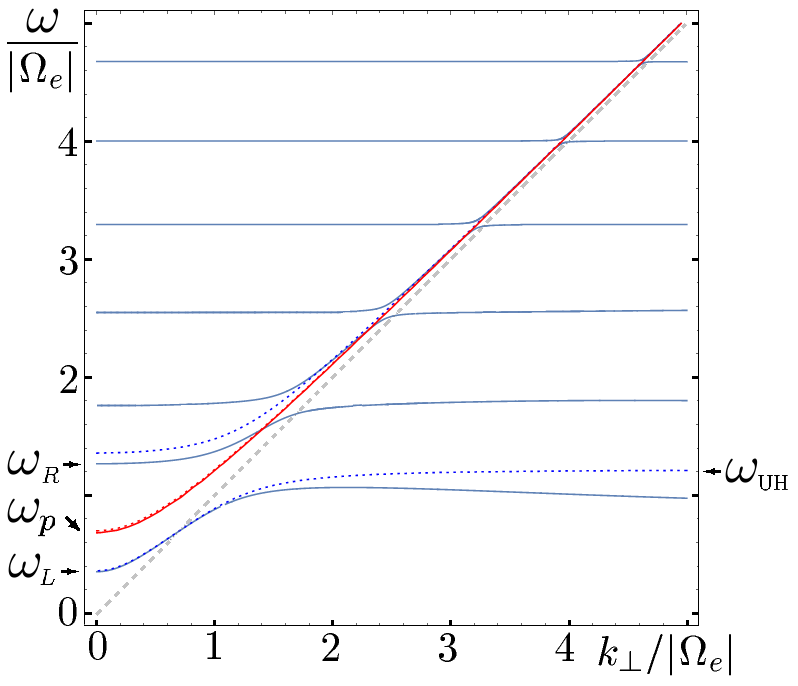}
	\caption{Perpendicular wave dispersion relations in a magnetized cold plasma, with immobile ions as neutralizing background. In a QED plasma (solid), the extraordinary wave (X, blue) hybridizes with cyclotron resonances, which are redshifted due to relativistic effects. Notice that in this example, where $\omega_{pe}/|\Omega_{e}|=0.7$ and $|\Omega_{e}|/m_e=0.1$, the 5th resonance occurs near $4|\Omega_e|$ instead of $5|\Omega_e|$. The hybridization is absent in a classical plasma (dashed), in which only the upper-hybrid (UH) wave remains. 
	While QED modifications to the X wave is significant, and lead to anharmonic cyclotron absorption features for x-ray pulsars, the 
	ordinary wave (O, red) is only slightly modified with a redshifted cutoff frequency. [Y. Shi, \textit{et al}. Phys. Rev. A \textbf{94}, 012124 (2016).]}
	\label{fig:Perp}
\end{figure}

The dispersion relation is simplified for wave propagation perpendicular to the magnetic field, in which case QED effects give rise to anharmonic cyclotron absorption features observed for x-ray pulsars \cite{Shi2016effective}. 
In this geometry, one eigenmode is the ordinary (O) wave (Fig.~\ref{fig:Perp}, red), whose electric field is polarized along $\mathbf{B}_0$. The O wave satisfies the simple dispersion relation $\omega^2=\omega_{p0}^2+k_\perp^2$, where $\omega_{p0}^2=\frac{m}{m_0}\omega_p^2$. For O wave, magnetization effects only enter through the modified ground state mass $m_0$: Due to Landau quantization, the zero-point energy in magnetic field increases the electron mass from $m$ to $m_0$, so the plasma frequency is reduced. The QED dispersion relation (solid) is close to the classical dispersion relation (dashed), unless the magnetic field is comparable to the Schwinger field. 
The other eigenmode is the extraordinary (X) wave (Fig.~\ref{fig:Perp}, blue), whose electric field is perpendicular to $\mathbf{B}_0$. 
Since the X wave imparts both energy and angular momentum to electrons, it causes electrons to jump between Landau levels. At nonresonant frequencies, the jump is virtual and the wave-particle coupling is weak. However, when the wave frequency matches the energy difference between two Landau levels, the $\omega\sim k$ transverse vacuum mode hybridizes with the $\omega\sim\omega_n$ longitudinal excitation, and a band gap is opened, giving rise to QED analogues of plasma Bernstein waves. 
Notice that in contrast to what happens in classical plasmas, finite band gaps exist even when the QED plasma is cold. Classically, electrons stop gyrating when $T_0\rightarrow 0$, in which case Bernstein waves vanish and only the UH wave remains. However, quantum mechanically, electrons cannot stop gyrating because $[p_1, p_2]=-ieB_0$ no longer commutes, where $p_\mu=-i\bar{D}_\mu$ is the kinetic momentum operator of electrons in background fields. Due to quantum uncertainty, cold electrons continue to gyrate at zero-point energy, and cyclotron resonances persist. 
In a cold plasma, the resonance frequency is $\omega_n=E_{n}-m_0$, where $E_{n}=\sqrt{m^2+|eB_0|(2n+1)}$ is the energy of the $n$-th Landau level. 
In weak magnetic field, $\omega_n\simeq n|\Omega_e|$ is harmonic. However, in strong magnetic fields, $\omega_n< n|\Omega_e|$ due to relativistic effects. The relativistic redshift is of order unity, namely, $\omega_n< (n-1)|\Omega|$ when $n>\sqrt{2m_e/|\Omega_e|}\approx9.4\times\sqrt{10^{12}\text{G}/B_0}$. We see the anharmonicity of cyclotron resonances is significant for neutron stars, and has already been observed for a number of x-ray pulsars \cite{Santangelo99,Heindl2000multiple,Pottschmidt05,Pottschmidt2012}.

\begin{figure}[h]
	\renewcommand{\figurename}{FIG.}
	\includegraphics[angle=0,width=0.48\textwidth]{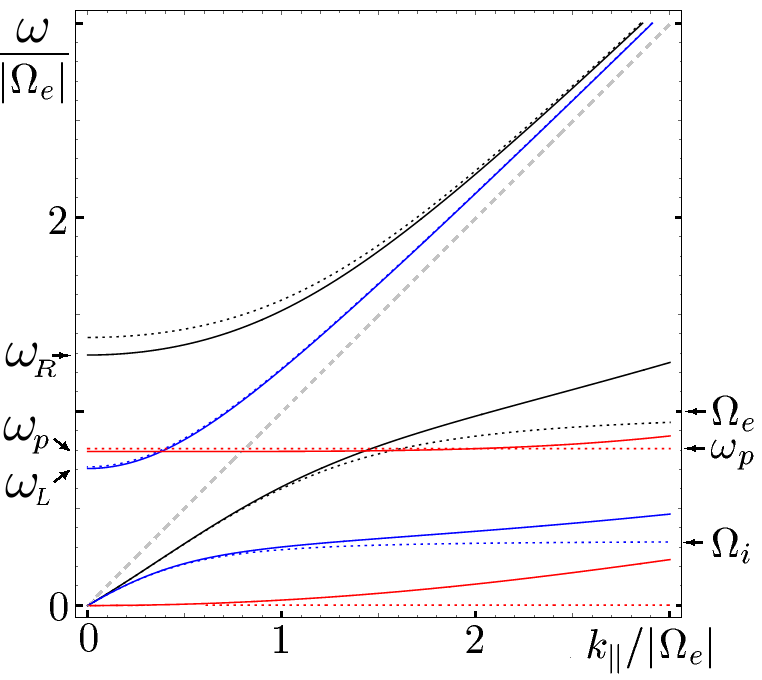}
	\caption{Parallel wave dispersion relations in a cold magnetized plasma. In this geometry, longitudinal plasma modes (red), which include both the Langmuir wave and the acoustic wave, decouple from the transverse modes, which include the right-handed (R, black) and left-handed (L, blue) circularly polarized waves. 
	For various effects to be visible on the same scale, we choose $\omega_{pe}/|\Omega_{e}|=0.7$, $|\Omega_{e}|/m_e=0.1$ and $m_i/m_e=3$ in this example. 
	The classical (dashed) and the QED (solid) dispersion relations differ due to two effects: ground-state mass shift, which affects the cutoff frequencies at $k_\parallel\rightarrow 0$, and relativistic-quantum recoil, which affects resonances when $k_\parallel\rightarrow \infty$.
    Since the R and L waves are modified differently, Faraday rotation is changed by QED effects. [Y. Shi, \textit{et al}. Phys. Rev. A \textbf{94}, 012124 (2016).]
    }
	\label{fig:Para}
\end{figure}

The wave dispersion relation is also simplified for wave propagation parallel to the magnetic field, in which case QED effects give rise to anomalous Faraday rotations \cite{Shi2016effective}. 
In this special geometry, transverse and longitudinal dynamics decouple, and the dispersion relation recovers the familiar form in plasma physics: 
The transverse modes are the right-handed (R) and left-handed (L) circularly polarized EM waves, with $n_\parallel^2=R$ (Fig.~\ref{fig:Para}, black) and $n_\parallel^2=L$ (Fig.~\ref{fig:Para}, blue); The longitudinal modes are the electrostatic plasma waves with $P=0$ (Fig.~\ref{fig:Para}, red). Here, we have used the conventional notation in plasma physics, where $R=S+D$ and $L=S-D$, and the permittivities are related to the plasma response tensor by $\omega^2(S-1)=\hat{\Sigma}^{11}=\hat{\Sigma}^{22}$, $-i\omega^2D=\hat{\Sigma}^{12}=-\hat{\Sigma}^{21}$, and $\omega^2(P-1)=\hat{\Sigma}^{33}$. 
For parallel propagation, QED effects enter through two mechanisms. The first is that magnetization increases the ground mass, so the plasma frequency $\omega_p\rightarrow\omega_{p0}$ is reduced, as we have seen for the perpendicular case. 
The second mechanism is recoil, namely, when a wave interacts with a particle, the momentum of the particle is transiently increased by $\hbar k$. This mechanism is also at play at other propagation angles, but is isolated from other effects for parallel propagation. To see how recoil comes into play, notice that for an electron with parallel momentum $k_\parallel$ its energy is $E_{n,k_\parallel}=\sqrt{m^2+|eB_0|(2n+1)+k_\parallel^2}$. A ground-state electron can resonantly interact with an R-wave photon when the photon frequency is $\omega=E_{1,k_\parallel}-m_0$. The resonance frequency can be written as $\omega_r=\frac{m}{m_0}|\Omega|+\kappa$, where $\kappa=(k_\parallel^2-\omega^2)/2m_0>0$ is the recoil momentum. We see recoil increases the photon energy that is required to excite the particle, and $\omega_r\rightarrow|\Omega|$ recovers the classical cyclotron resonance in the absence of QED effects.
Now, with QED effects included, the classical R-wave permittivity, which is given by $R_{c}=1-(\omega_p^2/\omega^2)/(1-|\Omega|/\omega)$ in a single-species plasma, is replaced by $R=1-(\omega_{p0}^2/\omega^2)(1-\kappa/\omega)/(1-\omega_r/\omega)$. This formula, which is derived rigorously, makes intuitive sense: The denominator gives rise to the expected resonance, and the numerator is such that $R\rightarrow P_c=1-\omega_{p0}^2/\omega^2$ recovers the unmagnetized limit. 
Similarly, a ground-state electron can resonantly interact with the L wave. Notice that resonant interaction not only requires the correct energy, but also the correct angular momentum. Unlike an R-wave photon, which has the correct angular momentum to excite gyrating electrons, the L wave has the opposite angular momentum. To flip the sign, an anti L-wave photon, namely, a negative frequency mode, is required. By replacing $\omega\rightarrow-\omega$, the L-wave permittivity is then $L=1-(\omega_{p0}^2/\omega^2)(1+\kappa/\omega)/(1+\omega_r/\omega)$. 
Finally, a ground-state electron can also resonantly interact with the P waves, which are longitudinally polarized and therefore have zero angular momentum. Since the P waves cannot change electron's perpendicular motion, the resonance condition becomes $\omega=E_{0,k_\parallel}-m_0$, which can be written as $\omega=\kappa$. Using the condition that $P(-\omega)=P(\omega)$ is time even, the rigorous formula $P=1-(\omega_{p0}^2/\omega^2)(1+\kappa/2m_0)/(1-\kappa^2/\omega^2)$ makes intuitive sense: It has the correct pole and recovers $P_c$ in the classical limit. To see why the numerator arises, we can use the condition that the wave cutoff frequency is unchanged by recoil, which vanishes at $k_\parallel=0$ where the wave carries no momentum.
An observable consequence of QED effects is the modification of Faraday rotation. Since the eigenmodes are R and L waves, the polarization angle $\theta$ of a linearly polarized EM wave rotates at the rate $\lambda d\theta/dz=\pi\Delta n$, where $\lambda=2\pi c/\omega$ is the vacuum wavelength and $\Delta n=n_L-n_R$ is the difference in refractive indexes between the L and R waves of the same frequency. In electron-ion plasmas where $m_i\gg m_e$, electron contributions dominate, and $n_{R/L}^2\simeq1-\frac{m\Omega}{\omega^2}-\frac{m\omega_p^2}{2m_0\omega^2}\mp\frac{m_0}{\omega}\pm\sqrt{\Big(\frac{m\Omega}{\omega^2}+\frac{m\omega_p^2}{2m_0\omega^2}\pm\frac{m_0}{\omega}\Big)^2\mp\frac{2m\omega_p^2}{\omega^3}}$. 
Since R and L waves are modified differently, the QED theory predicts a different Faraday rotation rate than the classical theory. 
The difference is more pronounced further away from the light cone, and is observable for typical radio pulsars \cite{Gueroult2019determining}. Moreover, at conditions feasible in laboratory, the difference is as large as $\sim10\%$ when one measures Faraday rotation in a plasma with density \mbox{$n_0\sim 10^{17}\,\text{cm}^{-3}$} and magnetic field \mbox{$B_0\sim 10^8$ G} using a 1-$\mu$m probe laser. The discrepancy becomes even larger when the ratio $|\Omega_e|/\omega_{pe}$ increases, and should be accounted for when using optical probes to diagnose strongly magnetized plasmas \cite{Shi2018laser}.

\subsection{\label{sec:lattice}Lattice QED simulation}
As is the case for most nonlinear theories, the QED plasma model is analytically solvable only in a few special cases, 
and numerical simulations are required to make predictions beyond the perturbative regime. 
For QED, the \textit{ab initio} numerical approach is lattice gauge theory, from which observables can be extracted by computing expectation values of relevant operators using numerical path integration. In usual lattice QED, what is of interest are fluctuations about the vacuum, so the numerical path integral mostly samples field configurations that are close to the classical vacuum $\phi_0=0$ and $\bar{A}_\mu=0$. 
In contrast, when studying plasmas, we are interested in nontrivial field configurations that satisfy the classical field equations: A particular solution to these equations corresponds to a particular situation of interest to plasma physics. 
For example, by specifying $\phi_0=\sqrt{n_0/2m}\exp(imt)$ and $\bar{A}_\mu=0$, one will be studying a uniform unmagnetized plasma with density $n_0$ where all particles are in the ground state. Having specified the classical backgrounds, one can then use numerical path integration to study quantum fluctuations in the background field theory [Eqs.~(\ref{eq:L_phi})-(\ref{eq:L_I})]. However, before we can do that, a nontrivial task is to determine the self-consistent background fields $\phi_0$ and $\bar{A}_\mu$, which usually requires numerically solving the partial differential equations [Eqs.~(\ref{eq:EOM_phi}) and (\ref{eq:EOM_A})] for given initial and boundary conditions.

While the classical gauge field $\bar{A}_\mu(x)$ can usually be regarded as a single-body wave function, the classical scalar field $\phi_0(x)$ is associated with a N-body wave function $\Phi_0$ as discussed earlier. Exactly solving the many-body problem is challenging, and some approximations are usually necessary. For example, one can take the commonly used ansatz that the bosonic many-body wave function is the symmetrized product of single-body wave functions: $\Phi_0(x_1,\dots,x_N)\propto\sum_{\sigma\in S_N}\psi_{\sigma(1)}(x_1)\dots\psi_{\sigma(N)}(x_N)$, where the summation is over the permutation group $S_N$. Similar constructions can be made for fermionic fields using the Slater determinant. Then, for given initial and boundary conditions, one solves for single-body wave functions $\psi_i$ under the influence of the mean field $\bar{A}_\mu$, and self-consistently advance $\bar{A}_\mu$ using the Maxwell's equations, in which the current density $\bar{J}^{\mu}$ is summed over all particles.

To numerically solve the coupled classical field equations, we develop a variational algorithm \cite{Shi2018simulations} based on discrete exterior calculus (DEC), which ensures good conservation properties. To simplify the notation, we denote $\psi_i\rightarrow\phi$ and $\bar{A}\rightarrow A$, with the implied understanding that all fields are single-body wave functions associated with the classical background fields.   
In DEC, the scalar field $\phi$ is a differential 0-form, and hence lives on vertices of the discretized spacetime manifold. In comparison, the gauge field $\bar{A}$, as well as the covariant derivative $D\phi$, are differential 1-forms, and hence live on edges of the spacetime grid. Finally, the field strength tensor $F=dA$ is the exterior derivative of $A$, and lives on faces of the spacetime grid as a differential 2-form. 
With the above discretization scheme, which respects the geometric structure of DEC, the algorithm automatically guarantees $d^2A=0$, namely, the two Maxwell's equations $\nabla\cdot\mathbf{B}=0$ and $\nabla\times\mathbf{E}=-\partial\mathbf{B}/\partial t$.
Moreover, the discretization scheme preserves the U(1)-gauge symmetry $\phi\rightarrow\phi e^{ie\chi}$ and $A_\mu\rightarrow A_\mu+\partial_\mu \chi$, so the discretized equations satisfy exactly local charge conservation.  
To obtain the discretized equations, we use the principle of least action to find field configurations that extremize the discretized action $S_d=\sum_c V_cL_c$. Here, the summation is over all cells of the discretized spacetime manifold, $V_c$ is the volume of the cell, and $L_c$ is the discretized Lagrangian density within each cell, which depends on the discretized $\phi_v$ and $A_e$, where the subscripts $v$ and $e$ denote vertices and edges, respectively.
The condition $\delta S_d/\delta \phi_v=0$ gives the discretized Klein-Gordon equation, which can be used to advance $\phi_v$ in time. Similarly, taking variation of $S_d$ with respect to $A_e$ on space-like edges, we obtain the discretized Amp\`ere's law, which can be used to advance $A_e$ in time. 
These two dynamical equations are coupled, and require self-consistent initial conditions. The condition is satisfied by solving the discretized Gauss's law, which is obtained by taking variation of $S_d$ with respect to $A_e$ on time-like edges. 
Finally, the two dynamical equations have an excess degree of freedom due to the U(1)-gauge symmetry. After gauge fixing, a unique solution for given initial and boundary conditions can then be found, which gives a self-consistent classical field configuration. 
It is noteworthy that the discretized equations satisfy exact local energy-momentum conservation \cite{Xiao2019lattice} in the limit where $\phi$ and $A$ decouple.

\subsection{\label{sec:lattice_example}Lattice simulations of laser-plasma interactions}
Using lattice QED simulations, we can study, for example, laser-plasma interactions \cite{Shi2018simulations}. Compared to PIC codes, which are perhaps the most important workhorse for short-pulse lasers nowadays, lattice simulations treat both collective plasma effects and high-energy QED processes self consistently within a single framework. 
Since PIC codes are designed to capture plasma effects only up to relativistic field strengths \cite{Birdsall1991particle,Birdsall2004plasma,Xiao2013variational,Xiao2015explicit,He2016hamiltonian,Jianyuan2018structure}, additional Monte-Carlo modules need to be turned on in order to capture specific QED processes \cite{Gonoskov15,Arber2015contemporary,Grismaye16,Gaudio19}. 
For example, to study laser induced electron-positron pair production, QED source terms are required. These source terms are usually computed from cross sections that are based on improved local-constant-field approximations (LCFA) \cite{Ritus1985quantum,Reiss1962absorption,Landau1971classical,Di2019improved}. The LCFA requires that the formation lengths of QED processes is much shorter than the characteristic length scales of EM fields. For an optical laser colliding with an electron beam, the condition is violated when the beam energy exceeds $\sim10$ GeV, whereby the Lorentz-boosted laser wavelength becomes comparable to the Compton wavelength. 
As another example, to capture processes that produce energetic photons, such as bremsstrahlung and inverse Compton scattering, PIC codes need to track photons as additional particles. This treatment becomes questionable when the spectrum of energetic photons is not well-separated from the spectrum of plasma fluctuations, in which case PIC codes face the difficult choice of whether to treat photons as subgrid-scale particles or as grid-scale waves.
Additionally, QED-PIC codes have difficulties in ensuring energy-momentum conservation: It is far from obvious how particles should recoil \cite{Cole2018experimental,Poder2018experimental,Shi2019radiation} and how EM fields should redistribute when QED processes occur in the PIC framework. 
These difficulties are overcome in lattice QED simulations by fully resolving all relevant scales. The required resolution, which is at sub-Compton scales, is of course very high and may be prohibitively expensive. However, the \textit{ab initio} approach is what it takes to capture the physics correctly when collective and QED processes have overlapping scales.

\begin{figure}[t]
	\renewcommand{\figurename}{FIG.}
	\includegraphics[angle=0,width=0.75\textwidth]{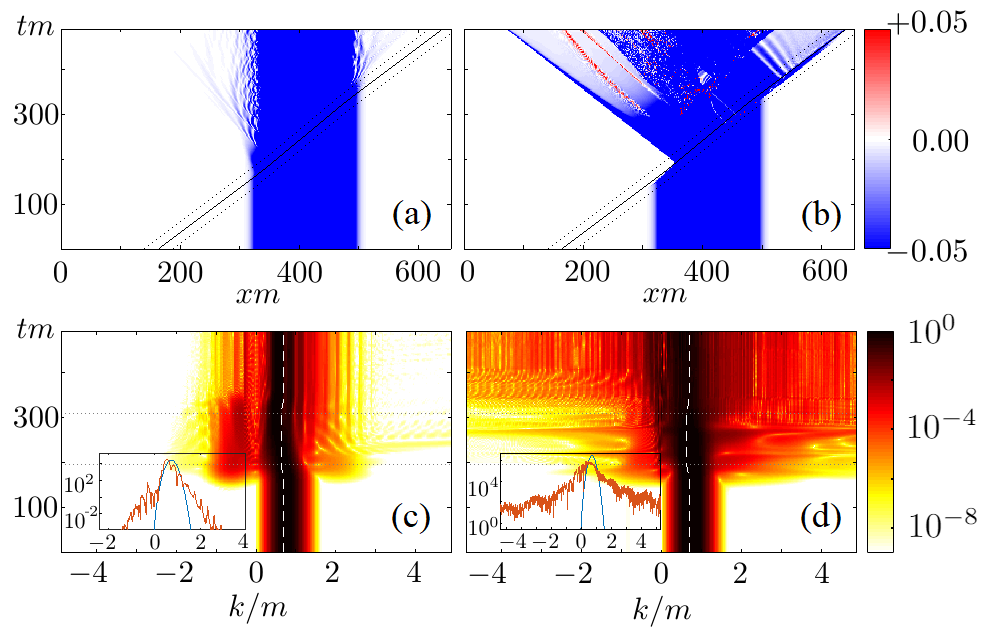}
	\caption{Charged density $\rho_e(x,t)/em^3$ (a, b) and spectral power of transverse fields $S(k,t)/S_\text{max}$ (c, d) in lattice QED simulations, where a plane-wave gamma-ray laser ($\omega=0.7m$) collides with an underdense 1D plasma slab ($\omega_p=0.3m$). 
	At relativistic intensity $a>1$ (a,c), familiar results are recovered: The laser compresses the plasma via pondermotive force, excites plasma waves via parametric processes, and accelerates electrons out of the plasma slab. At the same time, the laser is refracted, its frequency is redshifted due to decompression, and characteristic peaks are generated in the spectrum due to Raman scattering and harmonics generation. 
	At quantum intensity $a>mc^2/\hbar\omega$ (b, d), additional phenomena arise: Electron-positron pairs are produced during the interaction between EM and plasma waves. Notice that in this example, most positrons ($\rho_e>0$) emerge in the laser reflection direction. Additionally, positrons annihilate with electrons to produce gamma photons, and the EM spectrum is substantially broadened. 
	[Adapted from Y. Shi, \textit{et al}. Phys. Rev. E \textbf{97}, 053206 (2018).]}
	\label{fig:PartField}
\end{figure}

Fortunately, instead of requiring full path integrals, we can already learn a lot about laser-plasma interactions by solving the classical field equations. This is because laser and plasma contain a large number of on-shell particles that interact only weakly, so the classical field dynamics dominate quantum fluctuations \cite{Aarts02,Mueller04}. 
As an example, we consider a plane-wave laser colliding with a slab of neutral beam, and carry out 1D simulations in the beam frame \cite{Shi2018simulations}. Suppose the laser frequency is boosted to the gamma-ray range $\omega_0=0.7m$, which is not yet high enough for two photons to produce electron-positron pairs. 
Moreover, we use immobile ions whose only role is to provide a smooth electrostatic potential that confines the electrons, which are chosen to be initially in the ground state. By using a smooth potential well instead of a lattice of spiky Coulomb potentials, we ignore electron-ion collisions and hence subsequent effects such as bremsstrahlung. 
We choose the plasma density $n_0=m^3$, which corresponds to $\omega_p=0.3 m$. Although this ultra-high density is unlikely achievable in experiments, it allows interesting phenomena in 1D simulations that demonstrate the unique capabilities of lattice QED. 
When the laser intensity is relativistic, namely, when the normalized field strength $a=eE/m\omega_0 c>1$, lattice simulations recover well-known phenomena (Fig.~\ref{fig:PartField}a, c): When the laser hits the plasma, the ponderomotive force compresses the plasma, and laser photons are redshifted to $\omega<\omega_0$. Once the laser enters the plasma, it excites plasma waves, from which the laser is both forward and backward scattered, giving rise to characteristic peaks in the spectrum at $\omega+n\omega_p$, where $n$ are integers. In addition to Raman scattering, nonlinear plasma responses generate harmonics at $n\omega$, and a small fraction of electrons are accelerated irreversibly to high energy, leaving the plasma slab together with the laser. 
Beyond recovering well-known phenomena, lattice QED is capable of capturing effects at quantum intensity $a>mc^2/\hbar\omega$ (Fig.~\ref{fig:PartField}b, d): In the high-density plasma, the laser excites electrostatic waves whose amplitudes exceed the Schwinger field $E_c=m^2/e$, wherein electron-positron pairs are produced via the Schwinger mechanism. Additionally, interactions between ultra-intense electrostatic and EM waves also produce and accelerate pairs, causing a large fraction of positrons to leave the plasma slab in the backward direction. In the meantime, many positrons cannot make their way out of the plasma in 1D simulations, and annihilate with electrons to produce gamma photons. By annihilating accelerated electrons and positrons, energetic photons at energy much higher than $mc^2$ are produced, leading to a significantly broadened spectrum after the laser exists the plasma.

\section{\label{sec:conclusion}Concluding remarks}
In the early days of plasma physics, it was not uncommon to encounter researchers, who were trained in other branches of physics, to bring perspectives from the broader physics community and to contribute ideas in return \cite{Bohm1953collective,Landau1957oscillations,Klimontovich1958method,Platzman68}. However, with its core mission to deliver magnetic confinement fusion and to understand astrophysical phenomena, plasma physics was quickly narrowed down to focus on classical electrodynamics and nonequilibrium statistical mechanics. 
This situation was challenged by inertial confinement fusion and low temperature plasma physics, for which quantum mechanics is important. Moreover, with recent advances in HED laboratory drivers and high-intensity lasers, more assumptions made in the conventional plasma physics need to be scrutinized. 
For example, in this paper, we show that effects of magnetization on laser-plasma interactions can no longer be ignored for megagauss-level fields. Additionally, we show that QED effects on laser propagation already become observable in gigagauss fields. Moreover, even without strong static fields, QED processes can be induced by ultra-intense lasers interacting with plasmas, whose description requires ingredients outside the standard framework.
With advances in the field, we startlingly find ourselves return to the old days when plasma physics is more intertwined with other branches of physics. Having developed in relative isolation over many decades, the infusion of cross disciplinary ideas may enable us to answer fundamental scientific questions, such as the nature of the dark matter \cite{Terccas2018axion,Lawson19}, as well as to develop practical technologies, such as next-generation powerful lasers. 
This paper only touches a small fraction of what is possible, and much remains to be done in these exciting research directions.

\begin{acknowledgments}
This work was performed under the auspices of U.S. Department of Energy (DOE) by Lawrence Livermore National Laboratory (LLNL) under Contract DE-AC52-07NA27344, National Nuclear Security Administration Grant No. DE-NA0002948, Air Force Office of Scientific Research Grant No. FA9550-15-1-0391, and DOE Research Grant No. DEAC02-09CH11466. Y.S. is current supported by the Lawrence Fellowship
through LLNL Laboratory Directed Research and Development under Project No. 19-ERD-038.
\end{acknowledgments}


\begin{thebibliography}{101}%
	\makeatletter
	\providecommand \@ifxundefined [1]{%
		\@ifx{#1\undefined}
	}%
	\providecommand \@ifnum [1]{%
		\ifnum #1\expandafter \@firstoftwo
		\else \expandafter \@secondoftwo
		\fi
	}%
	\providecommand \@ifx [1]{%
		\ifx #1\expandafter \@firstoftwo
		\else \expandafter \@secondoftwo
		\fi
	}%
	\providecommand \natexlab [1]{#1}%
	\providecommand \enquote  [1]{``#1''}%
	\providecommand \bibnamefont  [1]{#1}%
	\providecommand \bibfnamefont [1]{#1}%
	\providecommand \citenamefont [1]{#1}%
	\providecommand \href@noop [0]{\@secondoftwo}%
	\providecommand \href [0]{\begingroup \@sanitize@url \@href}%
	\providecommand \@href[1]{\@@startlink{#1}\@@href}%
	\providecommand \@@href[1]{\endgroup#1\@@endlink}%
	\providecommand \@sanitize@url [0]{\catcode `\\12\catcode `\$12\catcode
		`\&12\catcode `\#12\catcode `\^12\catcode `\_12\catcode `\%12\relax}%
	\providecommand \@@startlink[1]{}%
	\providecommand \@@endlink[0]{}%
	\providecommand \url  [0]{\begingroup\@sanitize@url \@url }%
	\providecommand \@url [1]{\endgroup\@href {#1}{\urlprefix }}%
	\providecommand \urlprefix  [0]{URL }%
	\providecommand \Eprint [0]{\href }%
	\providecommand \doibase [0]{https://doi.org/}%
	\providecommand \selectlanguage [0]{\@gobble}%
	\providecommand \bibinfo  [0]{\@secondoftwo}%
	\providecommand \bibfield  [0]{\@secondoftwo}%
	\providecommand \translation [1]{[#1]}%
	\providecommand \BibitemOpen [0]{}%
	\providecommand \bibitemStop [0]{}%
	\providecommand \bibitemNoStop [0]{.\EOS\space}%
	\providecommand \EOS [0]{\spacefactor3000\relax}%
	\providecommand \BibitemShut  [1]{\csname bibitem#1\endcsname}%
	\let\auto@bib@innerbib\@empty
	\bibitem [{\citenamefont {Becker}(2009)}]{Becker2009x}%
	\BibitemOpen
	\bibinfo {editor} {\bibfnamefont {W.}~\bibnamefont {Becker}},\ ed.,\
	\href@noop {} {\emph {\bibinfo {title} {Neutron Stars and Pulsars}}}\
	(\bibinfo  {publisher} {Springer, Berlin, Heidelberg},\ \bibinfo {year}
	{2009})\BibitemShut {NoStop}%
	\bibitem [{\citenamefont {{National Research Council}}(2003)}]{NRC03}%
	\BibitemOpen
	\bibfield  {author} {\bibinfo {author} {\bibnamefont {{National Research
					Council}}},\ }\href {https://doi.org/10.17226/10544} {\emph {\bibinfo {title}
			{Frontiers in High Energy Density Physics: The X-Games of Contemporary
				Science}}}\ (\bibinfo  {publisher} {The National Academies Press},\ \bibinfo
	{address} {Washington, DC},\ \bibinfo {year} {2003})\BibitemShut {NoStop}%
	\bibitem [{\citenamefont {Burnell}(2017)}]{Burnell2017past}%
	\BibitemOpen
	\bibfield  {author} {\bibinfo {author} {\bibfnamefont {J.~B.}\ \bibnamefont
			{Burnell}},\ }\bibfield  {title} {\enquote {\bibinfo {title} {The past,
				present and future of pulsars},}\ }\href@noop {} {\bibfield  {journal}
		{\bibinfo  {journal} {Nature Astronomy}\ }\textbf {\bibinfo {volume} {1}},\
		\bibinfo {pages} {831--834} (\bibinfo {year} {2017})}\BibitemShut {NoStop}%
	\bibitem [{\citenamefont {Jennings}\ \emph {et~al.}(2010)\citenamefont
		{Jennings}, \citenamefont {Cuneo}, \citenamefont {Waisman}, \citenamefont
		{Sinars}, \citenamefont {Ampleford}, \citenamefont {Bennett}, \citenamefont
		{Stygar},\ and\ \citenamefont {Chittenden}}]{Jennings2010simulations}%
	\BibitemOpen
	\bibfield  {author} {\bibinfo {author} {\bibfnamefont {C.~A.}\ \bibnamefont
			{Jennings}}, \bibinfo {author} {\bibfnamefont {M.~E.}\ \bibnamefont {Cuneo}},
		\bibinfo {author} {\bibfnamefont {E.~M.}\ \bibnamefont {Waisman}}, \bibinfo
		{author} {\bibfnamefont {D.~B.}\ \bibnamefont {Sinars}}, \bibinfo {author}
		{\bibfnamefont {D.~J.}\ \bibnamefont {Ampleford}}, \bibinfo {author}
		{\bibfnamefont {G.~R.}\ \bibnamefont {Bennett}}, \bibinfo {author}
		{\bibfnamefont {W.~A.}\ \bibnamefont {Stygar}},\ and\ \bibinfo {author}
		{\bibfnamefont {J.~P.}\ \bibnamefont {Chittenden}},\ }\bibfield  {title}
	{\enquote {\bibinfo {title} {Simulations of the implosion and stagnation of
				compact wire arrays},}\ }\href@noop {} {\bibfield  {journal} {\bibinfo
			{journal} {Phys. Plasmas}\ }\textbf {\bibinfo {volume} {17}},\ \bibinfo
		{pages} {092703} (\bibinfo {year} {2010})}\BibitemShut {NoStop}%
	\bibitem [{\citenamefont {Thornhill}\ \emph {et~al.}(2015)\citenamefont
		{Thornhill}, \citenamefont {Giuliani}, \citenamefont {Jones}, \citenamefont
		{Apruzese}, \citenamefont {Dasgupta}, \citenamefont {Chong}, \citenamefont
		{Harvey-Thompson}, \citenamefont {Ampleford}, \citenamefont {Hansen},
		\citenamefont {Coverdale} \emph {et~al.}}]{Thornhill15}%
	\BibitemOpen
	\bibfield  {author} {\bibinfo {author} {\bibfnamefont {J.~W.}\ \bibnamefont
			{Thornhill}}, \bibinfo {author} {\bibfnamefont {J.~L.}\ \bibnamefont
			{Giuliani}}, \bibinfo {author} {\bibfnamefont {B.}~\bibnamefont {Jones}},
		\bibinfo {author} {\bibfnamefont {J.~P.}\ \bibnamefont {Apruzese}}, \bibinfo
		{author} {\bibfnamefont {A.}~\bibnamefont {Dasgupta}}, \bibinfo {author}
		{\bibfnamefont {Y.~K.}\ \bibnamefont {Chong}}, \bibinfo {author}
		{\bibfnamefont {A.~J.}\ \bibnamefont {Harvey-Thompson}}, \bibinfo {author}
		{\bibfnamefont {D.~J.}\ \bibnamefont {Ampleford}}, \bibinfo {author}
		{\bibfnamefont {S.~B.}\ \bibnamefont {Hansen}}, \bibinfo {author}
		{\bibfnamefont {C.~A.}\ \bibnamefont {Coverdale}}, \emph {et~al.},\
	}\bibfield  {title} {\enquote {\bibinfo {title} {Two-dimensional {RMHD}
				modeling assessment of current flow, plasma conditions, and {Doppler} effects
				in recent {Z} argon experiments},}\ }\href@noop {} {\bibfield  {journal}
		{\bibinfo  {journal} {IEEE Trans. Plasma Sci.}\ }\textbf {\bibinfo {volume}
			{43}},\ \bibinfo {pages} {2480} (\bibinfo {year} {2015})}\BibitemShut
	{NoStop}%
	\bibitem [{\citenamefont {Shi}, \citenamefont {Qin},\ and\ \citenamefont
		{Fisch}(2018)}]{Shi2018laser}%
	\BibitemOpen
	\bibfield  {author} {\bibinfo {author} {\bibfnamefont {Y.}~\bibnamefont
			{Shi}}, \bibinfo {author} {\bibfnamefont {H.}~\bibnamefont {Qin}},\ and\
		\bibinfo {author} {\bibfnamefont {N.~J.}\ \bibnamefont {Fisch}},\ }\bibfield
	{title} {\enquote {\bibinfo {title} {Laser-plasma interactions in magnetized
				environment},}\ }\href@noop {} {\bibfield  {journal} {\bibinfo  {journal}
			{Phys. Plasmas}\ }\textbf {\bibinfo {volume} {25}},\ \bibinfo {pages}
		{055706} (\bibinfo {year} {2018})}\BibitemShut {NoStop}%
	\bibitem [{\citenamefont {Wurden}\ \emph {et~al.}(2016)\citenamefont {Wurden},
		\citenamefont {Hsu}, \citenamefont {Intrator}, \citenamefont {Grabowski},
		\citenamefont {Degnan}, \citenamefont {Domonkos}, \citenamefont {Turchi},
		\citenamefont {Campbell}, \citenamefont {Sinars}, \citenamefont {Herrmann}
		\emph {et~al.}}]{Wurden2016magneto}%
	\BibitemOpen
	\bibfield  {author} {\bibinfo {author} {\bibfnamefont {G.~A.}\ \bibnamefont
			{Wurden}}, \bibinfo {author} {\bibfnamefont {S.~C.}\ \bibnamefont {Hsu}},
		\bibinfo {author} {\bibfnamefont {T.~P.}\ \bibnamefont {Intrator}}, \bibinfo
		{author} {\bibfnamefont {T.~C.}\ \bibnamefont {Grabowski}}, \bibinfo {author}
		{\bibfnamefont {J.~H.}\ \bibnamefont {Degnan}}, \bibinfo {author}
		{\bibfnamefont {M.}~\bibnamefont {Domonkos}}, \bibinfo {author}
		{\bibfnamefont {P.~J.}\ \bibnamefont {Turchi}}, \bibinfo {author}
		{\bibfnamefont {E.~M.}\ \bibnamefont {Campbell}}, \bibinfo {author}
		{\bibfnamefont {D.~B.}\ \bibnamefont {Sinars}}, \bibinfo {author}
		{\bibfnamefont {M.~C.}\ \bibnamefont {Herrmann}}, \emph {et~al.},\ }\bibfield
	{title} {\enquote {\bibinfo {title} {Magneto-inertial fusion},}\ }\href@noop
	{} {\bibfield  {journal} {\bibinfo  {journal} {J. Fusion Energ.}\ }\textbf
		{\bibinfo {volume} {35}},\ \bibinfo {pages} {69--77} (\bibinfo {year}
		{2016})}\BibitemShut {NoStop}%
	\bibitem [{\citenamefont {Bula}\ \emph {et~al.}(1996)\citenamefont {Bula},
		\citenamefont {McDonald}, \citenamefont {Prebys}, \citenamefont {Bamber},
		\citenamefont {Boege}, \citenamefont {Kotseroglou}, \citenamefont
		{Melissinos}, \citenamefont {Meyerhofer}, \citenamefont {Ragg}, \citenamefont
		{Burke}, \citenamefont {Field}, \citenamefont {Horton-Smith}, \citenamefont
		{Odian}, \citenamefont {Spencer}, \citenamefont {Walz}, \citenamefont
		{Berridge}, \citenamefont {Bugg}, \citenamefont {Shmakov},\ and\
		\citenamefont {Weidemann}}]{Bula96}%
	\BibitemOpen
	\bibfield  {author} {\bibinfo {author} {\bibfnamefont {C.}~\bibnamefont
			{Bula}}, \bibinfo {author} {\bibfnamefont {K.~T.}\ \bibnamefont {McDonald}},
		\bibinfo {author} {\bibfnamefont {E.~J.}\ \bibnamefont {Prebys}}, \bibinfo
		{author} {\bibfnamefont {C.}~\bibnamefont {Bamber}}, \bibinfo {author}
		{\bibfnamefont {S.}~\bibnamefont {Boege}}, \bibinfo {author} {\bibfnamefont
			{T.}~\bibnamefont {Kotseroglou}}, \bibinfo {author} {\bibfnamefont {A.~C.}\
			\bibnamefont {Melissinos}}, \bibinfo {author} {\bibfnamefont {D.~D.}\
			\bibnamefont {Meyerhofer}}, \bibinfo {author} {\bibfnamefont
			{W.}~\bibnamefont {Ragg}}, \bibinfo {author} {\bibfnamefont {D.~L.}\
			\bibnamefont {Burke}}, \bibinfo {author} {\bibfnamefont {R.~C.}\ \bibnamefont
			{Field}}, \bibinfo {author} {\bibfnamefont {G.}~\bibnamefont {Horton-Smith}},
		\bibinfo {author} {\bibfnamefont {A.~C.}\ \bibnamefont {Odian}}, \bibinfo
		{author} {\bibfnamefont {J.~E.}\ \bibnamefont {Spencer}}, \bibinfo {author}
		{\bibfnamefont {D.}~\bibnamefont {Walz}}, \bibinfo {author} {\bibfnamefont
			{S.~C.}\ \bibnamefont {Berridge}}, \bibinfo {author} {\bibfnamefont {W.~M.}\
			\bibnamefont {Bugg}}, \bibinfo {author} {\bibfnamefont {K.}~\bibnamefont
			{Shmakov}},\ and\ \bibinfo {author} {\bibfnamefont {A.~W.}\ \bibnamefont
			{Weidemann}},\ }\bibfield  {title} {\enquote {\bibinfo {title} {Observation
				of nonlinear effects in compton scattering},}\ }\href
	{https://doi.org/10.1103/PhysRevLett.76.3116} {\bibfield  {journal} {\bibinfo
			{journal} {Phys. Rev. Lett.}\ }\textbf {\bibinfo {volume} {76}},\ \bibinfo
		{pages} {3116--3119} (\bibinfo {year} {1996})}\BibitemShut {NoStop}%
	\bibitem [{\citenamefont {Baumann}\ \emph {et~al.}(2019)\citenamefont
		{Baumann}, \citenamefont {Nerush}, \citenamefont {Pukhov},\ and\
		\citenamefont {Kostyukov}}]{Baumann2019probing}%
	\BibitemOpen
	\bibfield  {author} {\bibinfo {author} {\bibfnamefont {C.}~\bibnamefont
			{Baumann}}, \bibinfo {author} {\bibfnamefont {E.}~\bibnamefont {Nerush}},
		\bibinfo {author} {\bibfnamefont {A.}~\bibnamefont {Pukhov}},\ and\ \bibinfo
		{author} {\bibfnamefont {I.~Y.}\ \bibnamefont {Kostyukov}},\ }\bibfield
	{title} {\enquote {\bibinfo {title} {Probing non-perturbative {QED} with
				electron-laser collisions},}\ }\href@noop {} {\bibfield  {journal} {\bibinfo
			{journal} {Sci. Rep.}\ }\textbf {\bibinfo {volume} {9}},\ \bibinfo {pages}
		{1--8} (\bibinfo {year} {2019})}\BibitemShut {NoStop}%
	\bibitem [{\citenamefont {Meuren}\ \emph {et~al.}(2020)\citenamefont {Meuren},
		\citenamefont {Bucksbaum}, \citenamefont {Fisch}, \citenamefont {Fi{\'u}za},
		\citenamefont {Glenzer}, \citenamefont {Hogan}, \citenamefont {Qu},
		\citenamefont {Reis}, \citenamefont {White},\ and\ \citenamefont
		{Yakimenko}}]{Meuren2020seminal}%
	\BibitemOpen
	\bibfield  {author} {\bibinfo {author} {\bibfnamefont {S.}~\bibnamefont
			{Meuren}}, \bibinfo {author} {\bibfnamefont {P.~H.}\ \bibnamefont
			{Bucksbaum}}, \bibinfo {author} {\bibfnamefont {N.~J.}\ \bibnamefont
			{Fisch}}, \bibinfo {author} {\bibfnamefont {F.}~\bibnamefont {Fi{\'u}za}},
		\bibinfo {author} {\bibfnamefont {S.}~\bibnamefont {Glenzer}}, \bibinfo
		{author} {\bibfnamefont {M.~J.}\ \bibnamefont {Hogan}}, \bibinfo {author}
		{\bibfnamefont {K.}~\bibnamefont {Qu}}, \bibinfo {author} {\bibfnamefont
			{D.~A.}\ \bibnamefont {Reis}}, \bibinfo {author} {\bibfnamefont
			{G.}~\bibnamefont {White}},\ and\ \bibinfo {author} {\bibfnamefont
			{V.}~\bibnamefont {Yakimenko}},\ }\bibfield  {title} {\enquote {\bibinfo
			{title} {On seminal {HEDP} research opportunities enabled by colocating
				multi-petawatt laser with high-density electron beams},}\ }\href@noop {}
	{\bibfield  {journal} {\bibinfo  {journal} {arXiv:2002.10051}\ } (\bibinfo
		{year} {2020})}\BibitemShut {NoStop}%
	\bibitem [{\citenamefont {Yakimenko}\ \emph {et~al.}(2019)\citenamefont
		{Yakimenko}, \citenamefont {Meuren}, \citenamefont {Del~Gaudio},
		\citenamefont {Baumann}, \citenamefont {Fedotov}, \citenamefont {Fiuza},
		\citenamefont {Grismayer}, \citenamefont {Hogan}, \citenamefont {Pukhov},
		\citenamefont {Silva} \emph {et~al.}}]{Yakimenko2019prospect}%
	\BibitemOpen
	\bibfield  {author} {\bibinfo {author} {\bibfnamefont {V.}~\bibnamefont
			{Yakimenko}}, \bibinfo {author} {\bibfnamefont {S.}~\bibnamefont {Meuren}},
		\bibinfo {author} {\bibfnamefont {F.}~\bibnamefont {Del~Gaudio}}, \bibinfo
		{author} {\bibfnamefont {C.}~\bibnamefont {Baumann}}, \bibinfo {author}
		{\bibfnamefont {A.}~\bibnamefont {Fedotov}}, \bibinfo {author} {\bibfnamefont
			{F.}~\bibnamefont {Fiuza}}, \bibinfo {author} {\bibfnamefont
			{T.}~\bibnamefont {Grismayer}}, \bibinfo {author} {\bibfnamefont {M.~J.}\
			\bibnamefont {Hogan}}, \bibinfo {author} {\bibfnamefont {A.}~\bibnamefont
			{Pukhov}}, \bibinfo {author} {\bibfnamefont {L.~O.}\ \bibnamefont {Silva}},
		\emph {et~al.},\ }\bibfield  {title} {\enquote {\bibinfo {title} {Prospect of
				studying nonperturbative {QED} with beam-beam collisions},}\ }\href@noop {}
	{\bibfield  {journal} {\bibinfo  {journal} {Phys. Rev. Lett.}\ }\textbf
		{\bibinfo {volume} {122}},\ \bibinfo {pages} {190404} (\bibinfo {year}
		{2019})}\BibitemShut {NoStop}%
	\bibitem [{\citenamefont {Shi}(2018)}]{Shi_thesis18}%
	\BibitemOpen
	\bibfield  {author} {\bibinfo {author} {\bibfnamefont {Y.}~\bibnamefont
			{Shi}},\ }\emph {\bibinfo {title} {Plasma physics in strong-field regimes}},\
	\href@noop {} {Ph.D. thesis},\ \bibinfo  {school} {Princeton University}
	(\bibinfo {year} {2018})\BibitemShut {NoStop}%
	\bibitem [{\citenamefont {Davidson}(1972)}]{Davidson2012methods}%
	\BibitemOpen
	\bibfield  {author} {\bibinfo {author} {\bibfnamefont {R.}~\bibnamefont
			{Davidson}},\ }\href@noop {} {\emph {\bibinfo {title} {Methods in nonlinear
				plasma theory}}}\ (\bibinfo  {publisher} {Elsevier, New York},\ \bibinfo
	{year} {1972})\BibitemShut {NoStop}%
	\bibitem [{\citenamefont {Montgomery}(2016)}]{Montgomery2016two}%
	\BibitemOpen
	\bibfield  {author} {\bibinfo {author} {\bibfnamefont {D.~S.}\ \bibnamefont
			{Montgomery}},\ }\bibfield  {title} {\enquote {\bibinfo {title} {Two decades
				of progress in understanding and control of laser plasma instabilities in
				indirect drive inertial fusion},}\ }\href@noop {} {\bibfield  {journal}
		{\bibinfo  {journal} {Phys. Plasmas}\ }\textbf {\bibinfo {volume} {23}},\
		\bibinfo {pages} {055601} (\bibinfo {year} {2016})}\BibitemShut {NoStop}%
	\bibitem [{\citenamefont {Stenflo}(1972)}]{Stenflo1972kinetic}%
	\BibitemOpen
	\bibfield  {author} {\bibinfo {author} {\bibfnamefont {L.}~\bibnamefont
			{Stenflo}},\ }\bibfield  {title} {\enquote {\bibinfo {title} {Kinetic theory
				of three-wave interaction in a magnetized plasma},}\ }\href@noop {}
	{\bibfield  {journal} {\bibinfo  {journal} {J. Plasma Phys.}\ }\textbf
		{\bibinfo {volume} {7}},\ \bibinfo {pages} {107--116} (\bibinfo {year}
		{1972})}\BibitemShut {NoStop}%
	\bibitem [{\citenamefont {Grebogi}\ and\ \citenamefont
		{Liu}(1980)}]{Grebogi1980brillouin}%
	\BibitemOpen
	\bibfield  {author} {\bibinfo {author} {\bibfnamefont {C.}~\bibnamefont
			{Grebogi}}\ and\ \bibinfo {author} {\bibfnamefont {C.}~\bibnamefont {Liu}},\
	}\bibfield  {title} {\enquote {\bibinfo {title} {{Brillouin and Raman}
				scattering of an extraordinary mode in a magnetized plasma},}\ }\href@noop {}
	{\bibfield  {journal} {\bibinfo  {journal} {Phys. Fluids}\ }\textbf {\bibinfo
			{volume} {23}},\ \bibinfo {pages} {1330--1335} (\bibinfo {year}
		{1980})}\BibitemShut {NoStop}%
	\bibitem [{\citenamefont {Barr}\ \emph {et~al.}(1984)\citenamefont {Barr},
		\citenamefont {Boyd}, \citenamefont {Gardner},\ and\ \citenamefont
		{Rankin}}]{Barr1984raman}%
	\BibitemOpen
	\bibfield  {author} {\bibinfo {author} {\bibfnamefont {H.~C.}\ \bibnamefont
			{Barr}}, \bibinfo {author} {\bibfnamefont {T.~J.~M.}\ \bibnamefont {Boyd}},
		\bibinfo {author} {\bibfnamefont {L.~T.}\ \bibnamefont {Gardner}},\ and\
		\bibinfo {author} {\bibfnamefont {R.}~\bibnamefont {Rankin}},\ }\bibfield
	{title} {\enquote {\bibinfo {title} {{Raman} and two-plasmon decay
				instabilities in a magnetized plasma},}\ }\href@noop {} {\bibfield  {journal}
		{\bibinfo  {journal} {Phys. fluids}\ }\textbf {\bibinfo {volume} {27}},\
		\bibinfo {pages} {2730--2737} (\bibinfo {year} {1984})}\BibitemShut {NoStop}%
	\bibitem [{\citenamefont {Wong}\ and\ \citenamefont
		{Goldstein}(1986)}]{Wong1986parametric}%
	\BibitemOpen
	\bibfield  {author} {\bibinfo {author} {\bibfnamefont {H.}~\bibnamefont
			{Wong}}\ and\ \bibinfo {author} {\bibfnamefont {M.}~\bibnamefont
			{Goldstein}},\ }\bibfield  {title} {\enquote {\bibinfo {title} {Parametric
				instabilities of circularly polarized {Alfv{\'e}n} waves including
				dispersion},}\ }\href@noop {} {\bibfield  {journal} {\bibinfo  {journal} {J.
				Geophys. Res. Space}\ }\textbf {\bibinfo {volume} {91}},\ \bibinfo {pages}
		{5617--5628} (\bibinfo {year} {1986})}\BibitemShut {NoStop}%
	\bibitem [{\citenamefont {Vi{\~n}as}\ and\ \citenamefont
		{Goldstein}(1991)}]{Vinas1991parametric}%
	\BibitemOpen
	\bibfield  {author} {\bibinfo {author} {\bibfnamefont {A.~F.}\ \bibnamefont
			{Vi{\~n}as}}\ and\ \bibinfo {author} {\bibfnamefont {M.~L.}\ \bibnamefont
			{Goldstein}},\ }\bibfield  {title} {\enquote {\bibinfo {title} {Parametric
				instabilities of circularly polarized large-amplitude dispersive {Alfv{\'e}n}
				waves: excitation of parallel-propagating electromagnetic daughter waves},}\
	}\href@noop {} {\bibfield  {journal} {\bibinfo  {journal} {J. plasma Phys.}\
		}\textbf {\bibinfo {volume} {46}},\ \bibinfo {pages} {107--127} (\bibinfo
		{year} {1991})}\BibitemShut {NoStop}%
	\bibitem [{\citenamefont {Brodin}\ and\ \citenamefont
		{Stenflo}(2012)}]{Brodin2012three}%
	\BibitemOpen
	\bibfield  {author} {\bibinfo {author} {\bibfnamefont {G.}~\bibnamefont
			{Brodin}}\ and\ \bibinfo {author} {\bibfnamefont {L.}~\bibnamefont
			{Stenflo}},\ }\bibfield  {title} {\enquote {\bibinfo {title} {Three-wave
				coupling coefficients for a magnetized plasma},}\ }\href@noop {} {\bibfield
		{journal} {\bibinfo  {journal} {Phys. Scripta}\ }\textbf {\bibinfo {volume}
			{85}},\ \bibinfo {pages} {035504} (\bibinfo {year} {2012})}\BibitemShut
	{NoStop}%
	\bibitem [{\citenamefont {Shi}, \citenamefont {Qin},\ and\ \citenamefont
		{Fisch}(2017{\natexlab{a}})}]{Shi2017three}%
	\BibitemOpen
	\bibfield  {author} {\bibinfo {author} {\bibfnamefont {Y.}~\bibnamefont
			{Shi}}, \bibinfo {author} {\bibfnamefont {H.}~\bibnamefont {Qin}},\ and\
		\bibinfo {author} {\bibfnamefont {N.~J.}\ \bibnamefont {Fisch}},\ }\bibfield
	{title} {\enquote {\bibinfo {title} {Three-wave scattering in magnetized
				plasmas: From cold fluid to quantized {Lagrangian}},}\ }\href@noop {}
	{\bibfield  {journal} {\bibinfo  {journal} {Phys. Rev. E}\ }\textbf {\bibinfo
			{volume} {96}},\ \bibinfo {pages} {023204} (\bibinfo {year}
		{2017}{\natexlab{a}})}\BibitemShut {NoStop}%
	\bibitem [{\citenamefont {Shi}(2019{\natexlab{a}})}]{Shi2019three}%
	\BibitemOpen
	\bibfield  {author} {\bibinfo {author} {\bibfnamefont {Y.}~\bibnamefont
			{Shi}},\ }\bibfield  {title} {\enquote {\bibinfo {title} {Three-wave
				interactions in magnetized warm-fluid plasmas: General theory with evaluable
				coupling coefficient},}\ }\href@noop {} {\bibfield  {journal} {\bibinfo
			{journal} {Phys. Rev. E}\ }\textbf {\bibinfo {volume} {99}},\ \bibinfo
		{pages} {063212} (\bibinfo {year} {2019}{\natexlab{a}})}\BibitemShut
	{NoStop}%
	\bibitem [{\citenamefont {Shi}\ and\ \citenamefont
		{Fisch}(2019)}]{Shi2019amplification}%
	\BibitemOpen
	\bibfield  {author} {\bibinfo {author} {\bibfnamefont {Y.}~\bibnamefont
			{Shi}}\ and\ \bibinfo {author} {\bibfnamefont {N.~J.}\ \bibnamefont
			{Fisch}},\ }\bibfield  {title} {\enquote {\bibinfo {title} {Amplification of
				mid-infrared lasers via backscattering in magnetized plasmas},}\ }\href@noop
	{} {\bibfield  {journal} {\bibinfo  {journal} {Phys. Plasmas}\ }\textbf
		{\bibinfo {volume} {26}},\ \bibinfo {pages} {072114} (\bibinfo {year}
		{2019})}\BibitemShut {NoStop}%
	\bibitem [{\citenamefont {Loudon}(2000)}]{Loudon2000quantum}%
	\BibitemOpen
	\bibfield  {author} {\bibinfo {author} {\bibfnamefont {R.}~\bibnamefont
			{Loudon}},\ }\href@noop {} {\emph {\bibinfo {title} {The quantum theory of
				light}}}\ (\bibinfo  {publisher} {Oxford University Press},\ \bibinfo {year}
	{2000})\BibitemShut {NoStop}%
	\bibitem [{\citenamefont {Shi}\ \emph {et~al.}(2020)\citenamefont {Shi},
		\citenamefont {Castelli}, \citenamefont {Joseph}, \citenamefont {Geyko},
		\citenamefont {Graziani}, \citenamefont {Libby}, \citenamefont {Parker},
		\citenamefont {Rosen},\ and\ \citenamefont {DuBois}}]{Shi2020quantum}%
	\BibitemOpen
	\bibfield  {author} {\bibinfo {author} {\bibfnamefont {Y.}~\bibnamefont
			{Shi}}, \bibinfo {author} {\bibfnamefont {A.~R.}\ \bibnamefont {Castelli}},
		\bibinfo {author} {\bibfnamefont {I.}~\bibnamefont {Joseph}}, \bibinfo
		{author} {\bibfnamefont {V.}~\bibnamefont {Geyko}}, \bibinfo {author}
		{\bibfnamefont {F.~R.}\ \bibnamefont {Graziani}}, \bibinfo {author}
		{\bibfnamefont {S.~B.}\ \bibnamefont {Libby}}, \bibinfo {author}
		{\bibfnamefont {J.~B.}\ \bibnamefont {Parker}}, \bibinfo {author}
		{\bibfnamefont {Y.~J.}\ \bibnamefont {Rosen}},\ and\ \bibinfo {author}
		{\bibfnamefont {J.~L.}\ \bibnamefont {DuBois}},\ }\bibfield  {title}
	{\enquote {\bibinfo {title} {Quantum computation of three-wave interactions
				with engineered cubic couplings},}\ }\href@noop {} {\bibfield  {journal}
		{\bibinfo  {journal} {arXiv:2004.06885}\ } (\bibinfo {year}
		{2020})}\BibitemShut {NoStop}%
	\bibitem [{\citenamefont {Shi}, \citenamefont {Qin},\ and\ \citenamefont
		{Fisch}(2017{\natexlab{b}})}]{Shi2017laser}%
	\BibitemOpen
	\bibfield  {author} {\bibinfo {author} {\bibfnamefont {Y.}~\bibnamefont
			{Shi}}, \bibinfo {author} {\bibfnamefont {H.}~\bibnamefont {Qin}},\ and\
		\bibinfo {author} {\bibfnamefont {N.~J.}\ \bibnamefont {Fisch}},\ }\bibfield
	{title} {\enquote {\bibinfo {title} {Laser-pulse compression using magnetized
				plasmas},}\ }\href@noop {} {\bibfield  {journal} {\bibinfo  {journal} {Phys.
				Rev. E}\ }\textbf {\bibinfo {volume} {95}},\ \bibinfo {pages} {023211}
		(\bibinfo {year} {2017}{\natexlab{b}})}\BibitemShut {NoStop}%
	\bibitem [{\citenamefont {Malkin}, \citenamefont {Shvets},\ and\ \citenamefont
		{Fisch}(1999)}]{Malkin99}%
	\BibitemOpen
	\bibfield  {author} {\bibinfo {author} {\bibfnamefont {V.~M.}\ \bibnamefont
			{Malkin}}, \bibinfo {author} {\bibfnamefont {G.}~\bibnamefont {Shvets}},\
		and\ \bibinfo {author} {\bibfnamefont {N.~J.}\ \bibnamefont {Fisch}},\
	}\bibfield  {title} {\enquote {\bibinfo {title} {Fast compression of laser
				beams to highly overcritical powers},}\ }\href
	{https://doi.org/10.1103/PhysRevLett.82.4448} {\bibfield  {journal} {\bibinfo
			{journal} {Phys. Rev. Lett.}\ }\textbf {\bibinfo {volume} {82}},\ \bibinfo
		{pages} {4448--4451} (\bibinfo {year} {1999})}\BibitemShut {NoStop}%
	\bibitem [{\citenamefont {Andreev}\ \emph {et~al.}(2006)\citenamefont
		{Andreev}, \citenamefont {Riconda}, \citenamefont {Tikhonchuk},\ and\
		\citenamefont {Weber}}]{Andreev2006short}%
	\BibitemOpen
	\bibfield  {author} {\bibinfo {author} {\bibfnamefont {A.~A.}\ \bibnamefont
			{Andreev}}, \bibinfo {author} {\bibfnamefont {C.}~\bibnamefont {Riconda}},
		\bibinfo {author} {\bibfnamefont {V.~T.}\ \bibnamefont {Tikhonchuk}},\ and\
		\bibinfo {author} {\bibfnamefont {S.}~\bibnamefont {Weber}},\ }\bibfield
	{title} {\enquote {\bibinfo {title} {Short light pulse amplification and
				compression by stimulated {Brillouin} scattering in plasmas in the strong
				coupling regime},}\ }\href@noop {} {\bibfield  {journal} {\bibinfo  {journal}
			{Phys. plasmas}\ }\textbf {\bibinfo {volume} {13}},\ \bibinfo {pages}
		{053110} (\bibinfo {year} {2006})}\BibitemShut {NoStop}%
	\bibitem [{\citenamefont {Mourou}, \citenamefont {Barty},\ and\ \citenamefont
		{Perry}(1998)}]{Mourou1998ultrahigh}%
	\BibitemOpen
	\bibfield  {author} {\bibinfo {author} {\bibfnamefont {G.~A.}\ \bibnamefont
			{Mourou}}, \bibinfo {author} {\bibfnamefont {C.~P.~J.}\ \bibnamefont
			{Barty}},\ and\ \bibinfo {author} {\bibfnamefont {M.~D.}\ \bibnamefont
			{Perry}},\ }\bibfield  {title} {\enquote {\bibinfo {title}
			{Ultrahigh-intensity lasers: Physics of the extreme on a tabletop},}\
	}\href@noop {} {\bibfield  {journal} {\bibinfo  {journal} {Physics Today}\
		}\textbf {\bibinfo {volume} {51}},\ \bibinfo {pages} {22--28} (\bibinfo
		{year} {1998})}\BibitemShut {NoStop}%
	\bibitem [{\citenamefont {Malkin}\ and\ \citenamefont
		{Fisch}(2016)}]{Malkin2016extended}%
	\BibitemOpen
	\bibfield  {author} {\bibinfo {author} {\bibfnamefont {V.~M.}\ \bibnamefont
			{Malkin}}\ and\ \bibinfo {author} {\bibfnamefont {N.~J.}\ \bibnamefont
			{Fisch}},\ }\bibfield  {title} {\enquote {\bibinfo {title} {Extended
				propagation of powerful laser pulses in focusing kerr media},}\ }\href@noop
	{} {\bibfield  {journal} {\bibinfo  {journal} {Phys. Rev. Lett.}\ }\textbf
		{\bibinfo {volume} {117}},\ \bibinfo {pages} {133901} (\bibinfo {year}
		{2016})}\BibitemShut {NoStop}%
	\bibitem [{\citenamefont {Ping}\ \emph {et~al.}(2000)\citenamefont {Ping},
		\citenamefont {Geltner}, \citenamefont {Fisch}, \citenamefont {Shvets},\ and\
		\citenamefont {Suckewer}}]{Ping00}%
	\BibitemOpen
	\bibfield  {author} {\bibinfo {author} {\bibfnamefont {Y.}~\bibnamefont
			{Ping}}, \bibinfo {author} {\bibfnamefont {I.}~\bibnamefont {Geltner}},
		\bibinfo {author} {\bibfnamefont {N.~J.}\ \bibnamefont {Fisch}}, \bibinfo
		{author} {\bibfnamefont {G.}~\bibnamefont {Shvets}},\ and\ \bibinfo {author}
		{\bibfnamefont {S.}~\bibnamefont {Suckewer}},\ }\bibfield  {title} {\enquote
		{\bibinfo {title} {Demonstration of ultrashort laser pulse amplification in
				plasmas by a counterpropagating pumping beam},}\ }\href
	{https://doi.org/10.1103/PhysRevE.62.R4532} {\bibfield  {journal} {\bibinfo
			{journal} {Phys. Rev. E}\ }\textbf {\bibinfo {volume} {62}},\ \bibinfo
		{pages} {R4532--R4535} (\bibinfo {year} {2000})}\BibitemShut {NoStop}%
	\bibitem [{\citenamefont {Ping}\ \emph {et~al.}(2004)\citenamefont {Ping},
		\citenamefont {Cheng}, \citenamefont {Suckewer}, \citenamefont {Clark},\ and\
		\citenamefont {Fisch}}]{Ping2004amplification}%
	\BibitemOpen
	\bibfield  {author} {\bibinfo {author} {\bibfnamefont {Y.}~\bibnamefont
			{Ping}}, \bibinfo {author} {\bibfnamefont {W.}~\bibnamefont {Cheng}},
		\bibinfo {author} {\bibfnamefont {S.}~\bibnamefont {Suckewer}}, \bibinfo
		{author} {\bibfnamefont {D.~S.}\ \bibnamefont {Clark}},\ and\ \bibinfo
		{author} {\bibfnamefont {N.~J.}\ \bibnamefont {Fisch}},\ }\bibfield  {title}
	{\enquote {\bibinfo {title} {Amplification of ultrashort laser pulses by a
				resonant {Raman} scheme in a gas-jet plasma},}\ }\href@noop {} {\bibfield
		{journal} {\bibinfo  {journal} {Phys. Rev. Lett.}\ }\textbf {\bibinfo
			{volume} {92}},\ \bibinfo {pages} {175007} (\bibinfo {year}
		{2004})}\BibitemShut {NoStop}%
	\bibitem [{\citenamefont {Cheng}\ \emph {et~al.}(2005)\citenamefont {Cheng},
		\citenamefont {Avitzour}, \citenamefont {Ping}, \citenamefont {Suckewer},
		\citenamefont {Fisch}, \citenamefont {Hur},\ and\ \citenamefont
		{Wurtele}}]{Cheng2005reaching}%
	\BibitemOpen
	\bibfield  {author} {\bibinfo {author} {\bibfnamefont {W.}~\bibnamefont
			{Cheng}}, \bibinfo {author} {\bibfnamefont {Y.}~\bibnamefont {Avitzour}},
		\bibinfo {author} {\bibfnamefont {Y.}~\bibnamefont {Ping}}, \bibinfo {author}
		{\bibfnamefont {S.}~\bibnamefont {Suckewer}}, \bibinfo {author}
		{\bibfnamefont {N.~J.}\ \bibnamefont {Fisch}}, \bibinfo {author}
		{\bibfnamefont {M.~S.}\ \bibnamefont {Hur}},\ and\ \bibinfo {author}
		{\bibfnamefont {J.~S.}\ \bibnamefont {Wurtele}},\ }\bibfield  {title}
	{\enquote {\bibinfo {title} {Reaching the nonlinear regime of {Raman}
				amplification of ultrashort laser pulses},}\ }\href@noop {} {\bibfield
		{journal} {\bibinfo  {journal} {Phys. Rev. Lett.}\ }\textbf {\bibinfo
			{volume} {94}},\ \bibinfo {pages} {045003} (\bibinfo {year}
		{2005})}\BibitemShut {NoStop}%
	\bibitem [{\citenamefont {Ren}\ \emph {et~al.}(2007)\citenamefont {Ren},
		\citenamefont {Cheng}, \citenamefont {Li},\ and\ \citenamefont
		{Suckewer}}]{Ren2007new}%
	\BibitemOpen
	\bibfield  {author} {\bibinfo {author} {\bibfnamefont {J.}~\bibnamefont
			{Ren}}, \bibinfo {author} {\bibfnamefont {W.}~\bibnamefont {Cheng}}, \bibinfo
		{author} {\bibfnamefont {S.}~\bibnamefont {Li}},\ and\ \bibinfo {author}
		{\bibfnamefont {S.}~\bibnamefont {Suckewer}},\ }\bibfield  {title} {\enquote
		{\bibinfo {title} {A new method for generating ultraintense and ultrashort
				laser pulses},}\ }\href@noop {} {\bibfield  {journal} {\bibinfo  {journal}
			{Nat. Phys.}\ }\textbf {\bibinfo {volume} {3}},\ \bibinfo {pages} {732--736}
		(\bibinfo {year} {2007})}\BibitemShut {NoStop}%
	\bibitem [{\citenamefont {Pai}\ \emph {et~al.}(2008)\citenamefont {Pai},
		\citenamefont {Lin}, \citenamefont {Ha}, \citenamefont {Huang}, \citenamefont
		{Tsou}, \citenamefont {Chu}, \citenamefont {Lin}, \citenamefont {Wang},\ and\
		\citenamefont {Chen}}]{Pai08}%
	\BibitemOpen
	\bibfield  {author} {\bibinfo {author} {\bibfnamefont {C.-H.}\ \bibnamefont
			{Pai}}, \bibinfo {author} {\bibfnamefont {M.-W.}\ \bibnamefont {Lin}},
		\bibinfo {author} {\bibfnamefont {L.-C.}\ \bibnamefont {Ha}}, \bibinfo
		{author} {\bibfnamefont {S.-T.}\ \bibnamefont {Huang}}, \bibinfo {author}
		{\bibfnamefont {Y.-C.}\ \bibnamefont {Tsou}}, \bibinfo {author}
		{\bibfnamefont {H.-H.}\ \bibnamefont {Chu}}, \bibinfo {author} {\bibfnamefont
			{J.-Y.}\ \bibnamefont {Lin}}, \bibinfo {author} {\bibfnamefont
			{J.}~\bibnamefont {Wang}},\ and\ \bibinfo {author} {\bibfnamefont {S.-Y.}\
			\bibnamefont {Chen}},\ }\bibfield  {title} {\enquote {\bibinfo {title}
			{Backward raman amplification in a plasma waveguide},}\ }\href
	{https://doi.org/10.1103/PhysRevLett.101.065005} {\bibfield  {journal}
		{\bibinfo  {journal} {Phys. Rev. Lett.}\ }\textbf {\bibinfo {volume} {101}},\
		\bibinfo {pages} {065005} (\bibinfo {year} {2008})}\BibitemShut {NoStop}%
	\bibitem [{\citenamefont {Ping}\ \emph {et~al.}(2009)\citenamefont {Ping},
		\citenamefont {Kirkwood}, \citenamefont {Wang}, \citenamefont {Clark},
		\citenamefont {Wilks}, \citenamefont {Meezan}, \citenamefont {Berger},
		\citenamefont {Wurtele}, \citenamefont {Fisch}, \citenamefont {Malkin} \emph
		{et~al.}}]{Ping2009development}%
	\BibitemOpen
	\bibfield  {author} {\bibinfo {author} {\bibfnamefont {Y.}~\bibnamefont
			{Ping}}, \bibinfo {author} {\bibfnamefont {R.}~\bibnamefont {Kirkwood}},
		\bibinfo {author} {\bibfnamefont {T.-L.}\ \bibnamefont {Wang}}, \bibinfo
		{author} {\bibfnamefont {D.~S.}\ \bibnamefont {Clark}}, \bibinfo {author}
		{\bibfnamefont {S.~C.}\ \bibnamefont {Wilks}}, \bibinfo {author}
		{\bibfnamefont {N.}~\bibnamefont {Meezan}}, \bibinfo {author} {\bibfnamefont
			{R.~L.}\ \bibnamefont {Berger}}, \bibinfo {author} {\bibfnamefont
			{J.}~\bibnamefont {Wurtele}}, \bibinfo {author} {\bibfnamefont {N.~J.}\
			\bibnamefont {Fisch}}, \bibinfo {author} {\bibfnamefont {V.~M.}\ \bibnamefont
			{Malkin}}, \emph {et~al.},\ }\bibfield  {title} {\enquote {\bibinfo {title}
			{Development of a nanosecond-laser-pumped {Raman} amplifier for short laser
				pulses in plasma},}\ }\href@noop {} {\bibfield  {journal} {\bibinfo
			{journal} {Phys. Plasmas}\ }\textbf {\bibinfo {volume} {16}},\ \bibinfo
		{pages} {123113} (\bibinfo {year} {2009})}\BibitemShut {NoStop}%
	\bibitem [{\citenamefont {Kirkwood}\ \emph {et~al.}(2011)\citenamefont
		{Kirkwood}, \citenamefont {Ping}, \citenamefont {Wilks}, \citenamefont
		{Meezan}, \citenamefont {Michel}, \citenamefont {Williams}, \citenamefont
		{Clark}, \citenamefont {Suter}, \citenamefont {Landen}, \citenamefont {Fisch}
		\emph {et~al.}}]{Kirkwood2011observation}%
	\BibitemOpen
	\bibfield  {author} {\bibinfo {author} {\bibfnamefont {R.~K.}\ \bibnamefont
			{Kirkwood}}, \bibinfo {author} {\bibfnamefont {Y.}~\bibnamefont {Ping}},
		\bibinfo {author} {\bibfnamefont {S.~C.}\ \bibnamefont {Wilks}}, \bibinfo
		{author} {\bibfnamefont {N.}~\bibnamefont {Meezan}}, \bibinfo {author}
		{\bibfnamefont {P.}~\bibnamefont {Michel}}, \bibinfo {author} {\bibfnamefont
			{E.}~\bibnamefont {Williams}}, \bibinfo {author} {\bibfnamefont
			{D.}~\bibnamefont {Clark}}, \bibinfo {author} {\bibfnamefont
			{L.}~\bibnamefont {Suter}}, \bibinfo {author} {\bibfnamefont
			{O.}~\bibnamefont {Landen}}, \bibinfo {author} {\bibfnamefont {N.~J.}\
			\bibnamefont {Fisch}}, \emph {et~al.},\ }\bibfield  {title} {\enquote
		{\bibinfo {title} {Observation of amplification of light by {Langmuir} waves
				and its saturation on the electron kinetic timescale},}\ }\href@noop {}
	{\bibfield  {journal} {\bibinfo  {journal} {J. Plasma Phys.}\ }\textbf
		{\bibinfo {volume} {77}},\ \bibinfo {pages} {521} (\bibinfo {year}
		{2011})}\BibitemShut {NoStop}%
	\bibitem [{\citenamefont {Lancia}\ \emph {et~al.}(2016)\citenamefont {Lancia},
		\citenamefont {Giribono}, \citenamefont {Vassura}, \citenamefont
		{Chiaramello}, \citenamefont {Riconda}, \citenamefont {Weber}, \citenamefont
		{Castan}, \citenamefont {Chatelain}, \citenamefont {Frank}, \citenamefont
		{Gangolf}, \citenamefont {Quinn}, \citenamefont {Fuchs},\ and\ \citenamefont
		{Marqu\`es}}]{Lancia16}%
	\BibitemOpen
	\bibfield  {author} {\bibinfo {author} {\bibfnamefont {L.}~\bibnamefont
			{Lancia}}, \bibinfo {author} {\bibfnamefont {A.}~\bibnamefont {Giribono}},
		\bibinfo {author} {\bibfnamefont {L.}~\bibnamefont {Vassura}}, \bibinfo
		{author} {\bibfnamefont {M.}~\bibnamefont {Chiaramello}}, \bibinfo {author}
		{\bibfnamefont {C.}~\bibnamefont {Riconda}}, \bibinfo {author} {\bibfnamefont
			{S.}~\bibnamefont {Weber}}, \bibinfo {author} {\bibfnamefont
			{A.}~\bibnamefont {Castan}}, \bibinfo {author} {\bibfnamefont
			{A.}~\bibnamefont {Chatelain}}, \bibinfo {author} {\bibfnamefont
			{A.}~\bibnamefont {Frank}}, \bibinfo {author} {\bibfnamefont
			{T.}~\bibnamefont {Gangolf}}, \bibinfo {author} {\bibfnamefont {M.~N.}\
			\bibnamefont {Quinn}}, \bibinfo {author} {\bibfnamefont {J.}~\bibnamefont
			{Fuchs}},\ and\ \bibinfo {author} {\bibfnamefont {J.-R.}\ \bibnamefont
			{Marqu\`es}},\ }\bibfield  {title} {\enquote {\bibinfo {title} {Signatures of
				the self-similar regime of strongly coupled stimulated brillouin scattering
				for efficient short laser pulse amplification},}\ }\href
	{https://doi.org/10.1103/PhysRevLett.116.075001} {\bibfield  {journal}
		{\bibinfo  {journal} {Phys. Rev. Lett.}\ }\textbf {\bibinfo {volume} {116}},\
		\bibinfo {pages} {075001} (\bibinfo {year} {2016})}\BibitemShut {NoStop}%
	\bibitem [{\citenamefont {Vieux}\ \emph {et~al.}(2017)\citenamefont {Vieux},
		\citenamefont {Cipiccia}, \citenamefont {Grant}, \citenamefont {Lemos},
		\citenamefont {Grant}, \citenamefont {Ciocarlan}, \citenamefont {Ersfeld},
		\citenamefont {Hur}, \citenamefont {Lepipas}, \citenamefont {Manahan} \emph
		{et~al.}}]{Vieux2017ultra}%
	\BibitemOpen
	\bibfield  {author} {\bibinfo {author} {\bibfnamefont {G.}~\bibnamefont
			{Vieux}}, \bibinfo {author} {\bibfnamefont {S.}~\bibnamefont {Cipiccia}},
		\bibinfo {author} {\bibfnamefont {D.}~\bibnamefont {Grant}}, \bibinfo
		{author} {\bibfnamefont {N.}~\bibnamefont {Lemos}}, \bibinfo {author}
		{\bibfnamefont {P.}~\bibnamefont {Grant}}, \bibinfo {author} {\bibfnamefont
			{C.}~\bibnamefont {Ciocarlan}}, \bibinfo {author} {\bibfnamefont
			{B.}~\bibnamefont {Ersfeld}}, \bibinfo {author} {\bibfnamefont
			{M.}~\bibnamefont {Hur}}, \bibinfo {author} {\bibfnamefont {P.}~\bibnamefont
			{Lepipas}}, \bibinfo {author} {\bibfnamefont {G.}~\bibnamefont {Manahan}},
		\emph {et~al.},\ }\bibfield  {title} {\enquote {\bibinfo {title} {An
				ultra-high gain and efficient amplifier based on raman amplification in
				plasma},}\ }\href@noop {} {\bibfield  {journal} {\bibinfo  {journal} {Sci.
				Rep.}\ }\textbf {\bibinfo {volume} {7}},\ \bibinfo {pages} {1--10} (\bibinfo
		{year} {2017})}\BibitemShut {NoStop}%
	\bibitem [{\citenamefont {Clark}\ and\ \citenamefont
		{Fisch}(2003)}]{Clark2003operating}%
	\BibitemOpen
	\bibfield  {author} {\bibinfo {author} {\bibfnamefont {D.~S.}\ \bibnamefont
			{Clark}}\ and\ \bibinfo {author} {\bibfnamefont {N.~J.}\ \bibnamefont
			{Fisch}},\ }\bibfield  {title} {\enquote {\bibinfo {title} {Operating regime
				for a backward {Raman} laser amplifier in preformed plasma},}\ }\href@noop {}
	{\bibfield  {journal} {\bibinfo  {journal} {Phys. Plasmas}\ }\textbf
		{\bibinfo {volume} {10}},\ \bibinfo {pages} {3363--3370} (\bibinfo {year}
		{2003})}\BibitemShut {NoStop}%
	\bibitem [{\citenamefont {Trines}\ \emph {et~al.}(2011)\citenamefont {Trines},
		\citenamefont {Fiuza}, \citenamefont {Bingham}, \citenamefont {Fonseca},
		\citenamefont {Silva}, \citenamefont {Cairns},\ and\ \citenamefont
		{Norreys}}]{Trines2011simulations}%
	\BibitemOpen
	\bibfield  {author} {\bibinfo {author} {\bibfnamefont {R.~M. G.~M.}\
			\bibnamefont {Trines}}, \bibinfo {author} {\bibfnamefont {F.}~\bibnamefont
			{Fiuza}}, \bibinfo {author} {\bibfnamefont {R.}~\bibnamefont {Bingham}},
		\bibinfo {author} {\bibfnamefont {R.~A.}\ \bibnamefont {Fonseca}}, \bibinfo
		{author} {\bibfnamefont {L.~O.}\ \bibnamefont {Silva}}, \bibinfo {author}
		{\bibfnamefont {R.~A.}\ \bibnamefont {Cairns}},\ and\ \bibinfo {author}
		{\bibfnamefont {P.~A.}\ \bibnamefont {Norreys}},\ }\bibfield  {title}
	{\enquote {\bibinfo {title} {Simulations of efficient raman amplification
				into the multipetawatt regime},}\ }\href@noop {} {\bibfield  {journal}
		{\bibinfo  {journal} {Nat. Phys.}\ }\textbf {\bibinfo {volume} {7}},\
		\bibinfo {pages} {87--92} (\bibinfo {year} {2011})}\BibitemShut {NoStop}%
	\bibitem [{\citenamefont {Jia}\ \emph {et~al.}(2017)\citenamefont {Jia},
		\citenamefont {Shi}, \citenamefont {Qin},\ and\ \citenamefont
		{Fisch}}]{Jia2017kinetic}%
	\BibitemOpen
	\bibfield  {author} {\bibinfo {author} {\bibfnamefont {Q.}~\bibnamefont
			{Jia}}, \bibinfo {author} {\bibfnamefont {Y.}~\bibnamefont {Shi}}, \bibinfo
		{author} {\bibfnamefont {H.}~\bibnamefont {Qin}},\ and\ \bibinfo {author}
		{\bibfnamefont {N.~J.}\ \bibnamefont {Fisch}},\ }\bibfield  {title} {\enquote
		{\bibinfo {title} {Kinetic simulations of laser parametric amplification in
				magnetized plasmas},}\ }\href@noop {} {\bibfield  {journal} {\bibinfo
			{journal} {Phys. Plasmas}\ }\textbf {\bibinfo {volume} {24}},\ \bibinfo
		{pages} {093103} (\bibinfo {year} {2017})}\BibitemShut {NoStop}%
	\bibitem [{\citenamefont {Li}\ \emph {et~al.}(2020)\citenamefont {Li},
		\citenamefont {Zuo}, \citenamefont {Su},\ and\ \citenamefont
		{Yang}}]{Li2020boosting}%
	\BibitemOpen
	\bibfield  {author} {\bibinfo {author} {\bibfnamefont {Z.}~\bibnamefont
			{Li}}, \bibinfo {author} {\bibfnamefont {Y.}~\bibnamefont {Zuo}}, \bibinfo
		{author} {\bibfnamefont {J.}~\bibnamefont {Su}},\ and\ \bibinfo {author}
		{\bibfnamefont {S.}~\bibnamefont {Yang}},\ }\bibfield  {title} {\enquote
		{\bibinfo {title} {Boosting backward {Raman} amplification performance using
				an external magnetic field},}\ }\href@noop {} {\bibfield  {journal} {\bibinfo
			{journal} {IEEE T. Plasma Sci.}\ } (\bibinfo {year} {2020})}\BibitemShut
	{NoStop}%
	\bibitem [{\citenamefont {Sagdeev}\ and\ \citenamefont
		{Shapiro}(1973)}]{Sagdeev73}%
	\BibitemOpen
	\bibfield  {author} {\bibinfo {author} {\bibfnamefont {R.~Z.}\ \bibnamefont
			{Sagdeev}}\ and\ \bibinfo {author} {\bibfnamefont {V.~D.}\ \bibnamefont
			{Shapiro}},\ }\bibfield  {title} {\enquote {\bibinfo {title} {Influence of
				transverse magnetic field on {Landau} damping},}\ }\href@noop {} {\bibfield
		{journal} {\bibinfo  {journal} {JETP Lett.}\ }\textbf {\bibinfo {volume}
			{17}},\ \bibinfo {pages} {279--282} (\bibinfo {year} {1973})}\BibitemShut
	{NoStop}%
	\bibitem [{\citenamefont {Karney}(1978)}]{Karney78}%
	\BibitemOpen
	\bibfield  {author} {\bibinfo {author} {\bibfnamefont {C.~F.~F.}\
			\bibnamefont {Karney}},\ }\bibfield  {title} {\enquote {\bibinfo {title}
			{Stochastic ion heating by a lower hybrid wave},}\ }\href
	{https://doi.org/http://dx.doi.org/10.1063/1.862406} {\bibfield  {journal}
		{\bibinfo  {journal} {Phys. Fluids}\ }\textbf {\bibinfo {volume} {21}},\
		\bibinfo {pages} {1584--1599} (\bibinfo {year} {1978})}\BibitemShut {NoStop}%
	\bibitem [{\citenamefont {Karney}(1979)}]{Karney79}%
	\BibitemOpen
	\bibfield  {author} {\bibinfo {author} {\bibfnamefont {C.~F.~F.}\
			\bibnamefont {Karney}},\ }\bibfield  {title} {\enquote {\bibinfo {title}
			{Stochastic ion heating by a lower hybrid wave: Ii},}\ }\href@noop {}
	{\bibfield  {journal} {\bibinfo  {journal} {Phys. Fluids}\ }\textbf {\bibinfo
			{volume} {22}},\ \bibinfo {pages} {2188--2209} (\bibinfo {year}
		{1979})}\BibitemShut {NoStop}%
	\bibitem [{\citenamefont {Dawson}\ \emph {et~al.}(1983)\citenamefont {Dawson},
		\citenamefont {Decyk}, \citenamefont {Huff}, \citenamefont {Jechart},
		\citenamefont {Katsouleas}, \citenamefont {Leboeuf}, \citenamefont {Lembege},
		\citenamefont {Martinez}, \citenamefont {Ohsawa},\ and\ \citenamefont
		{Ratliff}}]{Dawson83}%
	\BibitemOpen
	\bibfield  {author} {\bibinfo {author} {\bibfnamefont {J.~M.}\ \bibnamefont
			{Dawson}}, \bibinfo {author} {\bibfnamefont {V.~K.}\ \bibnamefont {Decyk}},
		\bibinfo {author} {\bibfnamefont {R.~W.}\ \bibnamefont {Huff}}, \bibinfo
		{author} {\bibfnamefont {I.}~\bibnamefont {Jechart}}, \bibinfo {author}
		{\bibfnamefont {T.}~\bibnamefont {Katsouleas}}, \bibinfo {author}
		{\bibfnamefont {J.~N.}\ \bibnamefont {Leboeuf}}, \bibinfo {author}
		{\bibfnamefont {B.}~\bibnamefont {Lembege}}, \bibinfo {author} {\bibfnamefont
			{R.~M.}\ \bibnamefont {Martinez}}, \bibinfo {author} {\bibfnamefont
			{Y.}~\bibnamefont {Ohsawa}},\ and\ \bibinfo {author} {\bibfnamefont {S.~T.}\
			\bibnamefont {Ratliff}},\ }\bibfield  {title} {\enquote {\bibinfo {title}
			{Damping of large-amplitude plasma waves propagating perpendicular to the
				magnetic field},}\ }\href {https://doi.org/10.1103/PhysRevLett.50.1455}
	{\bibfield  {journal} {\bibinfo  {journal} {Phys. Rev. Lett.}\ }\textbf
		{\bibinfo {volume} {50}},\ \bibinfo {pages} {1455--1458} (\bibinfo {year}
		{1983})}\BibitemShut {NoStop}%
	\bibitem [{\citenamefont {Winjum}, \citenamefont {Tsung},\ and\ \citenamefont
		{Mori}(2018)}]{Winjum2018mitigation}%
	\BibitemOpen
	\bibfield  {author} {\bibinfo {author} {\bibfnamefont {B.~J.}\ \bibnamefont
			{Winjum}}, \bibinfo {author} {\bibfnamefont {F.~S.}\ \bibnamefont {Tsung}},\
		and\ \bibinfo {author} {\bibfnamefont {W.~B.}\ \bibnamefont {Mori}},\
	}\bibfield  {title} {\enquote {\bibinfo {title} {Mitigation of stimulated
				raman scattering in the kinetic regime by external magnetic fields},}\
	}\href@noop {} {\bibfield  {journal} {\bibinfo  {journal} {Phys. Rev. E}\
		}\textbf {\bibinfo {volume} {98}},\ \bibinfo {pages} {043208} (\bibinfo
		{year} {2018})}\BibitemShut {NoStop}%
	\bibitem [{\citenamefont {Malkin}, \citenamefont {Fisch},\ and\ \citenamefont
		{Wurtele}(2007)}]{Malkin2007compression}%
	\BibitemOpen
	\bibfield  {author} {\bibinfo {author} {\bibfnamefont {V.~M.}\ \bibnamefont
			{Malkin}}, \bibinfo {author} {\bibfnamefont {N.~J.}\ \bibnamefont {Fisch}},\
		and\ \bibinfo {author} {\bibfnamefont {J.~S.}\ \bibnamefont {Wurtele}},\
	}\bibfield  {title} {\enquote {\bibinfo {title} {Compression of powerful
				x-ray pulses to attosecond durations by stimulated {Raman} backscattering in
				plasmas},}\ }\href@noop {} {\bibfield  {journal} {\bibinfo  {journal} {Phys.
				Rev. E}\ }\textbf {\bibinfo {volume} {75}},\ \bibinfo {pages} {026404}
		(\bibinfo {year} {2007})}\BibitemShut {NoStop}%
	\bibitem [{\citenamefont {Malkin}\ and\ \citenamefont
		{Fisch}(2009)}]{Malkin2009quasitransient}%
	\BibitemOpen
	\bibfield  {author} {\bibinfo {author} {\bibfnamefont {V.~M.}\ \bibnamefont
			{Malkin}}\ and\ \bibinfo {author} {\bibfnamefont {N.~J.}\ \bibnamefont
			{Fisch}},\ }\bibfield  {title} {\enquote {\bibinfo {title} {Quasitransient
				regimes of backward {Raman} amplification of intense x-ray pulses},}\
	}\href@noop {} {\bibfield  {journal} {\bibinfo  {journal} {Phys. Rev. E}\
		}\textbf {\bibinfo {volume} {80}},\ \bibinfo {pages} {046409} (\bibinfo
		{year} {2009})}\BibitemShut {NoStop}%
	\bibitem [{\citenamefont {Edwards}, \citenamefont {Mikhailova},\ and\
		\citenamefont {Fisch}(2017)}]{Edwards2017x}%
	\BibitemOpen
	\bibfield  {author} {\bibinfo {author} {\bibfnamefont {M.~R.}\ \bibnamefont
			{Edwards}}, \bibinfo {author} {\bibfnamefont {J.~M.}\ \bibnamefont
			{Mikhailova}},\ and\ \bibinfo {author} {\bibfnamefont {N.~J.}\ \bibnamefont
			{Fisch}},\ }\bibfield  {title} {\enquote {\bibinfo {title} {X-ray
				amplification by stimulated {Brillouin} scattering},}\ }\href@noop {}
	{\bibfield  {journal} {\bibinfo  {journal} {Phys. Rev. E}\ }\textbf {\bibinfo
			{volume} {96}},\ \bibinfo {pages} {023209} (\bibinfo {year}
		{2017})}\BibitemShut {NoStop}%
	\bibitem [{\citenamefont {Edwards}\ \emph {et~al.}(2019)\citenamefont
		{Edwards}, \citenamefont {Shi}, \citenamefont {Mikhailova},\ and\
		\citenamefont {Fisch}}]{Edwards2019laser}%
	\BibitemOpen
	\bibfield  {author} {\bibinfo {author} {\bibfnamefont {M.~R.}\ \bibnamefont
			{Edwards}}, \bibinfo {author} {\bibfnamefont {Y.}~\bibnamefont {Shi}},
		\bibinfo {author} {\bibfnamefont {J.~M.}\ \bibnamefont {Mikhailova}},\ and\
		\bibinfo {author} {\bibfnamefont {N.~J.}\ \bibnamefont {Fisch}},\ }\bibfield
	{title} {\enquote {\bibinfo {title} {Laser amplification in strongly
				magnetized plasma},}\ }\href@noop {} {\bibfield  {journal} {\bibinfo
			{journal} {Phys. Rev. Lett.}\ }\textbf {\bibinfo {volume} {123}},\ \bibinfo
		{pages} {025001} (\bibinfo {year} {2019})}\BibitemShut {NoStop}%
	\bibitem [{\citenamefont {Jia}\ \emph {et~al.}(2016)\citenamefont {Jia},
		\citenamefont {Barth}, \citenamefont {Edwards}, \citenamefont {Mikhailova},\
		and\ \citenamefont {Fisch}}]{Jia2016distinguishing}%
	\BibitemOpen
	\bibfield  {author} {\bibinfo {author} {\bibfnamefont {Q.}~\bibnamefont
			{Jia}}, \bibinfo {author} {\bibfnamefont {I.}~\bibnamefont {Barth}}, \bibinfo
		{author} {\bibfnamefont {M.~R.}\ \bibnamefont {Edwards}}, \bibinfo {author}
		{\bibfnamefont {J.~M.}\ \bibnamefont {Mikhailova}},\ and\ \bibinfo {author}
		{\bibfnamefont {N.~J.}\ \bibnamefont {Fisch}},\ }\bibfield  {title} {\enquote
		{\bibinfo {title} {Distinguishing raman from strongly coupled {Brillouin}
				amplification for short pulses},}\ }\href@noop {} {\bibfield  {journal}
		{\bibinfo  {journal} {Phys. Plasmas}\ }\textbf {\bibinfo {volume} {23}},\
		\bibinfo {pages} {053118} (\bibinfo {year} {2016})}\BibitemShut {NoStop}%
	\bibitem [{\citenamefont {Edwards}\ \emph {et~al.}(2016)\citenamefont
		{Edwards}, \citenamefont {Jia}, \citenamefont {Mikhailova},\ and\
		\citenamefont {Fisch}}]{Edwards2016short}%
	\BibitemOpen
	\bibfield  {author} {\bibinfo {author} {\bibfnamefont {M.~R.}\ \bibnamefont
			{Edwards}}, \bibinfo {author} {\bibfnamefont {Q.}~\bibnamefont {Jia}},
		\bibinfo {author} {\bibfnamefont {J.~M.}\ \bibnamefont {Mikhailova}},\ and\
		\bibinfo {author} {\bibfnamefont {N.~J.}\ \bibnamefont {Fisch}},\ }\bibfield
	{title} {\enquote {\bibinfo {title} {Short-pulse amplification by strongly
				coupled stimulated {Brillouin} scattering},}\ }\href@noop {} {\bibfield
		{journal} {\bibinfo  {journal} {Phys. Plasmas}\ }\textbf {\bibinfo {volume}
			{23}},\ \bibinfo {pages} {083122} (\bibinfo {year} {2016})}\BibitemShut
	{NoStop}%
	\bibitem [{\citenamefont {Wagner}\ \emph {et~al.}(2004)\citenamefont {Wagner},
		\citenamefont {Tatarakis}, \citenamefont {Gopal}, \citenamefont {Beg},
		\citenamefont {Clark}, \citenamefont {Dangor}, \citenamefont {Evans},
		\citenamefont {Haines}, \citenamefont {Mangles}, \citenamefont {Norreys}
		\emph {et~al.}}]{Wagner2004laboratory}%
	\BibitemOpen
	\bibfield  {author} {\bibinfo {author} {\bibfnamefont {U.}~\bibnamefont
			{Wagner}}, \bibinfo {author} {\bibfnamefont {M.}~\bibnamefont {Tatarakis}},
		\bibinfo {author} {\bibfnamefont {A.}~\bibnamefont {Gopal}}, \bibinfo
		{author} {\bibfnamefont {F.~N.}\ \bibnamefont {Beg}}, \bibinfo {author}
		{\bibfnamefont {E.~L.}\ \bibnamefont {Clark}}, \bibinfo {author}
		{\bibfnamefont {A.~E.}\ \bibnamefont {Dangor}}, \bibinfo {author}
		{\bibfnamefont {R.~G.}\ \bibnamefont {Evans}}, \bibinfo {author}
		{\bibfnamefont {M.~G.}\ \bibnamefont {Haines}}, \bibinfo {author}
		{\bibfnamefont {S.~P.~D.}\ \bibnamefont {Mangles}}, \bibinfo {author}
		{\bibfnamefont {P.~A.}\ \bibnamefont {Norreys}}, \emph {et~al.},\ }\bibfield
	{title} {\enquote {\bibinfo {title} {Laboratory measurements of 0.7 {GG}
				magnetic fields generated during high-intensity laser interactions with dense
				plasmas},}\ }\href@noop {} {\bibfield  {journal} {\bibinfo  {journal} {Phys.
				Rev. E}\ }\textbf {\bibinfo {volume} {70}},\ \bibinfo {pages} {026401}
		(\bibinfo {year} {2004})}\BibitemShut {NoStop}%
	\bibitem [{\citenamefont {Fujioka}\ \emph {et~al.}(2013)\citenamefont
		{Fujioka}, \citenamefont {Zhang}, \citenamefont {Ishihara}, \citenamefont
		{Shigemori}, \citenamefont {Hironaka}, \citenamefont {Johzaki}, \citenamefont
		{Sunahara}, \citenamefont {Yamamoto}, \citenamefont {Nakashima},
		\citenamefont {Watanabe} \emph {et~al.}}]{Fujioka2013kilotesla}%
	\BibitemOpen
	\bibfield  {author} {\bibinfo {author} {\bibfnamefont {S.}~\bibnamefont
			{Fujioka}}, \bibinfo {author} {\bibfnamefont {Z.}~\bibnamefont {Zhang}},
		\bibinfo {author} {\bibfnamefont {K.}~\bibnamefont {Ishihara}}, \bibinfo
		{author} {\bibfnamefont {K.}~\bibnamefont {Shigemori}}, \bibinfo {author}
		{\bibfnamefont {Y.}~\bibnamefont {Hironaka}}, \bibinfo {author}
		{\bibfnamefont {T.}~\bibnamefont {Johzaki}}, \bibinfo {author} {\bibfnamefont
			{A.}~\bibnamefont {Sunahara}}, \bibinfo {author} {\bibfnamefont
			{N.}~\bibnamefont {Yamamoto}}, \bibinfo {author} {\bibfnamefont
			{H.}~\bibnamefont {Nakashima}}, \bibinfo {author} {\bibfnamefont
			{T.}~\bibnamefont {Watanabe}}, \emph {et~al.},\ }\bibfield  {title} {\enquote
		{\bibinfo {title} {Kilotesla magnetic field due to a capacitor-coil target
				driven by high power laser},}\ }\href@noop {} {\bibfield  {journal} {\bibinfo
			{journal} {Sci. Rep.}\ }\textbf {\bibinfo {volume} {3}},\ \bibinfo {pages}
		{1170} (\bibinfo {year} {2013})}\BibitemShut {NoStop}%
	\bibitem [{\citenamefont {Chen}\ and\ \citenamefont
		{Xiao}(2019)}]{Chen2019gauge}%
	\BibitemOpen
	\bibfield  {author} {\bibinfo {author} {\bibfnamefont {Q.}~\bibnamefont
			{Chen}}\ and\ \bibinfo {author} {\bibfnamefont {J.}~\bibnamefont {Xiao}},\
	}\bibfield  {title} {\enquote {\bibinfo {title} {Gauge and {Poincar\'e}
				invariant canonical symplectic algorithms for real-time lattice strong-field
				quantum electrodynamics},}\ }\href@noop {} {\bibfield  {journal} {\bibinfo
			{journal} {arXiv:1910.09215}\ } (\bibinfo {year} {2019})}\BibitemShut
	{NoStop}%
	\bibitem [{\citenamefont {Wu}\ and\ \citenamefont
		{Zhang}(2020)}]{Wu2020background}%
	\BibitemOpen
	\bibfield  {author} {\bibinfo {author} {\bibfnamefont {S.}~\bibnamefont
			{Wu}}\ and\ \bibinfo {author} {\bibfnamefont {J.-y.}\ \bibnamefont {Zhang}},\
	}\bibfield  {title} {\enquote {\bibinfo {title} {Background field method in
				thermo field dynamics for wave propagation in unmagnetized spinor {QED}
				plasmas},}\ }\href@noop {} {\bibfield  {journal} {\bibinfo  {journal} {Phys.
				Plasmas}\ }\textbf {\bibinfo {volume} {27}},\ \bibinfo {pages} {112104}
		(\bibinfo {year} {2020})}\BibitemShut {NoStop}%
	\bibitem [{\citenamefont {Rojas}\ and\ \citenamefont
		{Shabad}(1979)}]{Rojas1979polarization}%
	\BibitemOpen
	\bibfield  {author} {\bibinfo {author} {\bibfnamefont {H.~P.}\ \bibnamefont
			{Rojas}}\ and\ \bibinfo {author} {\bibfnamefont {A.~E.}\ \bibnamefont
			{Shabad}},\ }\bibfield  {title} {\enquote {\bibinfo {title} {Polarization of
				relativistic electron and positron gas in a strong magnetic field.
				propagation of electromagnetic waves},}\ }\href@noop {} {\bibfield  {journal}
		{\bibinfo  {journal} {Ann. Phys.}\ }\textbf {\bibinfo {volume} {121}},\
		\bibinfo {pages} {432--455} (\bibinfo {year} {1979})}\BibitemShut {NoStop}%
	\bibitem [{\citenamefont {Inagaki}, \citenamefont {Kimura},\ and\ \citenamefont
		{Murata}(2005)}]{Inagaki2005proper}%
	\BibitemOpen
	\bibfield  {author} {\bibinfo {author} {\bibfnamefont {T.}~\bibnamefont
			{Inagaki}}, \bibinfo {author} {\bibfnamefont {D.}~\bibnamefont {Kimura}},\
		and\ \bibinfo {author} {\bibfnamefont {T.}~\bibnamefont {Murata}},\
	}\bibfield  {title} {\enquote {\bibinfo {title} {Proper-time formalism in a
				constant magnetic field at finite temperature and chemical potential},}\
	}\href@noop {} {\bibfield  {journal} {\bibinfo  {journal} {Int. J. Mod. Phys.
				A}\ }\textbf {\bibinfo {volume} {20}},\ \bibinfo {pages} {4995--5007}
		(\bibinfo {year} {2005})}\BibitemShut {NoStop}%
	\bibitem [{\citenamefont {Bezzerides}\ and\ \citenamefont
		{DuBois}(1972)}]{Bezzerides1972quantum}%
	\BibitemOpen
	\bibfield  {author} {\bibinfo {author} {\bibfnamefont {B.}~\bibnamefont
			{Bezzerides}}\ and\ \bibinfo {author} {\bibfnamefont {D.}~\bibnamefont
			{DuBois}},\ }\bibfield  {title} {\enquote {\bibinfo {title} {Quantum
				electrodynamics of nonthermal relativistic plasmas: Kinetic theory},}\
	}\href@noop {} {\bibfield  {journal} {\bibinfo  {journal} {Ann. Phys.}\
		}\textbf {\bibinfo {volume} {70}},\ \bibinfo {pages} {10--66} (\bibinfo
		{year} {1972})}\BibitemShut {NoStop}%
	\bibitem [{\citenamefont {Melrose}(2012)}]{Melrose2012quantum}%
	\BibitemOpen
	\bibfield  {author} {\bibinfo {author} {\bibfnamefont {D.}~\bibnamefont
			{Melrose}},\ }\href@noop {} {\emph {\bibinfo {title} {Quantum plasmadynamics:
				magnetized plasmas}}},\ Vol.\ \bibinfo {volume} {854}\ (\bibinfo  {publisher}
	{Springer},\ \bibinfo {year} {2012})\BibitemShut {NoStop}%
	\bibitem [{\citenamefont {Brodin}\ and\ \citenamefont
		{Marklund}(2008)}]{Brodin2008quantum}%
	\BibitemOpen
	\bibfield  {author} {\bibinfo {author} {\bibfnamefont {G.}~\bibnamefont
			{Brodin}}\ and\ \bibinfo {author} {\bibfnamefont {M.}~\bibnamefont
			{Marklund}},\ }\bibfield  {title} {\enquote {\bibinfo {title} {Quantum, spin
				and {QED} effects in plasmas},}\ }in\ \href@noop {} {\emph {\bibinfo
			{booktitle} {New Aspects Of Plasma Physics}}}\ (\bibinfo  {publisher} {World
		Scientific},\ \bibinfo {year} {2008})\ pp.\ \bibinfo {pages}
	{26--34}\BibitemShut {NoStop}%
	\bibitem [{\citenamefont {Haas}(2011)}]{Haas2011quantum}%
	\BibitemOpen
	\bibfield  {author} {\bibinfo {author} {\bibfnamefont {F.}~\bibnamefont
			{Haas}},\ }\href@noop {} {\emph {\bibinfo {title} {Quantum plasmas: An
				hydrodynamic approach}}},\ Vol.~\bibinfo {volume} {65}\ (\bibinfo
	{publisher} {Springer-Verlag, New York},\ \bibinfo {year} {2011})\BibitemShut
	{NoStop}%
	\bibitem [{\citenamefont {Shi}, \citenamefont {Fisch},\ and\ \citenamefont
		{Qin}(2016)}]{Shi2016effective}%
	\BibitemOpen
	\bibfield  {author} {\bibinfo {author} {\bibfnamefont {Y.}~\bibnamefont
			{Shi}}, \bibinfo {author} {\bibfnamefont {N.~J.}\ \bibnamefont {Fisch}},\
		and\ \bibinfo {author} {\bibfnamefont {H.}~\bibnamefont {Qin}},\ }\bibfield
	{title} {\enquote {\bibinfo {title} {Effective-action approach to wave
				propagation in scalar {QED} plasmas},}\ }\href@noop {} {\bibfield  {journal}
		{\bibinfo  {journal} {Phys. Rev. A}\ }\textbf {\bibinfo {volume} {94}},\
		\bibinfo {pages} {012124} (\bibinfo {year} {2016})}\BibitemShut {NoStop}%
	\bibitem [{\citenamefont {Shi}(2019{\natexlab{b}})}]{Shi2019force}%
	\BibitemOpen
	\bibfield  {author} {\bibinfo {author} {\bibfnamefont {Y.}~\bibnamefont
			{Shi}},\ }\bibfield  {title} {\enquote {\bibinfo {title} {Force, curvature,
				or mass: disambiguating causes of uniform gravity},}\ }\href@noop {}
	{\bibfield  {journal} {\bibinfo  {journal} {arXiv:1908.02159}\ } (\bibinfo
		{year} {2019}{\natexlab{b}})}\BibitemShut {NoStop}%
	\bibitem [{\citenamefont {Furry}(1951)}]{Furry51}%
	\BibitemOpen
	\bibfield  {author} {\bibinfo {author} {\bibfnamefont {W.~H.}\ \bibnamefont
			{Furry}},\ }\bibfield  {title} {\enquote {\bibinfo {title} {On bound states
				and scattering in positron theory},}\ }\href
	{https://doi.org/10.1103/PhysRev.81.115} {\bibfield  {journal} {\bibinfo
			{journal} {Phys. Rev.}\ }\textbf {\bibinfo {volume} {81}},\ \bibinfo {pages}
		{115--124} (\bibinfo {year} {1951})}\BibitemShut {NoStop}%
	\bibitem [{\citenamefont {Heisenberg}\ and\ \citenamefont
		{Euler}(1936)}]{Heisenberg1936consequences}%
	\BibitemOpen
	\bibfield  {author} {\bibinfo {author} {\bibfnamefont {W.}~\bibnamefont
			{Heisenberg}}\ and\ \bibinfo {author} {\bibfnamefont {H.}~\bibnamefont
			{Euler}},\ }\bibfield  {title} {\enquote {\bibinfo {title} {Consequences of
				{Dirac}'s theory of the positron},}\ }\href@noop {} {\bibfield  {journal}
		{\bibinfo  {journal} {Z. Phys.}\ }\textbf {\bibinfo {volume} {98}},\ \bibinfo
		{pages} {714} (\bibinfo {year} {1936})}\BibitemShut {NoStop}%
	\bibitem [{\citenamefont {Dunne}(2012)}]{Dunne2012heisenberg}%
	\BibitemOpen
	\bibfield  {author} {\bibinfo {author} {\bibfnamefont {G.~V.}\ \bibnamefont
			{Dunne}},\ }\bibfield  {title} {\enquote {\bibinfo {title} {The
				{Heisenberg-Euler} effective action: 75 years on},}\ }\href@noop {}
	{\bibfield  {journal} {\bibinfo  {journal} {Int. J. Mod. Phys. A}\ }\textbf
		{\bibinfo {volume} {27}},\ \bibinfo {pages} {1260004} (\bibinfo {year}
		{2012})}\BibitemShut {NoStop}%
	\bibitem [{\citenamefont {Santangelo}\ \emph {et~al.}(1999)\citenamefont
		{Santangelo}, \citenamefont {Segreto}, \citenamefont {Giarrusso},
		\citenamefont {Fiume}, \citenamefont {Orlandini}, \citenamefont {Parmar},
		\citenamefont {Oosterbroek}, \citenamefont {Bulik}, \citenamefont {Mihara},
		\citenamefont {Campana} \emph {et~al.}}]{Santangelo99}%
	\BibitemOpen
	\bibfield  {author} {\bibinfo {author} {\bibfnamefont {A.}~\bibnamefont
			{Santangelo}}, \bibinfo {author} {\bibfnamefont {A.}~\bibnamefont {Segreto}},
		\bibinfo {author} {\bibfnamefont {S.}~\bibnamefont {Giarrusso}}, \bibinfo
		{author} {\bibfnamefont {D.~D.}\ \bibnamefont {Fiume}}, \bibinfo {author}
		{\bibfnamefont {M.}~\bibnamefont {Orlandini}}, \bibinfo {author}
		{\bibfnamefont {A.~N.}\ \bibnamefont {Parmar}}, \bibinfo {author}
		{\bibfnamefont {T.}~\bibnamefont {Oosterbroek}}, \bibinfo {author}
		{\bibfnamefont {T.}~\bibnamefont {Bulik}}, \bibinfo {author} {\bibfnamefont
			{T.}~\bibnamefont {Mihara}}, \bibinfo {author} {\bibfnamefont
			{S.}~\bibnamefont {Campana}}, \emph {et~al.},\ }\bibfield  {title} {\enquote
		{\bibinfo {title} {A {BEPPOSAX} study of the pulsating transient {X}0115+63:
				The first {X}-ray spectrum with four cyclotron harmonic features},}\
	}\href@noop {} {\bibfield  {journal} {\bibinfo  {journal} {Astrophys. J.
				Lett.}\ }\textbf {\bibinfo {volume} {523}},\ \bibinfo {pages} {L85} (\bibinfo
		{year} {1999})}\BibitemShut {NoStop}%
	\bibitem [{\citenamefont {Heindl}\ \emph {et~al.}(2000)\citenamefont {Heindl},
		\citenamefont {Coburn}, \citenamefont {Gruber}, \citenamefont {Pelling},
		\citenamefont {Rothschild}, \citenamefont {Wilms}, \citenamefont
		{Pottschmidt},\ and\ \citenamefont {Staubert}}]{Heindl2000multiple}%
	\BibitemOpen
	\bibfield  {author} {\bibinfo {author} {\bibfnamefont {W.~A.}\ \bibnamefont
			{Heindl}}, \bibinfo {author} {\bibfnamefont {W.}~\bibnamefont {Coburn}},
		\bibinfo {author} {\bibfnamefont {D.~E.}\ \bibnamefont {Gruber}}, \bibinfo
		{author} {\bibfnamefont {M.}~\bibnamefont {Pelling}}, \bibinfo {author}
		{\bibfnamefont {R.~E.}\ \bibnamefont {Rothschild}}, \bibinfo {author}
		{\bibfnamefont {J.}~\bibnamefont {Wilms}}, \bibinfo {author} {\bibfnamefont
			{K.}~\bibnamefont {Pottschmidt}},\ and\ \bibinfo {author} {\bibfnamefont
			{R.}~\bibnamefont {Staubert}},\ }\bibfield  {title} {\enquote {\bibinfo
			{title} {Multiple cyclotron lines in the spectrum of {4U 0115+ 63}},}\
	}\href@noop {} {\bibfield  {journal} {\bibinfo  {journal} {AIP Conf. Proc.}\
		}\textbf {\bibinfo {volume} {510}},\ \bibinfo {pages} {173--177} (\bibinfo
		{year} {2000})}\BibitemShut {NoStop}%
	\bibitem [{\citenamefont {Pottschmidt}\ \emph {et~al.}(2005)\citenamefont
		{Pottschmidt}, \citenamefont {Kreykenbohm}, \citenamefont {Wilms},
		\citenamefont {Coburn}, \citenamefont {Rothschild}, \citenamefont
		{Kretschmar}, \citenamefont {McBride}, \citenamefont {Suchy},\ and\
		\citenamefont {Staubert}}]{Pottschmidt05}%
	\BibitemOpen
	\bibfield  {author} {\bibinfo {author} {\bibfnamefont {K.}~\bibnamefont
			{Pottschmidt}}, \bibinfo {author} {\bibfnamefont {I.}~\bibnamefont
			{Kreykenbohm}}, \bibinfo {author} {\bibfnamefont {J.}~\bibnamefont {Wilms}},
		\bibinfo {author} {\bibfnamefont {W.}~\bibnamefont {Coburn}}, \bibinfo
		{author} {\bibfnamefont {R.~E.}\ \bibnamefont {Rothschild}}, \bibinfo
		{author} {\bibfnamefont {P.}~\bibnamefont {Kretschmar}}, \bibinfo {author}
		{\bibfnamefont {V.}~\bibnamefont {McBride}}, \bibinfo {author} {\bibfnamefont
			{S.}~\bibnamefont {Suchy}},\ and\ \bibinfo {author} {\bibfnamefont
			{R.}~\bibnamefont {Staubert}},\ }\bibfield  {title} {\enquote {\bibinfo
			{title} {{RXTE} discovery of multiple cyclotron lines during the 2004
				{D}ecember outburst of {V}0332+53},}\ }\href@noop {} {\bibfield  {journal}
		{\bibinfo  {journal} {Astrophys. J. Lett.}\ }\textbf {\bibinfo {volume}
			{634}},\ \bibinfo {pages} {L97} (\bibinfo {year} {2005})}\BibitemShut
	{NoStop}%
	\bibitem [{\citenamefont {Pottschmidt}\ \emph {et~al.}(2012)\citenamefont
		{Pottschmidt}, \citenamefont {Suchy}, \citenamefont {Rivers}, \citenamefont
		{Rothschild}, \citenamefont {Marcu}, \citenamefont {Barrag{\'a}n},
		\citenamefont {K{\"u}hnel}, \citenamefont {F{\"u}rst}, \citenamefont
		{Schwarm}, \citenamefont {Kreykenbohm} \emph {et~al.}}]{Pottschmidt2012}%
	\BibitemOpen
	\bibfield  {author} {\bibinfo {author} {\bibfnamefont {K.}~\bibnamefont
			{Pottschmidt}}, \bibinfo {author} {\bibfnamefont {S.}~\bibnamefont {Suchy}},
		\bibinfo {author} {\bibfnamefont {E.}~\bibnamefont {Rivers}}, \bibinfo
		{author} {\bibfnamefont {R.~E.}\ \bibnamefont {Rothschild}}, \bibinfo
		{author} {\bibfnamefont {D.~M.}\ \bibnamefont {Marcu}}, \bibinfo {author}
		{\bibfnamefont {L.}~\bibnamefont {Barrag{\'a}n}}, \bibinfo {author}
		{\bibfnamefont {M.}~\bibnamefont {K{\"u}hnel}}, \bibinfo {author}
		{\bibfnamefont {F.}~\bibnamefont {F{\"u}rst}}, \bibinfo {author}
		{\bibfnamefont {F.}~\bibnamefont {Schwarm}}, \bibinfo {author} {\bibfnamefont
			{I.}~\bibnamefont {Kreykenbohm}}, \emph {et~al.},\ }\bibfield  {title}
	{\enquote {\bibinfo {title} {A {S}uzaku view of cyclotron line sources and
				candidates},}\ }\href@noop {} {\bibfield  {journal} {\bibinfo  {journal} {AIP
				Conf. Proc.}\ }\textbf {\bibinfo {volume} {1427}},\ \bibinfo {pages} {60}
		(\bibinfo {year} {2012})}\BibitemShut {NoStop}%
	\bibitem [{\citenamefont {Gueroult}\ \emph {et~al.}(2019)\citenamefont
		{Gueroult}, \citenamefont {Shi}, \citenamefont {Rax},\ and\ \citenamefont
		{Fisch}}]{Gueroult2019determining}%
	\BibitemOpen
	\bibfield  {author} {\bibinfo {author} {\bibfnamefont {R.}~\bibnamefont
			{Gueroult}}, \bibinfo {author} {\bibfnamefont {Y.}~\bibnamefont {Shi}},
		\bibinfo {author} {\bibfnamefont {J.-M.}\ \bibnamefont {Rax}},\ and\ \bibinfo
		{author} {\bibfnamefont {N.~J.}\ \bibnamefont {Fisch}},\ }\bibfield  {title}
	{\enquote {\bibinfo {title} {Determining the rotation direction in
				pulsars},}\ }\href@noop {} {\bibfield  {journal} {\bibinfo  {journal} {Nat.
				Commun.}\ }\textbf {\bibinfo {volume} {10}},\ \bibinfo {pages} {1--8}
		(\bibinfo {year} {2019})}\BibitemShut {NoStop}%
	\bibitem [{\citenamefont {Shi}\ \emph {et~al.}(2018)\citenamefont {Shi},
		\citenamefont {Xiao}, \citenamefont {Qin},\ and\ \citenamefont
		{Fisch}}]{Shi2018simulations}%
	\BibitemOpen
	\bibfield  {author} {\bibinfo {author} {\bibfnamefont {Y.}~\bibnamefont
			{Shi}}, \bibinfo {author} {\bibfnamefont {J.}~\bibnamefont {Xiao}}, \bibinfo
		{author} {\bibfnamefont {H.}~\bibnamefont {Qin}},\ and\ \bibinfo {author}
		{\bibfnamefont {N.~J.}\ \bibnamefont {Fisch}},\ }\bibfield  {title} {\enquote
		{\bibinfo {title} {Simulations of relativistic quantum plasmas using
				real-time lattice scalar {QED}},}\ }\href@noop {} {\bibfield  {journal}
		{\bibinfo  {journal} {Phys. Rev. E}\ }\textbf {\bibinfo {volume} {97}},\
		\bibinfo {pages} {053206} (\bibinfo {year} {2018})}\BibitemShut {NoStop}%
	\bibitem [{\citenamefont {Xiao}\ \emph {et~al.}(2019)\citenamefont {Xiao},
		\citenamefont {Qin}, \citenamefont {Shi}, \citenamefont {Liu},\ and\
		\citenamefont {Zhang}}]{Xiao2019lattice}%
	\BibitemOpen
	\bibfield  {author} {\bibinfo {author} {\bibfnamefont {J.}~\bibnamefont
			{Xiao}}, \bibinfo {author} {\bibfnamefont {H.}~\bibnamefont {Qin}}, \bibinfo
		{author} {\bibfnamefont {Y.}~\bibnamefont {Shi}}, \bibinfo {author}
		{\bibfnamefont {J.}~\bibnamefont {Liu}},\ and\ \bibinfo {author}
		{\bibfnamefont {R.}~\bibnamefont {Zhang}},\ }\bibfield  {title} {\enquote
		{\bibinfo {title} {A lattice {Maxwell} system with discrete space--time
				symmetry and local energy--momentum conservation},}\ }\href@noop {}
	{\bibfield  {journal} {\bibinfo  {journal} {Phys. Lett. A}\ }\textbf
		{\bibinfo {volume} {383}},\ \bibinfo {pages} {808--812} (\bibinfo {year}
		{2019})}\BibitemShut {NoStop}%
	\bibitem [{\citenamefont {Birdsall}(1991)}]{Birdsall1991particle}%
	\BibitemOpen
	\bibfield  {author} {\bibinfo {author} {\bibfnamefont {C.~K.}\ \bibnamefont
			{Birdsall}},\ }\bibfield  {title} {\enquote {\bibinfo {title}
			{{Particle-in-cell charged-particle simulations, plus Monte Carlo collisions
					with neutral atoms, PIC-MCC}},}\ }\href@noop {} {\bibfield  {journal}
		{\bibinfo  {journal} {IEEE Trans. Plasma Sci.}\ }\textbf {\bibinfo {volume}
			{19}},\ \bibinfo {pages} {65--85} (\bibinfo {year} {1991})}\BibitemShut
	{NoStop}%
	\bibitem [{\citenamefont {Birdsall}\ and\ \citenamefont
		{Langdon}(2005)}]{Birdsall2004plasma}%
	\BibitemOpen
	\bibfield  {author} {\bibinfo {author} {\bibfnamefont {C.~K.}\ \bibnamefont
			{Birdsall}}\ and\ \bibinfo {author} {\bibfnamefont {A.~B.}\ \bibnamefont
			{Langdon}},\ }\href@noop {} {\emph {\bibinfo {title} {Plasma physics via
				computer simulation}}}\ (\bibinfo  {publisher} {Taylor \& Francis},\ \bibinfo
	{year} {2005})\BibitemShut {NoStop}%
	\bibitem [{\citenamefont {Xiao}\ \emph {et~al.}(2013)\citenamefont {Xiao},
		\citenamefont {Liu}, \citenamefont {Qin},\ and\ \citenamefont
		{Yu}}]{Xiao2013variational}%
	\BibitemOpen
	\bibfield  {author} {\bibinfo {author} {\bibfnamefont {J.}~\bibnamefont
			{Xiao}}, \bibinfo {author} {\bibfnamefont {J.}~\bibnamefont {Liu}}, \bibinfo
		{author} {\bibfnamefont {H.}~\bibnamefont {Qin}},\ and\ \bibinfo {author}
		{\bibfnamefont {Z.}~\bibnamefont {Yu}},\ }\bibfield  {title} {\enquote
		{\bibinfo {title} {A variational multi-symplectic particle-in-cell algorithm
				with smoothing functions for the {Vlasov-Maxwell} system},}\ }\href@noop {}
	{\bibfield  {journal} {\bibinfo  {journal} {Phys. Plasmas}\ }\textbf
		{\bibinfo {volume} {20}},\ \bibinfo {pages} {102517} (\bibinfo {year}
		{2013})}\BibitemShut {NoStop}%
	\bibitem [{\citenamefont {Xiao}\ \emph {et~al.}(2015)\citenamefont {Xiao},
		\citenamefont {Qin}, \citenamefont {Liu}, \citenamefont {He}, \citenamefont
		{Zhang},\ and\ \citenamefont {Sun}}]{Xiao2015explicit}%
	\BibitemOpen
	\bibfield  {author} {\bibinfo {author} {\bibfnamefont {J.}~\bibnamefont
			{Xiao}}, \bibinfo {author} {\bibfnamefont {H.}~\bibnamefont {Qin}}, \bibinfo
		{author} {\bibfnamefont {J.}~\bibnamefont {Liu}}, \bibinfo {author}
		{\bibfnamefont {Y.}~\bibnamefont {He}}, \bibinfo {author} {\bibfnamefont
			{R.}~\bibnamefont {Zhang}},\ and\ \bibinfo {author} {\bibfnamefont
			{Y.}~\bibnamefont {Sun}},\ }\bibfield  {title} {\enquote {\bibinfo {title}
			{Explicit high-order non-canonical symplectic particle-in-cell algorithms for
				{Vlasov-Maxwell} systems},}\ }\href@noop {} {\bibfield  {journal} {\bibinfo
			{journal} {Phys. Plasmas}\ }\textbf {\bibinfo {volume} {22}},\ \bibinfo
		{pages} {112504} (\bibinfo {year} {2015})}\BibitemShut {NoStop}%
	\bibitem [{\citenamefont {He}\ \emph {et~al.}(2016)\citenamefont {He},
		\citenamefont {Sun}, \citenamefont {Qin},\ and\ \citenamefont
		{Liu}}]{He2016hamiltonian}%
	\BibitemOpen
	\bibfield  {author} {\bibinfo {author} {\bibfnamefont {Y.}~\bibnamefont
			{He}}, \bibinfo {author} {\bibfnamefont {Y.}~\bibnamefont {Sun}}, \bibinfo
		{author} {\bibfnamefont {H.}~\bibnamefont {Qin}},\ and\ \bibinfo {author}
		{\bibfnamefont {J.}~\bibnamefont {Liu}},\ }\bibfield  {title} {\enquote
		{\bibinfo {title} {{Hamiltonian particle-in-cell methods for Vlasov-Maxwell
					equations}},}\ }\href@noop {} {\bibfield  {journal} {\bibinfo  {journal}
			{Phys. Plasmas}\ }\textbf {\bibinfo {volume} {23}},\ \bibinfo {pages}
		{092108} (\bibinfo {year} {2016})}\BibitemShut {NoStop}%
	\bibitem [{\citenamefont {Xiao}, \citenamefont {Qin},\ and\ \citenamefont
		{Liu}(2018)}]{Jianyuan2018structure}%
	\BibitemOpen
	\bibfield  {author} {\bibinfo {author} {\bibfnamefont {J.}~\bibnamefont
			{Xiao}}, \bibinfo {author} {\bibfnamefont {H.}~\bibnamefont {Qin}},\ and\
		\bibinfo {author} {\bibfnamefont {J.}~\bibnamefont {Liu}},\ }\bibfield
	{title} {\enquote {\bibinfo {title} {Structure-preserving geometric
				particle-in-cell methods for {Vlasov-Maxwell} systems},}\ }\href@noop {}
	{\bibfield  {journal} {\bibinfo  {journal} {Plasma Sci. Technol.}\ }\textbf
		{\bibinfo {volume} {20}},\ \bibinfo {pages} {110501} (\bibinfo {year}
		{2018})}\BibitemShut {NoStop}%
	\bibitem [{\citenamefont {Gonoskov}\ \emph {et~al.}(2015)\citenamefont
		{Gonoskov}, \citenamefont {Bastrakov}, \citenamefont {Efimenko},
		\citenamefont {Ilderton}, \citenamefont {Marklund}, \citenamefont {Meyerov},
		\citenamefont {Muraviev}, \citenamefont {Sergeev}, \citenamefont {Surmin},\
		and\ \citenamefont {Wallin}}]{Gonoskov15}%
	\BibitemOpen
	\bibfield  {author} {\bibinfo {author} {\bibfnamefont {A.}~\bibnamefont
			{Gonoskov}}, \bibinfo {author} {\bibfnamefont {S.}~\bibnamefont {Bastrakov}},
		\bibinfo {author} {\bibfnamefont {E.}~\bibnamefont {Efimenko}}, \bibinfo
		{author} {\bibfnamefont {A.}~\bibnamefont {Ilderton}}, \bibinfo {author}
		{\bibfnamefont {M.}~\bibnamefont {Marklund}}, \bibinfo {author}
		{\bibfnamefont {I.}~\bibnamefont {Meyerov}}, \bibinfo {author} {\bibfnamefont
			{A.}~\bibnamefont {Muraviev}}, \bibinfo {author} {\bibfnamefont
			{A.}~\bibnamefont {Sergeev}}, \bibinfo {author} {\bibfnamefont
			{I.}~\bibnamefont {Surmin}},\ and\ \bibinfo {author} {\bibfnamefont
			{E.}~\bibnamefont {Wallin}},\ }\bibfield  {title} {\enquote {\bibinfo {title}
			{Extended particle-in-cell schemes for physics in ultrastrong laser fields:
				Review and developments},}\ }\href
	{https://doi.org/10.1103/PhysRevE.92.023305} {\bibfield  {journal} {\bibinfo
			{journal} {Phys. Rev. E}\ }\textbf {\bibinfo {volume} {92}},\ \bibinfo
		{pages} {023305} (\bibinfo {year} {2015})}\BibitemShut {NoStop}%
	\bibitem [{\citenamefont {Arber}\ \emph {et~al.}(2015)\citenamefont {Arber},
		\citenamefont {Bennett}, \citenamefont {Brady}, \citenamefont
		{Lawrence-Douglas}, \citenamefont {Ramsay}, \citenamefont {Sircombe},
		\citenamefont {Gillies}, \citenamefont {Evans}, \citenamefont {Schmitz},
		\citenamefont {Bell} \emph {et~al.}}]{Arber2015contemporary}%
	\BibitemOpen
	\bibfield  {author} {\bibinfo {author} {\bibfnamefont {T.~D.}\ \bibnamefont
			{Arber}}, \bibinfo {author} {\bibfnamefont {K.}~\bibnamefont {Bennett}},
		\bibinfo {author} {\bibfnamefont {C.~S.}\ \bibnamefont {Brady}}, \bibinfo
		{author} {\bibfnamefont {A.}~\bibnamefont {Lawrence-Douglas}}, \bibinfo
		{author} {\bibfnamefont {M.~G.}\ \bibnamefont {Ramsay}}, \bibinfo {author}
		{\bibfnamefont {N.~J.}\ \bibnamefont {Sircombe}}, \bibinfo {author}
		{\bibfnamefont {P.}~\bibnamefont {Gillies}}, \bibinfo {author} {\bibfnamefont
			{R.~G.}\ \bibnamefont {Evans}}, \bibinfo {author} {\bibfnamefont
			{H.}~\bibnamefont {Schmitz}}, \bibinfo {author} {\bibfnamefont {A.~R.}\
			\bibnamefont {Bell}}, \emph {et~al.},\ }\bibfield  {title} {\enquote
		{\bibinfo {title} {Contemporary particle-in-cell approach to laser-plasma
				modelling},}\ }\href@noop {} {\bibfield  {journal} {\bibinfo  {journal}
			{Plasma Phys. Contr. F.}\ }\textbf {\bibinfo {volume} {57}},\ \bibinfo
		{pages} {113001} (\bibinfo {year} {2015})}\BibitemShut {NoStop}%
	\bibitem [{\citenamefont {Grismayer}\ \emph {et~al.}(2016)\citenamefont
		{Grismayer}, \citenamefont {Vranic}, \citenamefont {Martins}, \citenamefont
		{Fonseca},\ and\ \citenamefont {Silva}}]{Grismaye16}%
	\BibitemOpen
	\bibfield  {author} {\bibinfo {author} {\bibfnamefont {T.}~\bibnamefont
			{Grismayer}}, \bibinfo {author} {\bibfnamefont {M.}~\bibnamefont {Vranic}},
		\bibinfo {author} {\bibfnamefont {J.~L.}\ \bibnamefont {Martins}}, \bibinfo
		{author} {\bibfnamefont {R.}~\bibnamefont {Fonseca}},\ and\ \bibinfo {author}
		{\bibfnamefont {L.}~\bibnamefont {Silva}},\ }\bibfield  {title} {\enquote
		{\bibinfo {title} {Laser absorption via quantum electrodynamics cascades in
				counter propagating laser pulses},}\ }\href@noop {} {\bibfield  {journal}
		{\bibinfo  {journal} {Phys. Plasmas}\ }\textbf {\bibinfo {volume} {23}},\
		\bibinfo {pages} {056706} (\bibinfo {year} {2016})}\BibitemShut {NoStop}%
	\bibitem [{\citenamefont {Del~Gaudio}\ \emph {et~al.}(2019)\citenamefont
		{Del~Gaudio}, \citenamefont {Grismayer}, \citenamefont {Fonseca},
		\citenamefont {Mori},\ and\ \citenamefont {Silva}}]{Gaudio19}%
	\BibitemOpen
	\bibfield  {author} {\bibinfo {author} {\bibfnamefont {F.}~\bibnamefont
			{Del~Gaudio}}, \bibinfo {author} {\bibfnamefont {T.}~\bibnamefont
			{Grismayer}}, \bibinfo {author} {\bibfnamefont {R.~A.}\ \bibnamefont
			{Fonseca}}, \bibinfo {author} {\bibfnamefont {W.~B.}\ \bibnamefont {Mori}},\
		and\ \bibinfo {author} {\bibfnamefont {L.~O.}\ \bibnamefont {Silva}},\
	}\bibfield  {title} {\enquote {\bibinfo {title} {Bright $\ensuremath{\gamma}$
				rays source and nonlinear {Breit-Wheeler} pairs in the collision of high
				density particle beams},}\ }\href
	{https://doi.org/10.1103/PhysRevAccelBeams.22.023402} {\bibfield  {journal}
		{\bibinfo  {journal} {Phys. Rev. Accel. Beams}\ }\textbf {\bibinfo {volume}
			{22}},\ \bibinfo {pages} {023402} (\bibinfo {year} {2019})}\BibitemShut
	{NoStop}%
	\bibitem [{\citenamefont {Ritus}(1985)}]{Ritus1985quantum}%
	\BibitemOpen
	\bibfield  {author} {\bibinfo {author} {\bibfnamefont {V.~I.}\ \bibnamefont
			{Ritus}},\ }\bibfield  {title} {\enquote {\bibinfo {title} {Quantum effects
				of the interaction of elementary particles with an intense electromagnetic
				field},}\ }\href@noop {} {\bibfield  {journal} {\bibinfo  {journal} {J. Sov.
				Laser Res.}\ }\textbf {\bibinfo {volume} {6}},\ \bibinfo {pages} {497--617}
		(\bibinfo {year} {1985})}\BibitemShut {NoStop}%
	\bibitem [{\citenamefont {Reiss}(1962)}]{Reiss1962absorption}%
	\BibitemOpen
	\bibfield  {author} {\bibinfo {author} {\bibfnamefont {H.~R.}\ \bibnamefont
			{Reiss}},\ }\bibfield  {title} {\enquote {\bibinfo {title} {Absorption of
				light by light},}\ }\href@noop {} {\bibfield  {journal} {\bibinfo  {journal}
			{J. Math. Phys.}\ }\textbf {\bibinfo {volume} {3}},\ \bibinfo {pages}
		{59--67} (\bibinfo {year} {1962})}\BibitemShut {NoStop}%
	\bibitem [{\citenamefont {Landau}\ and\ \citenamefont
		{Lifshitz}(1971)}]{Landau1971classical}%
	\BibitemOpen
	\bibfield  {author} {\bibinfo {author} {\bibfnamefont {L.~D.}\ \bibnamefont
			{Landau}}\ and\ \bibinfo {author} {\bibfnamefont {E.~M.}\ \bibnamefont
			{Lifshitz}},\ }\bibfield  {title} {\enquote {\bibinfo {title} {The classical
				theory of fields},}\ }\href@noop {} {\  (\bibinfo {year} {1971})}\BibitemShut
	{NoStop}%
	\bibitem [{\citenamefont {Di~Piazza}\ \emph {et~al.}(2019)\citenamefont
		{Di~Piazza}, \citenamefont {Tamburini}, \citenamefont {Meuren},\ and\
		\citenamefont {Keitel}}]{Di2019improved}%
	\BibitemOpen
	\bibfield  {author} {\bibinfo {author} {\bibfnamefont {A.}~\bibnamefont
			{Di~Piazza}}, \bibinfo {author} {\bibfnamefont {M.}~\bibnamefont
			{Tamburini}}, \bibinfo {author} {\bibfnamefont {S.}~\bibnamefont {Meuren}},\
		and\ \bibinfo {author} {\bibfnamefont {C.~H.}\ \bibnamefont {Keitel}},\
	}\bibfield  {title} {\enquote {\bibinfo {title} {Improved
				local-constant-field approximation for strong-field qed codes},}\ }\href@noop
	{} {\bibfield  {journal} {\bibinfo  {journal} {Phys. Rev. A}\ }\textbf
		{\bibinfo {volume} {99}},\ \bibinfo {pages} {022125} (\bibinfo {year}
		{2019})}\BibitemShut {NoStop}%
	\bibitem [{\citenamefont {Cole}\ \emph {et~al.}(2018)\citenamefont {Cole},
		\citenamefont {Behm}, \citenamefont {Gerstmayr}, \citenamefont {Blackburn},
		\citenamefont {Wood}, \citenamefont {Baird}, \citenamefont {Duff},
		\citenamefont {Harvey}, \citenamefont {Ilderton}, \citenamefont {Joglekar}
		\emph {et~al.}}]{Cole2018experimental}%
	\BibitemOpen
	\bibfield  {author} {\bibinfo {author} {\bibfnamefont {J.~M.}\ \bibnamefont
			{Cole}}, \bibinfo {author} {\bibfnamefont {K.~T.}\ \bibnamefont {Behm}},
		\bibinfo {author} {\bibfnamefont {E.}~\bibnamefont {Gerstmayr}}, \bibinfo
		{author} {\bibfnamefont {T.~G.}\ \bibnamefont {Blackburn}}, \bibinfo {author}
		{\bibfnamefont {J.~C.}\ \bibnamefont {Wood}}, \bibinfo {author}
		{\bibfnamefont {C.~D.}\ \bibnamefont {Baird}}, \bibinfo {author}
		{\bibfnamefont {M.~J.}\ \bibnamefont {Duff}}, \bibinfo {author}
		{\bibfnamefont {C.}~\bibnamefont {Harvey}}, \bibinfo {author} {\bibfnamefont
			{A.}~\bibnamefont {Ilderton}}, \bibinfo {author} {\bibfnamefont {A.~S.}\
			\bibnamefont {Joglekar}}, \emph {et~al.},\ }\bibfield  {title} {\enquote
		{\bibinfo {title} {Experimental evidence of radiation reaction in the
				collision of a high-intensity laser pulse with a laser-wakefield accelerated
				electron beam},}\ }\href@noop {} {\bibfield  {journal} {\bibinfo  {journal}
			{Phys. Rev. X}\ }\textbf {\bibinfo {volume} {8}},\ \bibinfo {pages} {011020}
		(\bibinfo {year} {2018})}\BibitemShut {NoStop}%
	\bibitem [{\citenamefont {Poder}\ \emph {et~al.}(2018)\citenamefont {Poder},
		\citenamefont {Tamburini}, \citenamefont {Sarri}, \citenamefont {Di~Piazza},
		\citenamefont {Kuschel}, \citenamefont {Baird}, \citenamefont {Behm},
		\citenamefont {Bohlen}, \citenamefont {Cole}, \citenamefont {Corvan} \emph
		{et~al.}}]{Poder2018experimental}%
	\BibitemOpen
	\bibfield  {author} {\bibinfo {author} {\bibfnamefont {K.}~\bibnamefont
			{Poder}}, \bibinfo {author} {\bibfnamefont {M.}~\bibnamefont {Tamburini}},
		\bibinfo {author} {\bibfnamefont {G.}~\bibnamefont {Sarri}}, \bibinfo
		{author} {\bibfnamefont {A.}~\bibnamefont {Di~Piazza}}, \bibinfo {author}
		{\bibfnamefont {S.}~\bibnamefont {Kuschel}}, \bibinfo {author} {\bibfnamefont
			{C.~D.}\ \bibnamefont {Baird}}, \bibinfo {author} {\bibfnamefont
			{K.}~\bibnamefont {Behm}}, \bibinfo {author} {\bibfnamefont {S.}~\bibnamefont
			{Bohlen}}, \bibinfo {author} {\bibfnamefont {J.~M.}\ \bibnamefont {Cole}},
		\bibinfo {author} {\bibfnamefont {D.~J.}\ \bibnamefont {Corvan}}, \emph
		{et~al.},\ }\bibfield  {title} {\enquote {\bibinfo {title} {Experimental
				signatures of the quantum nature of radiation reaction in the field of an
				ultraintense laser},}\ }\href@noop {} {\bibfield  {journal} {\bibinfo
			{journal} {Phys. Rev. X}\ }\textbf {\bibinfo {volume} {8}},\ \bibinfo {pages}
		{031004} (\bibinfo {year} {2018})}\BibitemShut {NoStop}%
	\bibitem [{\citenamefont {Shi}(2019{\natexlab{c}})}]{Shi2019radiation}%
	\BibitemOpen
	\bibfield  {author} {\bibinfo {author} {\bibfnamefont {Y.}~\bibnamefont
			{Shi}},\ }\bibfield  {title} {\enquote {\bibinfo {title} {Radiation reaction
				of classical hyperbolic oscillator: Experimental signatures},}\ }\href@noop
	{} {\bibfield  {journal} {\bibinfo  {journal} {Ann. Phys.}\ }\textbf
		{\bibinfo {volume} {405}},\ \bibinfo {pages} {130--154} (\bibinfo {year}
		{2019}{\natexlab{c}})}\BibitemShut {NoStop}%
	\bibitem [{\citenamefont {Aarts}\ and\ \citenamefont {Berges}(2002)}]{Aarts02}%
	\BibitemOpen
	\bibfield  {author} {\bibinfo {author} {\bibfnamefont {G.}~\bibnamefont
			{Aarts}}\ and\ \bibinfo {author} {\bibfnamefont {J.}~\bibnamefont {Berges}},\
	}\bibfield  {title} {\enquote {\bibinfo {title} {Classical aspects of quantum
				fields far from equilibrium},}\ }\href
	{https://doi.org/10.1103/PhysRevLett.88.041603} {\bibfield  {journal}
		{\bibinfo  {journal} {Phys. Rev. Lett.}\ }\textbf {\bibinfo {volume} {88}},\
		\bibinfo {pages} {041603} (\bibinfo {year} {2002})}\BibitemShut {NoStop}%
	\bibitem [{\citenamefont {Mueller}\ and\ \citenamefont
		{Son}(2004)}]{Mueller04}%
	\BibitemOpen
	\bibfield  {author} {\bibinfo {author} {\bibfnamefont {A.~H.}\ \bibnamefont
			{Mueller}}\ and\ \bibinfo {author} {\bibfnamefont {D.~T.}\ \bibnamefont
			{Son}},\ }\bibfield  {title} {\enquote {\bibinfo {title} {On the equivalence
				between the {B}oltzmann equation and classical field theory at large
				occupation numbers},}\ }\href@noop {} {\bibfield  {journal} {\bibinfo
			{journal} {Phys. Lett. B}\ }\textbf {\bibinfo {volume} {582}},\ \bibinfo
		{pages} {279} (\bibinfo {year} {2004})}\BibitemShut {NoStop}%
	\bibitem [{\citenamefont {Bohm}\ and\ \citenamefont
		{Pines}(1953)}]{Bohm1953collective}%
	\BibitemOpen
	\bibfield  {author} {\bibinfo {author} {\bibfnamefont {D.}~\bibnamefont
			{Bohm}}\ and\ \bibinfo {author} {\bibfnamefont {D.}~\bibnamefont {Pines}},\
	}\bibfield  {title} {\enquote {\bibinfo {title} {A collective description of
				electron interactions: {III. Coulomb} interactions in a degenerate electron
				gas},}\ }\href@noop {} {\bibfield  {journal} {\bibinfo  {journal} {Phys.
				Rev.}\ }\textbf {\bibinfo {volume} {92}},\ \bibinfo {pages} {609} (\bibinfo
		{year} {1953})}\BibitemShut {NoStop}%
	\bibitem [{\citenamefont {Landau}(1957)}]{Landau1957oscillations}%
	\BibitemOpen
	\bibfield  {author} {\bibinfo {author} {\bibfnamefont {L.}~\bibnamefont
			{Landau}},\ }\bibfield  {title} {\enquote {\bibinfo {title} {Oscillations in
				a {Fermi} liquid},}\ }\href@noop {} {\bibfield  {journal} {\bibinfo
			{journal} {Sov. Phys. JETP}\ }\textbf {\bibinfo {volume} {5}},\ \bibinfo
		{pages} {101--108} (\bibinfo {year} {1957})}\BibitemShut {NoStop}%
	\bibitem [{\citenamefont {Klimontovich}(1958)}]{Klimontovich1958method}%
	\BibitemOpen
	\bibfield  {author} {\bibinfo {author} {\bibfnamefont {Y.~L.}\ \bibnamefont
			{Klimontovich}},\ }\bibfield  {title} {\enquote {\bibinfo {title} {On the
				method of â€˜second quantizationâ€™ in phase space},}\ }\href@noop {}
	{\bibfield  {journal} {\bibinfo  {journal} {Sov. Phys. JETP}\ }\textbf
		{\bibinfo {volume} {6}},\ \bibinfo {pages} {753} (\bibinfo {year}
		{1958})}\BibitemShut {NoStop}%
	\bibitem [{\citenamefont {Platzman}, \citenamefont {Wolff},\ and\ \citenamefont
		{Tzoar}(1968)}]{Platzman68}%
	\BibitemOpen
	\bibfield  {author} {\bibinfo {author} {\bibfnamefont {P.~M.}\ \bibnamefont
			{Platzman}}, \bibinfo {author} {\bibfnamefont {P.~A.}\ \bibnamefont
			{Wolff}},\ and\ \bibinfo {author} {\bibfnamefont {N.}~\bibnamefont {Tzoar}},\
	}\bibfield  {title} {\enquote {\bibinfo {title} {Light scattering from a
				plasma in a magnetic field},}\ }\href
	{https://doi.org/10.1103/PhysRev.174.489} {\bibfield  {journal} {\bibinfo
			{journal} {Phys. Rev.}\ }\textbf {\bibinfo {volume} {174}},\ \bibinfo {pages}
		{489--494} (\bibinfo {year} {1968})}\BibitemShut {NoStop}%
	\bibitem [{\citenamefont {Ter{\c{c}}as}, \citenamefont {Rodrigues},\ and\
		\citenamefont {Mendon{\c{c}}a}(2018)}]{Terccas2018axion}%
	\BibitemOpen
	\bibfield  {author} {\bibinfo {author} {\bibfnamefont {H.}~\bibnamefont
			{Ter{\c{c}}as}}, \bibinfo {author} {\bibfnamefont {J.~D.}\ \bibnamefont
			{Rodrigues}},\ and\ \bibinfo {author} {\bibfnamefont {J.~T.}\ \bibnamefont
			{Mendon{\c{c}}a}},\ }\bibfield  {title} {\enquote {\bibinfo {title}
			{Axion-plasmon polaritons in strongly magnetized plasmas},}\ }\href@noop {}
	{\bibfield  {journal} {\bibinfo  {journal} {Phys. Rev. Lett.}\ }\textbf
		{\bibinfo {volume} {120}},\ \bibinfo {pages} {181803} (\bibinfo {year}
		{2018})}\BibitemShut {NoStop}%
	\bibitem [{\citenamefont {Lawson}\ \emph {et~al.}(2019)\citenamefont {Lawson},
		\citenamefont {Millar}, \citenamefont {Pancaldi}, \citenamefont
		{Vitagliano},\ and\ \citenamefont {Wilczek}}]{Lawson19}%
	\BibitemOpen
	\bibfield  {author} {\bibinfo {author} {\bibfnamefont {M.}~\bibnamefont
			{Lawson}}, \bibinfo {author} {\bibfnamefont {A.~J.}\ \bibnamefont {Millar}},
		\bibinfo {author} {\bibfnamefont {M.}~\bibnamefont {Pancaldi}}, \bibinfo
		{author} {\bibfnamefont {E.}~\bibnamefont {Vitagliano}},\ and\ \bibinfo
		{author} {\bibfnamefont {F.}~\bibnamefont {Wilczek}},\ }\bibfield  {title}
	{\enquote {\bibinfo {title} {Tunable axion plasma haloscopes},}\ }\href
	{https://doi.org/10.1103/PhysRevLett.123.141802} {\bibfield  {journal}
		{\bibinfo  {journal} {Phys. Rev. Lett.}\ }\textbf {\bibinfo {volume} {123}},\
		\bibinfo {pages} {141802} (\bibinfo {year} {2019})}\BibitemShut {NoStop}%
\end{thebibliography}
\end{document}